\pdfoutput=1
\documentclass[aps,amsmath,amssymb,twocolumn,showpacs,pre,nofootinbib]{revtex4}
\usepackage{times}
\usepackage{graphicx,color}

\newlength{\figwidth}
\newcommand{\fig}[2]{\includegraphics[width=#1\figwidth]{#2}}

\newlength{\bilderlength}
\newcommand{\bilderscale}{0.35}
\newcommand{\storebilderscale}{\bilderscale}
\newcommand{\bilderskip}{\hspace*{0.8ex}}
\newcommand{\textdiagram}[1]{\renewcommand{\bilderscale}{0.25}\diagram{#1}\renewcommand{\bilderscale}{\storebilderscale}}
\newcommand{\diagram}[1]{\settowidth{\bilderlength}{\bilderskip\includegraphics[scale=\bilderscale]{#1}\bilderskip}\parbox{\bilderlength}{\bilderskip\includegraphics[scale=\bilderscale]{#1}\bilderskip}}

\graphicspath{{./}{./figures/}}

\newcommand{\be}{\begin{equation}}
\newcommand{\ee}{\end{equation}}
\newcommand{\beq}{\begin{equation}}
\newcommand{\eeq}{\end{equation}}
\newcommand{\bea}{\begin{eqnarray}}
\newcommand{\eea}{\end{eqnarray}}

\newcommand{\nn}{\nonumber }
\newcommand{\Tr}{{\rm Tr}}

\newcommand{\rme}{\mathrm{e}}
\newcommand{\rmd}{\mathrm{d}}
\newcommand{\tr}{\mathrm{tr}}

\begin{document}

\title{
First-principle derivation of static avalanche-size distribution
\parbox{0mm}{\raisebox{8mm}[0mm][0mm]{\hspace*{-1.8cm}\normalsize \mbox{LPTENS-11/17, NSF-KITP-11-055}}}}
\author{\bf Pierre Le Doussal and Kay J\"org Wiese} \affiliation{CNRS-Laboratoire
de Physique Th{\'e}orique de l'Ecole Normale Sup{\'e}rieure, 24 rue
Lhomond,75005 Paris, France} 
\date{November 14, 2011}
\begin{abstract}
We study the energy minimization problem for an elastic interface in
a random potential plus a quadratic well. As the position of the
well is varied, the ground state undergoes jumps, called shocks or
static avalanches. We introduce an efficient and systematic method to compute the statistics 
of avalanche sizes and manifold displacements. The tree-level calculation, i.e.\ mean-field
limit, is obtained by solving a  saddle-point  equation. Graphically, it can be interpreted as a the sum  of all tree graphs. The
1-loop corrections are computed using results from the functional
renormalization group. At the upper critical dimension
the shock statistics is  described by the Brownian Force model (BFM),
the static version of the so-called ABBM model in the non-equilibrium context of 
depinning. This model can itself be treated exactly in any dimension and 
its shock statistics is that of a Levy process. Contact is
made with classical results in probability theory on the Burgers equation with 
Brownian initial conditions. In particular we obtain a functional extension of an
evolution equation introduced  by Carraro and Duchon, which 
recursively constructs the tree diagrams in the field theory.
\end{abstract}
\maketitle

\section{Introduction: Model and Method} 
Complex systems, as well as systems with quenched disorder,
often respond non-smoothly, with jumps or avalanches, to a change in external parameters, as e.g.\ an
applied field.  This is seen as Barkhausen noise in magnets
\cite{UrbachMadisonMarkert1995}, earthquakes in the
motion of tectonic plates
\cite{FisherDahmenRamanathanBen-Zion1997,DSFisher1998}, wetting of a
disordered substrate
\cite{MoulinetGuthmannRolley2002,MoulinetRossoKrauthRolley2004},  dry
friction \cite{CuleHwa1998}, cracks in brittle material \cite{BonamySantucciPonson2008,SchmittbuhlMaloy1997}, 
vortices in superconductors \cite{BlatterFeigelmanGeshkenbeinLarkinVinokur1994,LeDoussal2010Book}
and many more. 

Quite generally, these systems can be modeled by an elastic interface
pinned by disorder
\cite{DSFisher1986, BlatterFeigelmanGeshkenbeinLarkinVinokur1994,WieseLeDoussal2006, NarayanDSFisher1993a,
NattermannStepanowTangLeschhorn1992, DSFisher1998,ChauveLeDoussalWiese2000a,LeDoussalWieseChauve2002,LeDoussalWieseChauve2003,KoltonRossoGiamarchi2005,LeDoussal2010Book,DobrinevskiLeDoussalWiese2011}.
In a previous work \cite{LeDoussalMiddletonWiese2008,LeDoussalWiese2008c} we have obtained the probability distribution of the sizes of static avalanches
for an elastic interface in a random pinning potential. The interface is parameterized by a one-component displacement field $u(x)$, where $x$ is the $d$-dimensional internal coordinate. The interface is submitted to an additional quadratic well centered at $w$
(e.g.\ a spring acting on it) and the ground state of the interface, $u(x;w)$ experiences discontinuous jumps as the 
center-of-well position $w$ is varied. These static avalanches, also called shocks because of interesting connections
with the Burgers equation in the limit $d=0$ \cite{Burgers74,BecKhanin2007,BalentsBouchaudMezard1996,LeDoussal2006b,LeDoussal2008,LeDoussalMuellerWiese2010,LeDoussalMuellerWiese2011}, are characterized by their size, 
i.e.\ $S:= \int_x u(x;w)$ where $\int_x \equiv \int \rmd^d x$. In
\cite{LeDoussalMiddletonWiese2008,LeDoussalWiese2008c} we obtained the
distribution $P(S)$ from a combination of graphical and analytical
methods, first at tree, i.e.\ mean-field level, valid for $d \geq
d_{\mathrm{uc}}$, where the upper critical dimension is
$d_{\mathrm{uc}}=4$ for the usual short-range elasticity, and then to
first order in a dimensional expansion in $\epsilon=4-d$, by resumming
all 1-loop corrections. This calculation was technically rather
complicated as it required to sum an infinite set of diagrams, both at
the tree and 1-loop level.  Further their non-analytic dependence on $w$ had to be extracted. Thus \cite{LeDoussalMiddletonWiese2008,LeDoussalWiese2008c}  contains some amount of heuristics in extrapolating formulas from small moments of $P(S)$ to arbitrary ones, while the final structure is a relatively simple self-consistent equation. This suggests that a simpler method should exist. 

In this paper we present such a simple, complementary  method. It is powerful and versatile enough to
apply to many situations. It is extended in companion papers to (i) the depinning transition, for which the
avalanche-size distribution was also studied numerically in \cite{RossoLeDoussalWiese2009a}: There we
predict and measure the distribution of velocities inside an avalanche
\cite{LeDoussalWiese2011a,KoltonLeDoussalWiesetobe,LeDoussalWiesetobe,DobrinevskiLeDoussalWiesetobe}.
(ii) Elastic objects where the displacements $u(x)$ have more than one component \cite{LeDoussalWiese2010c,LeDoussalRossoWiese2011}. This new method accounts for the (relatively) simple structures
unveiled in \cite{LeDoussalWiese2008c}, via a saddle-point  equation and dressed
propagators. It also allows to derive a more precise picture of 
the structure of avalanches around the upper critical dimension $d_{\mathrm{uc}}$. In particular
we find that their statistics at $d_{\mathrm{uc}}$ is given in the statics by
a Brownian force model (BFM) which we study here, closely related to the
so-called ABBM model \cite{ABBM} that we also recently showed to
describe interfaces at $d_{\mathrm{uc}}$ near depinning \cite{LeDoussalWiese2011a}. 

In the second part of this article  we make  connection to the work by
Carraro and Duchon \cite{CarraroDuchon1998}, as well as Bertoin \cite{Bertoin1998}.  These authors use methods of probability theory  to study the Burgers
equation with Brownian initial conditions, which is the $d=0$ limit of the BFM for the interface.
Their description in terms of Levy processes is  extended to interfaces and
we unveil a new connection between evolution equations for these Levy processes
in Burgers dynamics, and the mean-field theory for pinned interfaces.

Our model is defined by the standard energy for a disordered elastic interface:
\begin{eqnarray}
{\cal  H}[u] &=&{\cal  H}_{\mathrm{el}}[u] + \int_x V(u(x),x)  \\
{\cal  H}_{\mathrm{el}}[u] &=& \frac{1}{2} \int_{xx'} g^{-1}_{xx'} [u(x)-w_x][u(x')-w_{x'}]  \ ,
\label{def1}
\end{eqnarray}
where $g_{xx'}=\int_q g_q \rme^{i q(x-x')}$ and the elastic energy kernel is, in the simplest case of short-range elasticity, $g^{-1}_q = q^2 + m^2$ (we denote $\int_q \equiv \frac{\rmd^d q}{(2 \pi)^d}$). 
A quadratic external potential of curvature $m^2$ and centered at $w_x$ has been added and acts as a large-scale (infrared) cutoff. In all cases 
$g_{q=0}=\int_{x'} g_{xx'} = m^{-2}$. $V(u,x)$ is a centered Gaussian random potential with correlator 
\begin{eqnarray}  \label{bare}
\overline{V(u,x) V(u',x')}=R_0(u-u') \delta^d(x-x')\ .
\end{eqnarray}
At finite temperature one considers the canonical partition sum ${\cal Z}=\int {\cal D}[u]\, \rme^{-{\cal H}[u]/T}$ in a given disorder realization (sample). Disorder averages are denoted by $\overline{\cdots}$, and thermal ones by $\langle \cdots \rangle $. 

To study the statics of this model, one introduces replica $u_a(x)$, $a=1,\ldots,n$ and considers the replicated action 
functional denoted ${\cal S}_{R_0}[u] \equiv {\cal S}[u] \equiv {\cal S}[\{u_a(x)\}]$:
\begin{eqnarray} \label{action0} 
\overline{{\cal Z}^n} &=&  \int \prod_a {\cal D}[ u_a] \rme^{- {\cal S}[u]}   \\
{\cal  S}[u] &=& \frac{1}{T} \sum_a {\cal H}_{\mathrm{el}}[u_a] - \frac{1}{2 T^2} \sum_{ab} \int_x R_0\big(u_a(x)-u_b(x)\big) \nonumber 
\end{eqnarray}
The correlation functions of the disordered model are obtained from those of the replicated
theory in the limit of $n\to 0$, implicit in all formula below. 

Let us now sketch the principle of the method, starting with a simple example. Since we are interested in probability distributions of observables, we need to compute averages of the form
\begin{equation}\label{av1}
 \overline{ \langle \rme^{ \int_x \lambda_x u(x) } \rangle } = \lim_{n \to 0} \left< \rme^{ \int_x \lambda_x u_1(x) } \right>_{\cal S} \ ,
\end{equation}
where $u_1(x)$ designates one of the $n$ replica. One defines
\begin{equation}
\left<O[u]\right>_{\cal S} := \frac{\int \prod_a {\cal D}[ u_a] O[u] \rme^{-
{\cal S}[u]}}{\int \prod_a {\cal D}[ u_a] \rme^{- {\cal S}[u]}}\ ,
\end{equation}
and since for $n=0$ the denominator equals one, it can be
dropped. Eq.~(\ref{av1}) is a generating function from which one can,
at least in principle, extract via Laplace inversion a probability distribution, here the distribution of the displacement field, i.e.\ ${\cal P}[{\sf u}] = \overline{\langle \prod_x \delta({\sf u}(x) - u(x)) \rangle }$ with a double average over sample and thermal realizations. Note that averages such as (\ref{av1}), and their multi-point generalizations discussed below, are
also frequently studied as generating functions of the distribution of the velocity field in turbulence. Burgers turbulence e.g.\ maps exactly to the present model at $d=0$ with time $t = 1/m^2$ and velocity $u$ \cite{Burgers74,BecKhanin2007,Polyakov1995,BouchaudMezardParisi1995,BernardGawedzki1998,Bertoin1998,LeDoussal2008,LeDoussalMuellerWiese2010,LeDoussalMuellerWiese2011,CarraroDuchon1998,Valageas2009,ChabanolDuchon2004, ChabanolDuchon2009}. 

We now recall a few basic facts from field theory. Let us consider $\Gamma[u]$, the {\it effective action} functional associated to the action ${\cal S}[u]$. Then the above average (\ref{av1}) can be expressed as
\begin{equation}\label{ex1}
\left< \rme^{ \int_x \lambda_x u_1(x) } \right>_{\cal S}  = \rme^{ \int_x \lambda_x u^\lambda(x) - \Gamma[u^\lambda]}\ ,
\end{equation}
where $u^\lambda(x)$ extremizes the exponential, i.e.\ is solution of the equation
\begin{equation}\label{ex2}
 \partial_{u_a(x)} \Gamma[u]\Big|_{u=u^\lambda} = \lambda_x \delta_{a1}\ .
\end{equation}
This property follows from the definition of the effective action as the Legendre transform of $W[\lambda]=\left< \rme^{ \int_x \lambda^a_x u_a(x) } \right>_S$ i.e.\ from the relation $W[\lambda]+ \Gamma[u]=\sum_a \int_x u_a(x) \lambda^a_x$. 

The calculation of $\Gamma[u]$ can  be performed  in an $\epsilon=d_{\mathrm{uc}}-d$ expansion around the upper critical dimension $d_{\mathrm{uc}}$. To lowest order in this expansion one replaces $\Gamma[u]$ by the action ${\cal S}[u]$. The corresponding calculation yields the {\it tree-level} result
\begin{eqnarray}\label{treegen}
 \left< \rme^{ \int_x \lambda_x u_1(x) } \right>^{\mathrm{tree}}_{\cal S} & =& \rme^{ \int_x \lambda_x u^\lambda(x) - {\cal S}[u^\lambda]} \\
 \partial_{u_a(x)} {\cal S}[u]\Big|_{u=u^\lambda} &=& \lambda_x \delta_{a1}\ . \label{mintree}
\end{eqnarray}
It is written in terms of a tree-level extremum field $u^{\lambda}\equiv u^{\lambda,\mathrm{tree} }$ which extremizes
(\ref{treegen}). This precisely amounts to resum {\it all tree diagrams} in the perturbation expansion in the non-linear part of the action ${\cal S}_{R_0}$, i.e.\ in the disorder $R_0$, also known as the {\em mean-field} calculation. This is discussed in section \ref{s:graph} and appendix \ref{s:diagrammatic}. For the problem at hand it gives the correct result for probability distributions for $d \geq d_{\mathrm{uc}}$, if the renormalized disorder $R$ is used in the action, rather than the bare one $R_0$, as discussed in \cite{LeDoussalWiese2008c} and again below. The corresponding action ${\cal S}_{R}$ is called the improved action. The precise definition of $R$ is recalled in Section \ref{s:eafpi},
and useful equivalent definitions can be found in Sections II and III of Ref.\ \cite{LeDoussalWiese2008c} (with which present
definitions and notations aim to be consistent). 

Here we start with the tree calculation in section \ref {s:tree}, by
first defining the proper observable to compute, a generalization of
(\ref{treegen}), and deriving the saddle-point  equation which resums
all tree diagrams. In the following section \ref{s:graph}, we give a graphical derivation and illustration of the saddle-point equations. We then 
compute in section \ref {s:1-loop} $\Gamma[u]$ to first order in $\epsilon$, and analyze the
resulting saddle-point  equation, from which the
avalanche-size distribution to 1-loop order is obtained.
In Section \ref{sec:bfm} we study a simpler model where
the force landscape is a Brownian motion, and we make the
connection to Levy processes and Burgers equation.
In Section \ref{sec:generalduchon} we derive the
generalized Carraro-Duchon equation which encodes
the mean-field theory of interfaces. 

The appendices contain details and extensions:  In appendix \ref{appgen} we study a non-uniform deformation $w_x$. Appendix \ref{app:diag} derives useful formulas for the diagonalization of replica-matrices. Appendix \ref{s:appB} calculates $\Gamma_1[u,v]$. Appendix \ref{s:diagrammatic} gives a diagrammatic interpretation of the loop corrections. 
Appendix \ref{app:bfm} contains a detailed derivation of many-point correlations in the
BFM. Appendices \ref{a:Carraro-Duchon} and \ref{app:erg} recall the derivation of the
Carraro-Duchon formula  and its connection to the exact RG equations.
In Appendix \ref{a:FRG4BFM}  we discuss the (near absence of)
loop corrections in the BFM model and we prove that it is an attractive
fixed point of the RG. Finally in Appendix \ref{poisson} we recall
how the statistics of shocks depends on their correlations.

\section{Tree-Level (Mean-Field) Calculation}
\label{s:tree}
\subsection{Avalanche Observables}
 
Let us recall the avalanche observables introduced in \cite{LeDoussalWiese2008c}. Unless 
stated otherwise the considerations below are valid in all generality 
(i.e.\ beyond mean field). 

At $T=0$, the minimal energy configuration of the interface $u(x;w)$, and its center of mass
$u(w):=L^{-d} \int_x u(x;w)$, advances by jumps as the position of the
center of the parabola, $w=w_x$ (taken uniform for now), is increased:
\begin{eqnarray}
 u(x;w) &=& \sum_i S_i(x) \theta(w-w_i)  \\
 u(w) &=& L^{-d} \sum_i S_i \theta(w-w_i)\ .
\end{eqnarray}
Here $w_i$ is the position, and $S_i := \int_x S_i(x)$ the total size of the
$i$-th shock. One defines $\rho(S):=\rho_0 \,P(S)=\sum_i
\overline{\delta(S-S_i) \delta(w-w_i)}$, where $\rho(S)$ is the
shock-size density, $P(S)$ the shock-size probability distribution
(normalized to unity) and $\rho_0$ the total shock density. From
$\overline{u(w)}=w$ it follows that $ \rho_0\left<S\right> = L^{d}$
whenever all motion occurs in shocks (which is the case here in the limit of interest, $m \to 0$). We denote size moments as $\left<S^p\right>:=\int \rmd S\, S^p P(S)$. It was
also shown in \cite{LeDoussalWiese2008c} that for $S_0 \ll S$, $P(S)$ takes the general form
\begin{eqnarray}\label{p(S)}
 P(S) &=& \frac{\left<S\right>}{S_m^2} \,p(S/S_m) \\
\label{Sm}
 S_m&:=&\frac{\left<S^2\right>}{2\left<S\right>} = \frac{R'''(0^+)}{m^4} \ .
\end{eqnarray}
Here $S_m \sim m^{-d-\zeta}$ is the scale of the large avalanches,
and $S_0 \ll S_m$ a microscopic cutoff. The function $p(s)$ is universal
with $\int_0^\infty \rmd s\, s p(s)=1$ and $\int_0^\infty \rmd s\, s^2 p(s)=2$. 

As detailed in \cite{LeDoussalWiese2008c} the shock-size moments can be extracted from the 
generating function:
\begin{eqnarray}\label{generat1}
 \hat Z(\lambda)& :=& L^{-d} \partial_{w} \,\overline{\rme^{\lambda L^d [u(w/2)-w-u(-w/2)]}}\Big|_{w=0^+} \nonumber \\
& =& \frac{ \langle \rme^{\lambda S} \rangle - 1 - \lambda \langle S
\rangle}{\langle S \rangle}\ . 
\end{eqnarray}
This formula follows from the fact that in a small window of width $w >0$ the probability that there is a shock is $\rho_0 w$, in which case the field $u(w)-w$ jumps by $S$. We used that due to statistical translation invariance, we can without loss of generality consider the interval \mbox{$]-w/2,w/2[$}. Equation (\ref{generat1}) can be compared to (\ref{av1}) with a uniform $\lambda_x=\lambda$: however, while 
(\ref{av1}) encodes only the {\it one point} probability of $u(w)$ (say at $w=0$), Eq.\ (\ref{generat1}) depends on the {\it two-point} joint probability distribution of the field $u(w)$ at two values of $w$ (denoted $-w/2$ and $w/2$), as required to study shocks. Furthermore, in $d=0$, $m^2(w-u(w))$ is the velocity field of the decaying Burgers equation at space point $w$ \cite{Burgers74,BecKhanin2007,LeDoussal2008} and
Eq.\ (\ref{generat1}) is thus the generating function of the distribution of velocity differences at two points in space distant by $w$. 
We retain below this terminology of $p$-point distributions. 

We now recall the results obtained at mean-field level in \cite{LeDoussalWiese2008c} by
resumming the tree diagrams. There it was shown that
$Z(\lambda):=\lambda + \hat Z(\lambda)$ satisfies the self-consistent equation
\begin{equation}\label{sc1}
Z(\lambda) = \lambda + \frac{R'''(0^+)}{m^4} Z(\lambda)^2 \ .
\end{equation}
This quadratic equation is easily solved, 
\begin{equation}\label{16}
Z_{\mathrm{tree}} (\lambda)\equiv  Z_{\mathrm{MF}} (\lambda) =  \frac{1}{2 S_{m}} \left ( 1-\sqrt{1-4 S_{m} \lambda } \right) \ ,
\end{equation}
where 
\bea
S_{m} = \frac{R''' (0^{+})}{m^{4}}
\eea is the characteristic
avalanche size
introduced in Eq.~(\ref{Sm}). Taylor-expanding Eq.~(\ref{16}) in $\lambda$, 
$Z (\lambda) = \lambda +S_{m} \lambda^{2} + \dots $, and comparing to
the definition (\ref{generat1}),  allows to identify $S_{m} =\frac{1}{2}
{\left< S^{2} \right>}/\left< S \right>$,  also
stated in Eq.\  (\ref{Sm}). 

Inverse Laplace transforming Eq.\ (\ref{16}),  one obtains the mean-field avalanche-size distribution\footnote{{These formula correspond to an infinite  density of avalanches. For discrete displacements $u$, as
illustrated in Appendix \ref{app:poisson}, $\rho_0$ is finite, and $Z(\lambda)$ is cut at  large negative $\lambda$, equivalent to
a small $S=S_{\mathrm{min}}$ for $P(S)$. This is further discussed in \cite{LeDoussalWiese2008c}.}}, valid 
for $d \geq d_{\mathrm{uc}}$ 
\begin{equation}\label{pmf}
 p_{\mathrm{tree}}(s)\equiv  p_{\mathrm{MF}}(s) = \frac{1}{2 \sqrt{\pi} s^{3/2} } \rme^{-s/4}\ .
\end{equation}
We now recover these results, and more, by introducing a method which does not use a graphical expansion. 

\subsection{Saddle-Point Equation} 
Here we show how to evaluate, at tree level, the slightly more general generating function for the joint probability 
of the field $u(x;w)$ at two ``points'' $w_x/2$ and $- w_x/2$. It corresponds to moving the
center of the parabola from $- w_x/2$ to $w_x/2$. While this is not the most general non-uniform move, its symmetry simplifies the analysis below. For future convenience, we denote here the full disorder correlator (which contains loop corrections to all orders), by $R$,  i.e.\ we consider the improved action ${\cal S}_{R}$. This is an improved tree approximation, i.e.\ it is the sum of all tree diagrams in $R$;  in $R_0$ it is the sum of all tree diagrams plus those loop diagrams in $R_0$ correcting $R$ itself. 

Generalizing Eq.~(\ref{treegen}) requires to introduce two sets of $n$ replicated
fields denoted $u_a,v_a$, $a=1,\ldots,n$, subject to the same disorder. We
find that\footnote{Here $u(x;w)$ is a functional of the field $w_x$. For simplicity however
we do not use the square bracket notation.}
\begin{equation} \label{tree2}
\overline{ \left< \rme^{ \int_x \lambda_x [u(x;w/2) - w_x - u(x;-w/2)]} \right> }^{\mathrm{tree}}  = \rme^{- {\cal S}_{\lambda}[u^\lambda,v^\lambda]} \ ,
\end{equation}
where (dropping the superscript $\lambda$):
\begin{eqnarray}
\lefteqn{ - {\cal S}_{\lambda}[u,v] = \int_x \lambda_x [u_1(x) - w_x - v_1(x)]} \nn \\
&& - \sum_a  \int_{xx'} \frac{g^{-1}_{xx'} }{2 T} \Big\{\left[u_a(x)-\frac{w_x}{2} \right]\!\left[u_a(x')-\frac{w_{x'}}{2}\right] \nn \\
&& \hphantom{ - \sum_a  \int_{xx'} \frac{g^{-1}_{xx'} }{2 T} \Big\{}
+ \left[v_a(x)+\frac{w_x}{2} \right]\!\left[v_a(x')+\frac{w_{x'}}{2} \right]\Big\} \nonumber  \\
&& + \frac{1}{2 T^2} \sum_{ab} \int_x \Big[R\big(u_a(x)-u_b(x)\big)+ R\big(v_a(x)-v_b(x)\big) \nn \\
&&\hphantom{ + \frac{1}{2 T^2} \sum_{ab} \int_x \Big[}  + 2 R\big(u_a(x)-v_b(x)\big) \Big] \label{Slambdauv} \ .
\end{eqnarray}
$u_a(x)$ and $v_a(x)$ are extrema of ${\cal S}_{\lambda}$, i.e.\
solution of 
\begin{align} \label{sp1}
&T \lambda_x \delta_{a1}  = g^{-1}_{xx'} \left[ u_a(x') - \frac{w_{x'}}{2} \right] \\
&\qquad  -  \frac{1}{T} \sum_{c}\left[ R'\big(u_a(x)-u_c(x)\big) + R'\big(u_a(x)-v_c(x)\big)  \right] \nonumber 
\end{align}
together with a similar  equation for $v$, obtained by  $(u,\lambda,w) \leftrightarrow (v,-\lambda,-w)$. One can
also write the functional derivative of  (\ref{tree2}),
\begin{align}\label{a1}  
\lefteqn{\partial_{w_y} \overline{\left<  \rme^{ \int_x \lambda_x [u(x;w/2) - w_x - u(x;-w/2)]} \right> }^{\mathrm{tree}}} \\
 &= \!\left\{ - \lambda_y +  \sum_a\int_{y'} \frac{g^{-1}_{yy'} }{2 T}
 \Big[u_a(y'){-}w_{y'}{-}v_a(y') \Big]  \right\} \! \rme^{- {\cal S}_{\lambda}[u^\lambda,v^\lambda]} \nn 
\end{align}
Due to the saddle-point  equations (\ref{sp1}), only the explicit dependence on $w$ appears. 

By parity, the solution of the saddle-point  equation  satisfies $v_a(x)=-u_a(x)$, which we use from now on. 
Since only replica $1$ is singled out, we look for a (replica symmetric) solution where all replica $a \neq 1$ assume
the same value. We denote
\begin{equation}\label{def1bis}
u_a(x)=u_1(x)- T U(x) \quad , \quad a \neq 1 \ ,
\end{equation}
which at this stage is just a definition. The saddle-point  equations
for the  two functions
$u_1(x)$ and $U(x)$ become
\begin{align} \label{sp2}
 &\!\!\!\lambda_x T = g^{-1}_{xx'} \left[u_1(x') - \frac{w_{x'}}{2}  \right] \\
&{  +  \frac{1}{T} \Big[ R'\big(T U(x)\big) + R'\big(2 u_1(x)-T U(x)\big) - R'\big(2
u_1(x)\big) \Big] }
 \nonumber \\
& \lambda_x T = T \int_{x'}g^{-1}_{xx'} U(x') - \frac{1}{T} R'\big(2 u_1(x) - 2 T U(x)\big) \label{sp3} \\
 & \hphantom{\lambda_x T =}+ \frac{1}{T} \left[2 R'\big(2 u_1(x)-T U(x)\big) - R'\big(2 u_1(x)\big) \right]\ . \nonumber
\end{align}
Note that the second equation has been obtained by subtracting in (\ref{sp1}) the equation for $a \neq 1$ from the one for  $a=1$.
In both equations the sum over replica indices has been performed
using (\ref{def1bis}), i.e $\sum_c F(u_c) = F(u_1)+(n-1) F(u_1-T
U_1)$, and then setting $n\to 0$. We have also used that $R'(u)$ is an odd function with $R'(0)=0$. Once 
these equations are solved, the solution can be used to compute the generating functions:
\begin{eqnarray}\label{tree3}
&& \overline{ \left< \rme^{ \int_x \lambda_x [u(x;w/2) - w_x - u(x;-w/2)]}\right> }^{\mathrm{tree}}  = \rme^{- {\cal S}_{\lambda}} \nonumber \\
&& \partial_{w_y} \overline{ \left< \rme^{ \int_x \lambda_x [u(x;w/2) - w_x - u(x;-w/2)]}  \right>}^{\mathrm{tree}}  \nonumber \\
&&\qquad  =  \Big[ - \lambda_y + \int_{y'}g^{-1}_{yy'} U(y')   \Big]  \rme^{- {\cal S}_{\lambda}}
\end{eqnarray}
with \begin{eqnarray}\label{a2}
\lefteqn{- {\cal S}_{\lambda} :=  - {\cal S}_{\lambda}[u,-u] } \\
\!&=& \int_x  \lambda_x (2 u_1(x) - w_x) \nn \\
&&+ \int_{xx'}g^{-1}_{xx'} \Big[ - U(x) \Big(2 u_1(x') -w_{x'}\Big) + T
U(x) U(x') \Big]\nn   \\
&& + \frac{1}{T^2} \int_{x} \Big[2 R(0) - 2 R\big(T U(x)\big)+ R\big(2 u_1(x){-} 2 T U(x)\big)
 \nn \\
&& \hphantom{ + \frac{1}{T^2} \int_{x} \Big[}
+ R\big(2 u_1(x)\big) - 2 R\big(2 u_1(x){-} T U(x)\big)\Big]\ .  \nonumber 
\end{eqnarray}The limit $n=0$ has been taken everywhere. This result is  equivalent to the graphical
summation of all tree diagrams, in terms of either $R_{0}$ or $R$,
depending on whether bare or renormalized perturbation theory is
used. It is valid for any $T$ and $w_x$, hence in principle it allows to
compute at tree level a rather general 2-point correlation function of
the field $u(x;w)$  at any temperature $T$.

In the absence of disorder, the saddle point is $u_1(x)-w_x/2=T U(x) = T \int_{x'}g_{xx'} \lambda_{x'}$, and
one obtains $- {\cal S}_{\lambda} = T \int_{x,x'}g_{xx'} \lambda_{x} \lambda_{x'}$.  The tree formula (\ref{tree3}) 
is then exact and corresponds to two copies with uncorrelated thermal
fluctuations. In presence of disorder, but for $\lambda_x=0$, one must have ${\cal S}_{\lambda}=0$. This is indeed the case, as the saddle point is then $U(x)=0$ and
$u_1(x)-\frac{w_x}{2}=0$. When there could be several solutions to the saddle-point  equation, the correct one
should reduce to that one in the small-$\lambda$ limit. The saddle-point  solution can also be obtained order by order in $\lambda$ from perturbation theory, i.e.\ a well-defined expansion of $u_1(x)$ and $U(x)$ in powers of $\lambda$ must exist.

\subsection{$T=0$ limit of the saddle-point  equations} 
 
We can now study the system at $T=0$. Then $R(u)$ is non-analytic, more
precisely it exhibits a linear cusp in its second derivative, $R'''(0^+) >0$. This cusp is related to the second moment of avalanche
sizes, as shown in \cite{LeDoussalWiese2008c}, via $R'''(0^+) = m^4 \left<S^2\right>/2 \left<S\right>$. We will recover this relation
here. 

In the $T\to 0$ limit we obtain a consistent solution assuming that  $U$ and
$u_1$ are going to a finite limit, as we show now. 
Expanding (\ref{sp2}) and (\ref{sp3}) in powers of $T$ we obtain to lowest order:
\begin{align}
& \int_{x'}g^{-1}_{xx'} \Big[u_1(x') -\frac{w_{x'}}2 \Big] +  \left[R''(0) - R''\big(2 u_1(x)\big)\right] U(x)\nonumber \\
&\qquad  =0 \label{a3}  \\
&\int_{x'}  g^{-1}_{xx'} U(x')   - R'''\big(2 u_1(x)\big) U(x)^2  = \lambda_x  \label{eq1}
\end{align}
The generating functions are, omitting the thermal averages $\left<
\dots \right>$, since we are studying $T=0$, 
\begin{align} \label{sp4}
& \overline{  \rme^{ \int_x \lambda_x [u ( x;w/2) - w_{x} - u (x; -w/2)]}
}^{\mathrm{tree}} =
\rme^{-{\cal S}_{\lambda}} \\ \label{sp5}
& \partial_{w_y} \overline{  \rme^{ \int_x \lambda_x [u ( x;w/2) - w_{x} - u (x; -w/2)]}  }^{\mathrm{tree}}  \nn\\& \quad \,=
 \big[ - \lambda_y +  \int_{y'} {g_{yy'}^{-1} U (y')}  \big]  \rme^{-{\cal S}_{\lambda}}  \\
 &
- {\cal S}_\lambda = \int_x   \Big[ u_1(x) - \frac{w_x}2\Big]  \!
\left[2  \lambda_x - \int_{x'} g^{-1}_{xx'} U(x') \right] \label{a5}\ ,
\end{align}
where (\ref{a3}) has been used to simplify ${\cal S}_{\lambda}$. These formula are
valid for arbitrary $w_{x}$. 

We now analyze these equations in several cases. 

\subsection{Uniform case: avalanches of center of mass}
\label{s:uniform}
Let us start with the simplest case of both $\lambda_x=\lambda$  and
$w_x=w$ uniform, corresponding to the generating function (\ref{generat1}). Then $u_1(x)=u_1$
and $U(x)=U$  satisfy
\begin{align} \label{Uu1}
& m^2 (u_1-w/2) +  \left[R''(0) - R''(2 u_1)\right] U  =0 \\
& m^2 U  - R'''(2 u_1) U^2  = \lambda \label{Uu2} \\
& L^{-d} \partial_{w}\, \overline{  \rme^{  \lambda L^d [u(w/2) - w - u(-w/2)]}  }^{\mathrm{tree}} \nn \\
& \qquad = [ - \lambda + m^2 U  ]  \rme^{- {\cal S}_\lambda}\label{32}  \\
& - {\cal S}_\lambda =  L^d (2 \lambda -  m^2 U) \left( u_1 -\frac{
{w}}2\right) \ .   \label{33}
\end{align}
These equations can be studied for any $w$. 

We now consider the limit $w \to 0^+$ from which avalanche observables can be extracted.
We look for a solution of the form
\begin{equation}
u_1 = y \frac{w}{2} + O(w^2)\ ,
\end{equation}
which implies that ${\cal S}_{\lambda} = O(w)$.  Hence 
\begin{equation}\label{y1}
L^{-d} \partial_{w}\, \overline{ \rme^{ \int \lambda L^d [u(w/2) - w - u(-w/2)]} }^{\mathrm{tree}}\Big|_{w=0^+} = - \lambda + m^2 U   \ .
\end{equation}
Assuming $y >0$  we obtain from (\ref{Uu1}) and (\ref{Uu2})
\begin{eqnarray}\label{y2}
 y   \left(1-\frac{2 R'''(0^+)}{m^2} U\right)&=&1 \\
m^2 U  - R'''(0^+) U^2  &=& \lambda  \ . \label{y2n}
\end{eqnarray}
Comparing the definition (\ref{generat1}) and the result (\ref{y1})
we see that we can identify
\beq
 Z(\lambda) \equiv  \lambda + \hat Z(\lambda)  = m^2 U  \ .
\eeq
Our self-consistent equation (\ref{y2n}) is indeed
the same as the one obtained in \cite{LeDoussalWiese2008c}, 
namely (\ref{sc1}). Its physical solution, which vanishes at $\lambda=0$,
is $m^2 U = Z(\lambda) = Z_{\mathrm{MF}} (\lambda)$
given in (\ref{16}). Note that
\beq
y = \frac{1}{\sqrt{1- 4 S_m \lambda}} 
\eeq
is indeed positive for this solution. The
breakdown at $\lambda \ge 1/(4 S_m)$ signals that the
Laplace transform of $P(S)$ does not exist beyond that value of $\lambda$,
due to the exponential tail at large $S$ in (\ref{pmf}).

\subsection{Non-uniform case: Local structure of avalanches} 
To obtain spatial information about avalanches one may consider 
both  $\lambda_x$  and $w_x$ non-uniform. We specify $w_x= \tilde w f(x)$ and vary $\tilde w$ from $\tilde w=-\frac{w}{2}$ to $\tilde w=+\frac{w}{2}$ for a fixed $f(x)$. In the limit of small 
positive $w$ we look for a
solution of the form
\begin{equation}
u_1(x)=y(x) \frac{w}{2} + O(w^2) \ .
\end{equation}
Again one finds ${\cal {\cal S}_{\lambda}} = O(w)$, hence
\begin{eqnarray}
\lefteqn{\partial_{w}\, \overline{ \rme^{ \int_x \lambda_x [u_x(w/2) - w_x - u_x(-w/2) ]} }^{\mathrm{tree}}\Big|_{w=0^+}}  \label{2pt} \\
& =&
\int_x f(x) \left[- \lambda_x + \int_{x'} g^{-1}_{xx'} U(x')  \right]\ . \qquad \qquad \nn
\end{eqnarray}
The field $U(x)$ satisfies Eq.\ (\ref{eq1}), i.e.\ 
\begin{equation}\label{41}
\int_{x'}g^{-1}_{xx'} U(x')   - R'''\big(y(x) w\big) U(x)^2  = \lambda_x  \ .
\end{equation}
Here we assume that $y(x)>0$, a more general discussion is given in Appendix \ref{appgen}. Then in the limit of $w \to 0^+$, one can replace $R'''\big(y(x) w\big) \to R'''(0^+)$ and (\ref{41})
becomes a closed equation for $U(x)$, {\em independent} of $y (x)$,
\begin{equation}\label{y3}
\int_{x'}g^{-1}_{xx'} U(x')   - R''' (0^{+}) U(x)^2  = \lambda_x  \ .
\end{equation} 
This is a classical equation for a cubic field theory, which admits instanton
solutions, from which local size distributions can be extracted, as discussed in \cite{LeDoussalWiese2008c}. 
Here we will not study again these applications, but simply make contact with the
notations used there. For that purpose we identify
\begin{equation}
Z_x(\lambda) =  \int_{x'}g^{-1}_{xx'} U(x')\ , 
\end{equation}
in terms of which the self-consistent equation becomes
\begin{equation}
Z_x(\lambda) = \lambda_x + R'''(0^+) \int_{x_{1},x_{2}}g_{x x_1} g_{x x_2} Z_{x_1}(\lambda) Z_{x_2}(\lambda)\ .
\end{equation}
These are Eqs.~(204) (in unrescaled form) and (F8) of \cite{LeDoussalWiese2008c}. The space-dependent generating function can then be written as
\begin{eqnarray}
&& \partial_{w_y} \overline{ \rme^{ \int_x \lambda_x[ u_x ( w/2) -
w_x - u_x ( -w/2)]} }^{\mathrm{tree}}\Big|_{w_y=0^+}   \nn \\
&&= Z_y(\lambda) - \lambda_y := \hat Z_y(\lambda)\ . \quad \quad \quad 
\end{eqnarray}
Note that a rescaled version of $U(x)$ was denoted $Y(x)$ in \cite{LeDoussalWiese2008c}. 

$Z_y(\lambda)$ is connected to local avalanche-size distributions. Assuming that
for fixed $L$ as $w \to 0^+$ the probability of a shock
during a change of $w_x$ is $\rho_0^f w$, we can write from Eq.\ (\ref{2pt})
\begin{eqnarray}  \label{local}
&& \rho_0^f \left< \rme^{\int_x \lambda_x S_x} -1 - \int_x \lambda_x S_x\right>_f \nn \\
&& = \int_x f(x) \left[ - \lambda_x + \int_{x'}g^{-1}_{xx'} U(x') \right] \ .
\end{eqnarray}
The subscript $f$ reminds that we use a non trivial $f(x)$. 
In addition, one can show that $\overline{u_x(w/2)-u_x(-w/2)} = w_x$ hence $\rho_0^f \left<S_x\right>_f=1$, which also
implies $\rho_0^f \left<S\right>_f = L^d$. Note that the size distribution, whose Laplace transform is
given by (\ref{local}), a priori depends on the function $f(x)$; the case $f(x)=1$ 
was studied in \cite{LeDoussalWiese2008c}. 

\subsection{Multi-point correlations of center-of-mass displacement}

\subsubsection{Discrete version}

We now indicate how to compute, at tree level (i.e.\ as a sum of all tree diagrams),
the correlations of the center-of-mass displacement field $u(w_i)-w_i$ at an arbitrary number of
discrete points $w_i$. For this we introduce a generating function, parameterized by $\lambda_i$:
\begin{align}
&\overline{ \rme^{  L^{d}  \sum_i \lambda_i [u(w_i)-w_i ] }}^{\mathrm{tree}} \\
&= \rme^{L^d [ \sum_i \lambda_i (u_{1i} - w_i )  - \sum_{ai} \frac{ m^2}{2 T} ( u_{ai} - w_i)^2+  \frac{1}{2 T^2} \sum_{abij} R(u_{ai}-u_{bj}) ] } \nn
\end{align}
The fields $u_{ai}$ are solutions of the saddle-point  equations
\begin{equation}
 m^2 (u_{ai} - w_i) -  \frac{1}{T} \sum_{cj} R'(u_{ai}-u_{cj})   = T \lambda_i \delta_{a1}\ .
\end{equation} 
As above, we look for a replica-symmetric saddle point $u_{ai}=u_i$ for $a \neq 1$.  Define 
$u_i:=u_{1i}-T U_i$, and subtract the equation  for  $a=2$ from the one for $a=1$:
\begin{eqnarray}
&& m^2 (u_{1i} - w_i ) \\
&&+  \frac{1}{T} \sum_{j} \Big[ R'(u_{1i}-u_{1j}+ T U_j) - R'(u_{1i}-u_{1j}) \Big]
  = T \lambda_i \nn \\
&& m^2 T U_i  + \frac{1}{T} \sum_{j} \Big[ R'(u_{1i}-u_{1j}+ T U_j) -  R'(u_{1i}-u_{1j}) \nn\\&&
+   R'(u_{1i}-u_{1j}- T U_i) 
-   R'(u_{1i}-u_{1j} - TU_{i} +T U_j)\Big]\nn\\
&& = T \lambda_i  
\end{eqnarray} 
Note that in the second equation the terms $i=j$ can be excluded. Taking the limit $T\to 0$, we find
\begin{eqnarray}
 m^2 (u_{1i} - w_i) + \sum_{j} R''(u_{1i}-u_{1j}) U_j  &=& 0 \qquad \nn \\
 m^2 U_i  + \sum_{j \neq i} R'''(u_{1i}-u_{1j}) U_i U_j  &=&  \lambda_i  \label{multi}\ .
\end{eqnarray} 
One must solve these equations for $U_i$ and $u_{1i}$ and insert the
result into
\begin{equation}
 \overline{ \rme^{  L^d \sum_i \lambda_i [u(w_i)- w_i] }}^{\mathrm{tree}}   
   = \rme^{ L^d \sum_i (\lambda_i - \frac{1}{2} m^2 U_i) (u_{1i} - w_i) }\ ,
\end{equation}
which has been simplified using the saddle-point  equations. Note that
one recovers the 2-point equations (\ref{Uu2}) from the solution $u_{12}=-u_{11}$ 
and $U_2=-U_1$ valid for $\lambda_2=-\lambda_1$. 

\subsubsection{Continuous version}
It is instructive to perform the same calculation in a continuum framework. 
We again restrict to the center of mass and compute
\beq
\overline{\left<\rme^{  L^{d} \int_w \lambda(w) [u(w) - w)] } \right>}^{\mathrm{tree}} = \rme^{- S_\lambda} 
\eeq 
for a test function $\lambda(w)$, depending on $w$ but uniform in space. Here and below $\int_w= \int_{-\infty}^{\infty} \rmd w$. 
We now introduce\footnote{The generalization
with space dependence would involve $\lambda_x(w)$ and replica fields $u_a(x,w)$.} replica fields $u_{ai} \to u_a(w)$ 
and then extremize the action:
\begin{eqnarray}
 L^{-d} S_\lambda &=& \int_w  \lambda(w) [u_1(w) - w] \nn\\
&& -  \int_w \sum_a \frac{m^2}{2 T} [u_a(w)-w]^2 \nn\\
&&+ \frac{1}{2 T^2}  \int_{w,w'} \sum_{ab} R\big(u_a(w) - u_b(w')\big)\ . \ \ \ 
\end{eqnarray} 
We will not repeat all the above manipulations. The 
saddle-point  equations w.r.t.\ $u_a(w)$ in the limit $T=0$ lead to
\begin{align}
& m^2 [u_1(w)-w] + \int_{w'} R''(u_1(w)-u_1(w')) U(w') = 0 \nn \\
& m^2 U(w) + \int_{w'} R'''(u_1(w)-u_1(w')) U(w) U(w') = \lambda(w) \nn\ .\qquad \\
& \label{multicont} 
\end{align}
Its solution is inserted into
\begin{equation}  \label{Scont} 
 -  L^{-d} S_\lambda = \int_{w} \Big[\lambda(w) - \frac{m^2}{2} U(w)\Big] [u_1(w)-w] \ .
\end{equation}
The general analysis of these multi-point correlations, and their discrete
analog (\ref{multi}) is left for the future. Below we 
study them, in Section \ref{sec:bfm} and Appendix \ref{app:bfm}, 
for a simpler model, where $R'''(u)$ is a constant, and
in Section \ref{sec:periodic} for  periodic disorder. We also discuss, in Section
\ref{sec:generalduchon} another powerful method to generate multi-point correlations.

\subsection{Displacement correlations for finite $w$: Absence of
correlations for $d \geq d_{\mathrm{uc}}$ }
\label{s:IIG}

The two-point correlation function of the center-of-mass displacement,
\begin{eqnarray}
\overline{ \rme^{ L^d \lambda [u(w/2) - w - u(-w/2)]}
 }^{\mathrm{tree}} = \rme^{-{\cal S}_{\lambda}}
\end{eqnarray}
can also be evaluated within tree level at finite $w>0$ by setting $u_1 =y w/2$, and 
solving the saddle-point equations (\ref{Uu2}) for $y$ and $U$, denoting
$y_w$ and $U_w$ these solutions. This can be done, e.g.\ order by order in $w$. Here  we give 
the small-$w$ expansion up to $O(w^4)$. It is convenient to introduce the
rescaled (dimensionless) variables $\tilde w$ and $\tilde \lambda$, 
\begin{eqnarray} 
w = m^d S_m \tilde w \quad , \quad \lambda = \tilde \lambda/S_m \ .
\end{eqnarray} 
Defining $\tilde Z_w = m^2 S_m U_w$, one finds
\begin{eqnarray}  \nn
&& - \frac{{\cal S}_{\lambda}}{(m L)^d} = \tilde w (\tilde Z - \tilde \lambda) 
+ m^{-\epsilon} R''''(0)  \frac{\tilde w^2 \tilde Z^2}{2(1-2 \tilde Z)^2} \\
&& \  \ \ \ \ + \tilde w^3 \tilde Z^2 m^{-2\epsilon} \left[ R''''(0)^2 
\frac{\tilde Z (1-\tilde  Z)}{(1-2 \tilde Z)^5} 
+ \frac{R'''(0) R^{(5)}(0)}{6(1-2 \tilde Z)^3} \right] \nn \\
&& \  \ \ \ \  +O(\tilde w^4) \label{expand}\\
&& \tilde Z \equiv \tilde Z_{\mathrm{tree}}(\tilde \lambda)= \frac{1}{2} \left(1- \sqrt{1-4 \tilde \lambda}\right)
\end{eqnarray}
More generally, one can introduce the dimensionless rescaled renormalized disorder correlator,
\begin{eqnarray} \label{rescR}
R''(u)=A_d m^{\epsilon-2 \zeta} \tilde R''(u m^{\zeta})\ ,
\end{eqnarray}
where $A_d=1/(\epsilon \tilde I_2)$, and for short-ranged disorder $\epsilon=4-d$ and 
\begin{equation}\label{I2}
\tilde I_2=\int_k \frac1{(1+k^2)^{2}}
\end{equation}
 is an amplitude; for details see \cite{LeDoussalWieseChauve2003}, and for generalization to long-range disorder \cite{LeDoussalWieseChauve2002,LeDoussalWieseMoulinetRolley2009,LeDoussalWiese2009a}. It is known that as $m \to 0$,  
$\tilde R''(u)$ goes to a fixed-point function in any $d$, measured in \cite{MiddletonLeDoussalWiese2006}. 
Then, 
in any $d$, and within the tree approximation,
\begin{equation}
 - \frac{{\cal S}_{\lambda}}{(m L)^d} = F(\tilde \lambda, \tilde w) := (y-1)(\tilde \lambda - \frac{1}{2} \tilde Z) \tilde w
\end{equation}
takes a scaling form, obtained by eliminating $y \equiv y_w$ and $\tilde Z\equiv \tilde Z_w$ in the 
rescaled saddle-point  equations:
\begin{eqnarray} \label{rescsp}
&& (y -1) A_d \tilde R'''(0) \frac{\tilde w}{2} + \frac{\tilde R''(0) - \tilde R''(y A_d \tilde R'''(0) \tilde w)}{\tilde R'''(0)} \tilde Z = 0 \nn \\
&& \tilde Z - \frac{\tilde R'''(y A_d \tilde R'''(0) \tilde w)}{
\tilde R'''(0)} \tilde Z^2 = \tilde \lambda \ .
\end{eqnarray} 
While these equations for the tree approximation  can be written for any $d<d_{\mathrm{uc}}$, they are  expected to become
exact as $d \to d_{\mathrm{uc}}$ and in $d >d_{\mathrm{uc}}$. For $d=d_{\mathrm{uc}}-\epsilon$, $\epsilon>0$, the rescaled disorder 
$\tilde R''(u)$ is uniformly of order $\epsilon$. In (\ref{rescsp}) the argument of the functions
is $O(\epsilon)$ which justifies an expansion of the system for small $w$. This is because
the scaling variable is
\begin{equation}
 w = A_d \tilde R'''(0) m^{-\zeta} \tilde w \equiv S_m m^{d} \tilde w\ .
\end{equation}
Hence near $d=d_{\mathrm{uc}}$ one can focus on (\ref{expand}). 
Since 
\begin{eqnarray}
&& m^{-\epsilon} R''''(0) = A_d \tilde R''''(0) = O(\epsilon) \\
&& m^{-2 \epsilon} R'''(0) R^{(5)}(0) = O(\epsilon^2)\ ,
\end{eqnarray}
we arrive (in terms of the unrescaled variables) for $w>0$ at the result
\begin{equation}
\overline{ \rme^{ L^d \lambda [u(w/2) - w - u(-w/2)]}
 }^{\mathrm{tree}} = \rme^{ L^d w \hat Z(\lambda) + O(\epsilon) }\ .
\end{equation}
The interpretation of this result is that, for
$d \geq d_{\mathrm{uc}}$ {\it the increments in the displacements in the center of mass
become uncorrelated}. It will further be discussed in Section \ref{s:Levy} in terms of
{\em Levy}  processes. 

Below $d_{\mathrm{uc}}$, in $d=d_{\mathrm{uc}}-\epsilon$, we expect correlations. They  only exist on a distance $\tilde w=O(1)$, 
i.e.\ $w = O(\epsilon) m^{-\zeta}$, a very small layer as $\epsilon \to 0$. The above result
(\ref{expand}) allows to compute the first correction in $\epsilon$; however 
one may also get contributions at one loop of the same order. This
calculation is performed elsewhere. 

\section{Graphical interpretation}\label{s:graph}

Here, we sketch a short graphical interpretation of the mean-field saddle-point
equations. We work in $d=0$ for simplicity. The results 
apply to the center-of-mass variable in any $d$, after restoring the necessary
factors of $L^d$. For easier comparison with section \ref{sec:generalduchon} we
use the notations
\bea
\Delta(u) = - R''(u) \quad , \quad t = \frac{1}{m^2} \ .
\eea 
We define
\begin{equation}\label{defZmath}
\rme^{\mathbb{Z} (\lambda ,w)} := \overline{\rme^{\lambda \left[u (w)-u (0) \right]}}\ .
\end{equation}
Thus $\mathbb{\hat Z} (\lambda ,w)= \mathbb{Z} (\lambda ,w)-w \lambda$ is the 
generating function of the connected moments $\overline{[u (w)-u (0) - w]^n}^c =  t^n K^{(n)}(w)$, which  in Eq.~(41) of \cite{LeDoussalWiese2008c}  were called 
the Kolmogorov-cumulants. In \cite{LeDoussalWiese2008c}, a graphical derivation of the recursion
relation  for the $O(w)$ term  $Z(\lambda)$ was given, noting
\begin{equation}\label{mathZ}
\mathbb{Z} (\lambda ,w) ={Z} (\lambda ) w + O (w^{2}) \ .
\end{equation}
At tree level, $\mathbb{Z}^{\mathrm{MF}} (\lambda ,w)$ is the
sum of {\em all} connected tree diagrams. The generating function $Z^{\mathrm{MF}}(\lambda)$ as a function of the bare
action can be written as a sum of {\em particularly simple} tree graphs, namely the ones of the form
\begin{equation}\label{eqg}
Z^{\mathrm{MF}}(\lambda) = {\cal K} \sum  \parbox{0.74\figwidth}{\fig{.74}{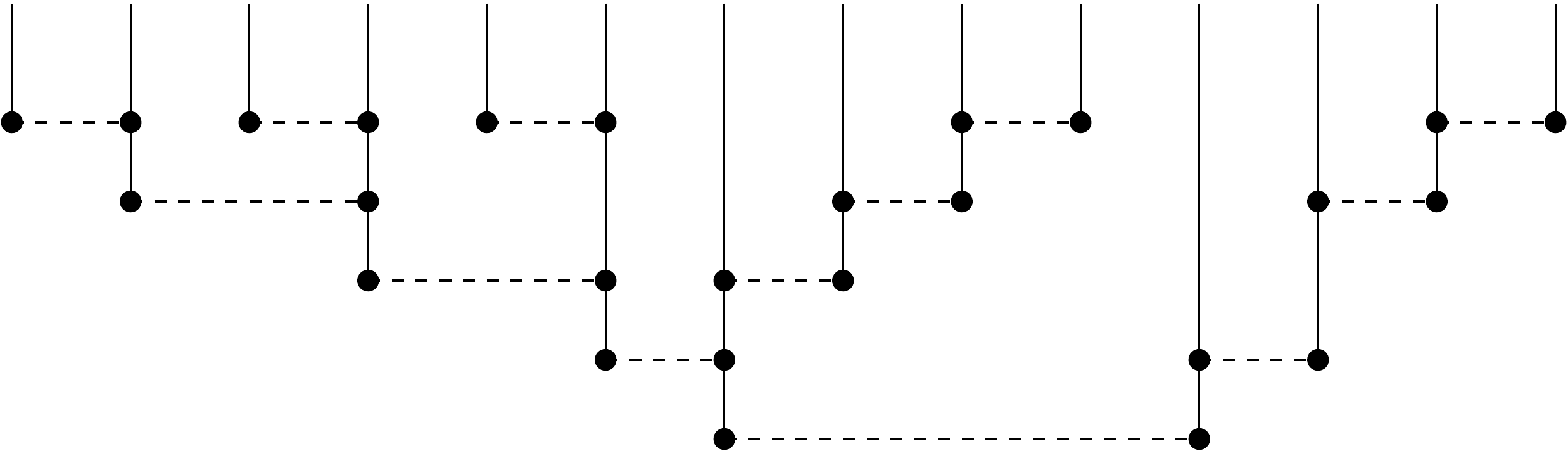}},
\end{equation}
where dotted lines represent the bare disorder $\Delta_0$. Each graph represents
correlations of the $\lambda[u(w_i)-w_i]$ fields at different points $w_i$ 
(external lines on the top, each coming with a factor of $\lambda$). The linear operator ${\cal K}$ 
identifies $w_i$ with $w$ or $0$, in order to build the Kolmogorov cumulants. The disorder vertex on the bottom contributes 
 a factor of $\Delta_0(w)-\Delta_0(0)$ to $\mathbb{Z} (\lambda ,w)$, which has
been expanded to first order in $w$ to obtain $Z(\lambda)$, hence it must be
counted as $\Delta_0'(0^+)$ in (\ref{eqg}). Equivalently,  one can include a
factor of $\partial_w|_{w=0}$ in ${\cal K}$. For more details of these
graphical rules see \cite{LeDoussalWiese2008c},
section V.C. They are also further used below in Section \ref{sec:generalduchon} and appendix \ref{s:diagrammatic}. 
The improved generating function  $Z^{\mathrm{MF},R}(\lambda)$
is the same sum of tree diagrams with $\Delta$ at each vertex,
hence, if reexpressed in terms of the bare disorder $\Delta_0$,
it is now a sum of graphs of the form:
\begin{equation}\label{x1}
Z^{\mathrm{MF},R}(\lambda) = {\cal K} \sum  \parbox{0.74\figwidth}{\fig{.74}{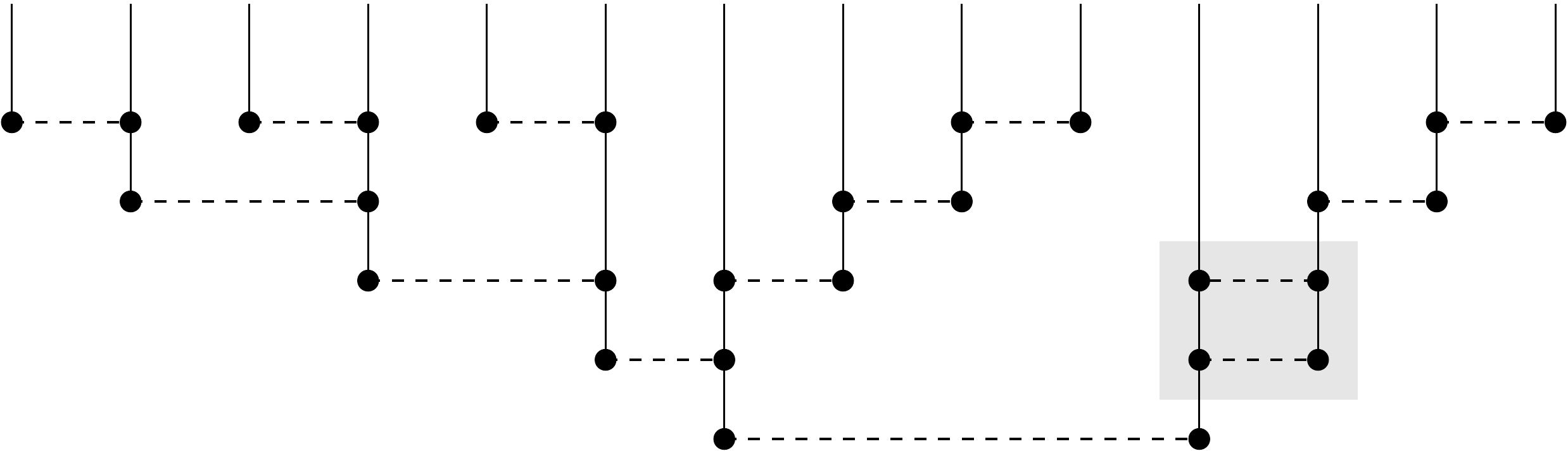}}.
\end{equation}
It contains all loop corrections to the $2$-point disorder vertex,
while loop corrections to higher-point disorder vertices are  neglected.
Explicit formulas for
low-order contributions can be found in
\cite{LeDoussalWiese2008c}.

The mean-field self-consistency equation for $Z(\lambda)=Z^{\mathrm{MF},R}(\lambda)$ reads
\begin{equation}\label{x2}
Z (\lambda) = \lambda - t^{2 } \Delta' (0^{+}) Z (\lambda)^{2}\ .
\end{equation} 
It is graphically written as \cite{LeDoussalWiese2008c}
\begin{equation}\label{74}
Z (\lambda ) =\   \parbox{0.8\figwidth}{\fig{.8}{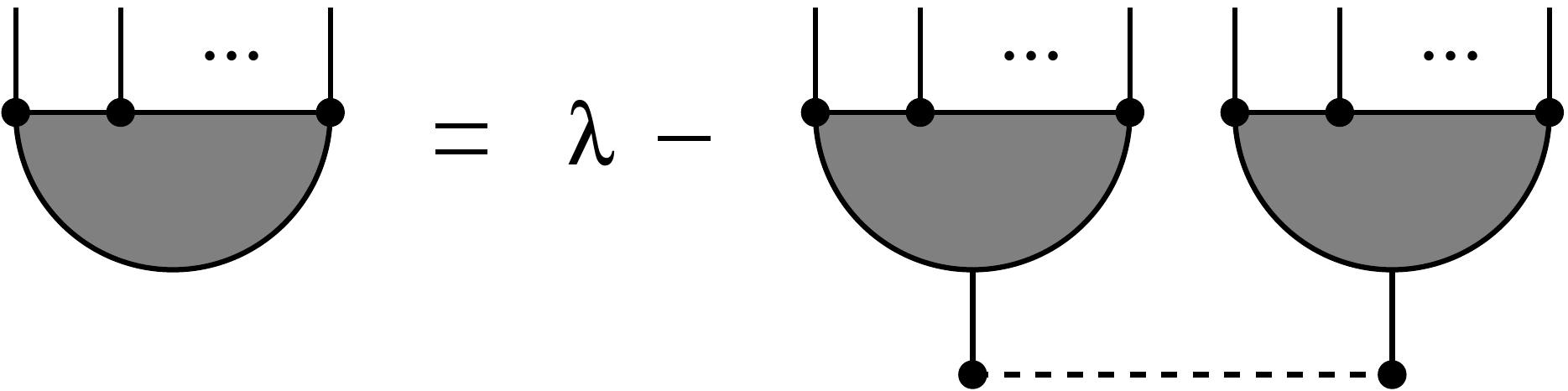}}\ .
\end{equation}
As indicated, the blob denotes $Z(\lambda)$ itself, while the
lowest disorder vertex  counts as a $\Delta'(0^+)$ and 
the lines entering the two blobs from below do not come
with differentiations\footnote{Note that in  \cite{LeDoussalWiese2008c} a factor of $|\Delta'(0^+)|=-\Delta'(0^+)$ has been absorbed in $\lambda$, whereas here it is explicitly written, resulting in a seemingly opposite sign.}. This 
self-consistent equation yields the desired sum of tree
diagrams {\it with only one lower disorder vertex},
i.e.\ a term $\Delta(w)-\Delta(0)$ expanded into $w \Delta'(0^+)$,
which is sufficient to obtain the $O(w)$ part in (\ref{mathZ})
as explained in \cite{LeDoussalWiese2008c}.

We now want to construct a recursion relation for 
$\mathbb{Z} (\lambda ,w)$, which yields its {\em complete} $w$ dependence.
We thus need to sum {\it all} tree diagrams. To generate them, it
{\em seems natural} to write
\begin{equation}\label{74b}
\mathbb{Z} (\lambda ,w) =\   \parbox{0.8\figwidth}{\fig{.8}{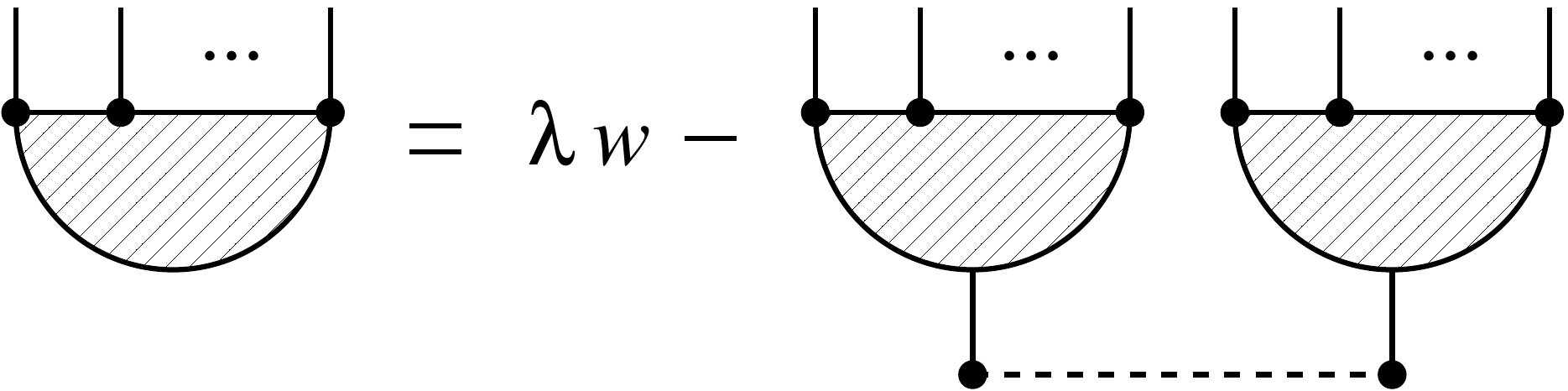}}\ .
\end{equation}
Now the lower vertex is $\Delta(w)-\Delta(0)$ and the line
entering a blob $\mathbb{Z}(\lambda,w)$ from below acts as derivative w.r.t.\ $w$.

However, a new difficulty arises:  One may have two or more lower vertices, as e.g.\ in 
\begin{equation} \textdiagram{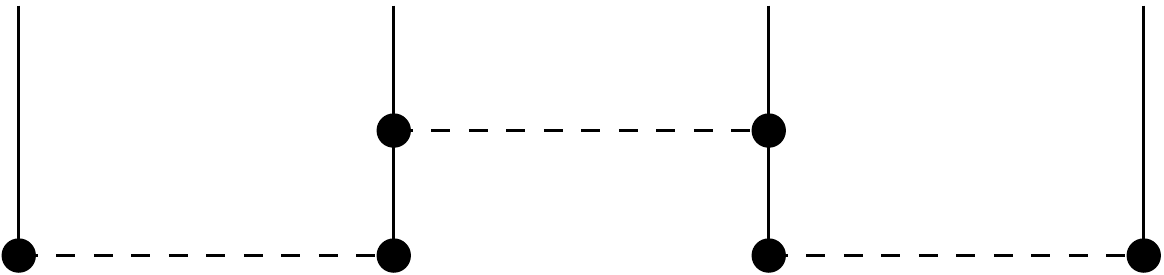}\ .\end{equation}
Unfortunately,  the self-consistency equation (\ref{74b}) is then incorrect, as it  leads to
an over-counting since there are several ways to construct the
same graph. Fortunately this can be corrected. 
Let us  explain the source of the problem, and its correction on an example:
\begin{eqnarray}\label{gasp}
\textdiagram{fig1} &=&  + \textdiagram{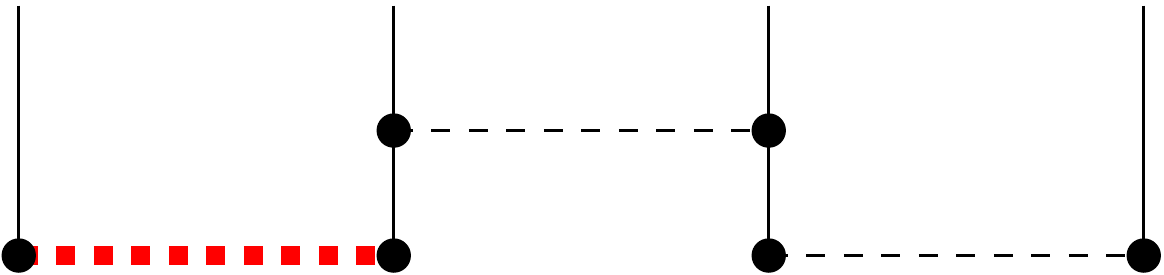} \nonumber \\
&& +\textdiagram{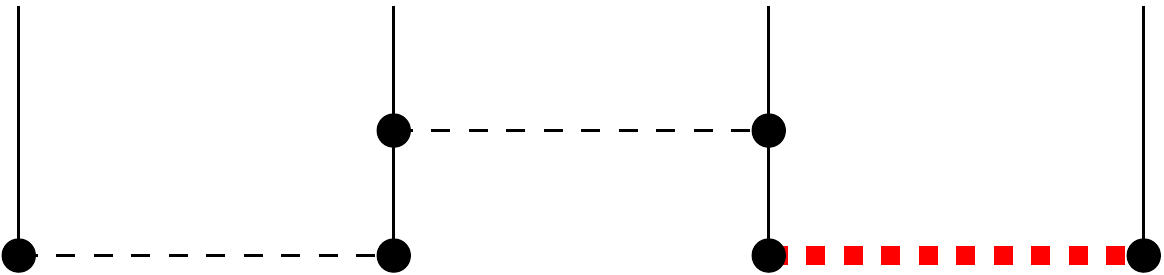}\nonumber \\
 &&-\textdiagram{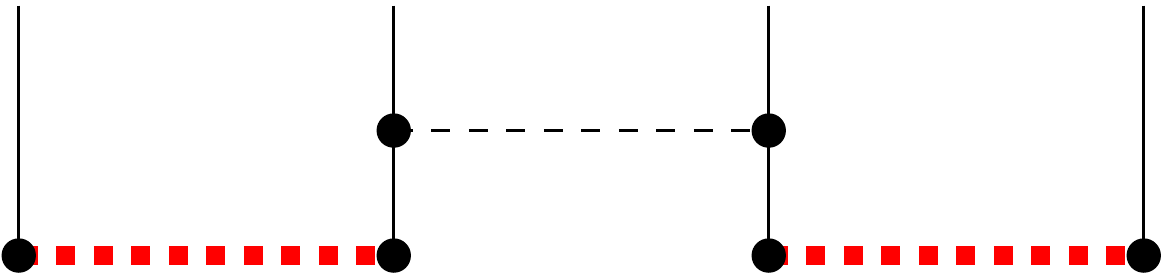}\ .\qquad 
\end{eqnarray}
On the l.h.s.\ we have plotted the contribution to $\mathbb{Z}(\lambda ,w)$, with the correct combinatorial factor. The  first two terms on the r.h.s.\ appear in a
recursion relation of the form (\ref{74b}), plotting the added vertex in Eq.\
(\ref{74b}) red (fat in black-and-white). This leads to an over-counting, which can be corrected by subtracting the last term, which has {\em two}  marked (red/fat) vertices. 

In the case of three lower vertices, the recursion reads
\begin{align}\label{x3}
&\!\!\!\!\!\!\!\!\!\textdiagram{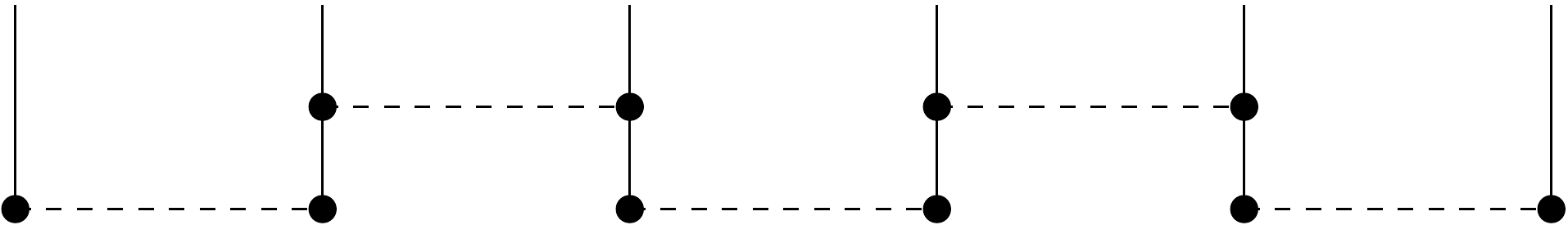}\nonumber \\
=&+ \textdiagram{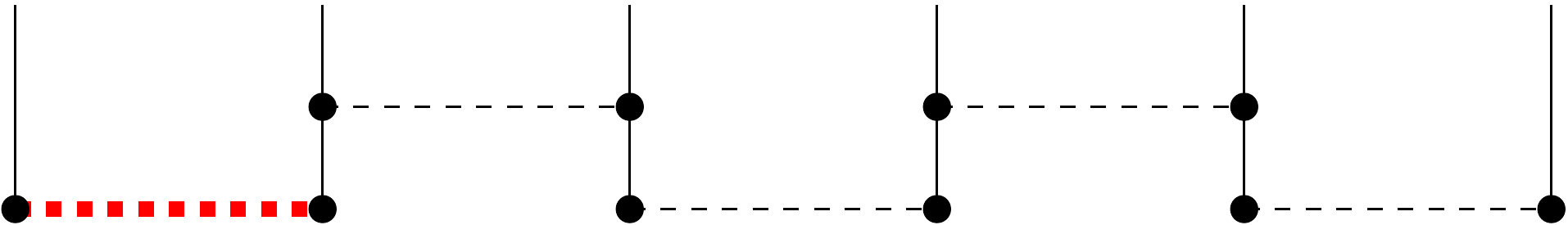}\nonumber \\
&+\textdiagram{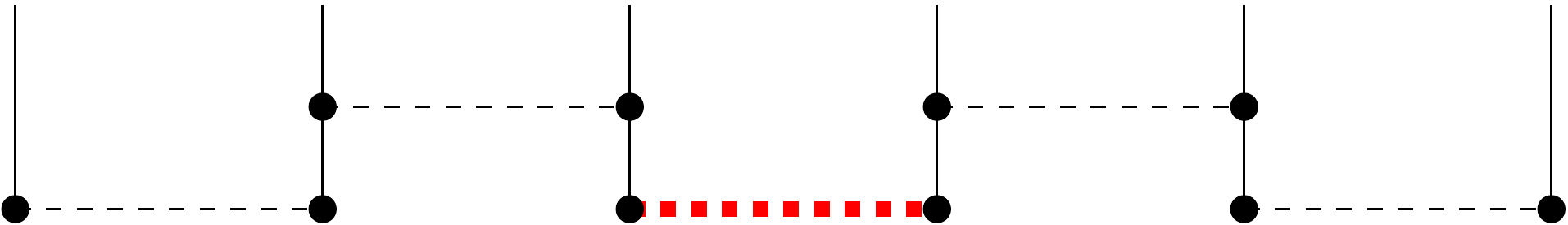}\nonumber \\
&+\textdiagram{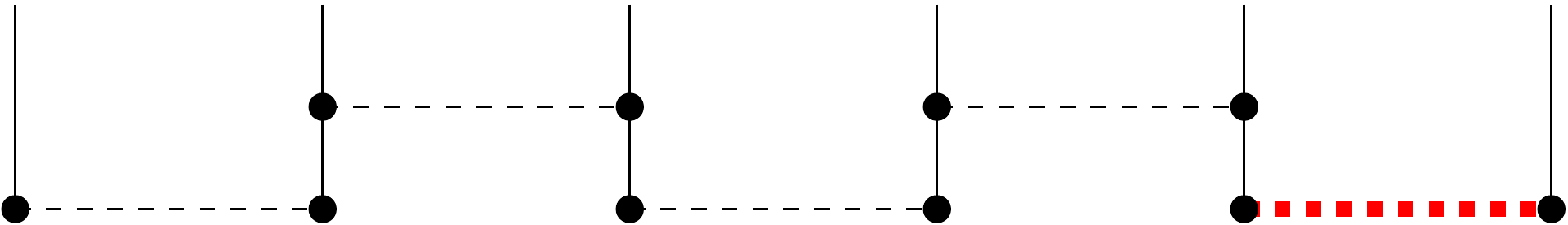}\nonumber \\
& -\textdiagram{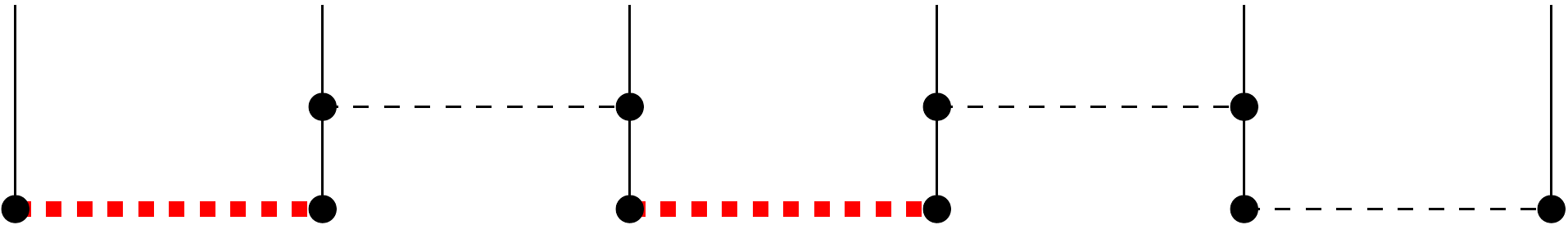}\nonumber \\
&-\textdiagram{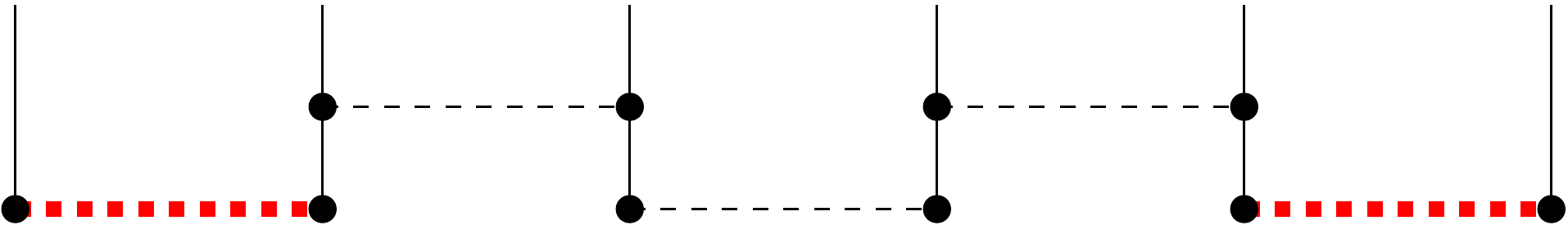}\nonumber \\
&-\textdiagram{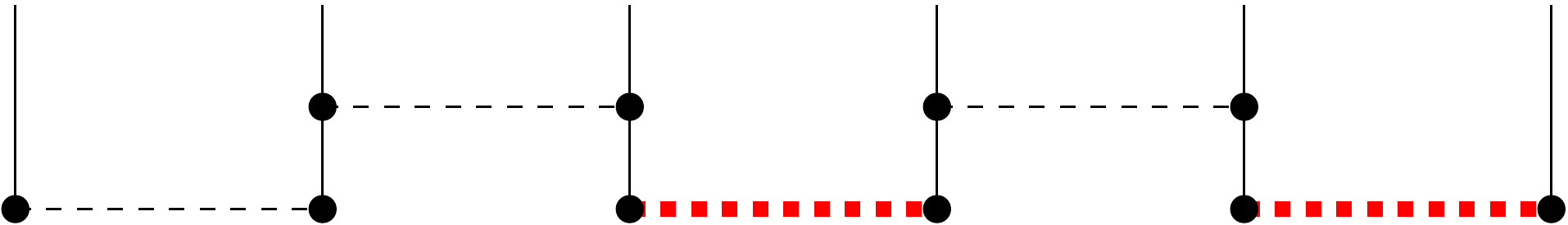}\nn \\
&+\textdiagram{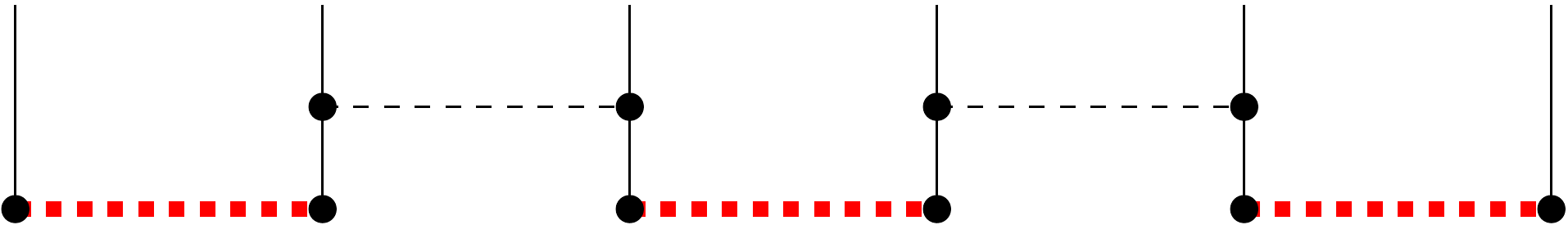}
\end{align}
Now pairs of lower vertices are subtracted, leading to a cancellation of all terms, and consequently the
{\em triplet} of lower vertices has to be added. This can be generalized,  
replacing the self-consistency condition (\ref{74b}) by 
\begin{equation}\label{tree-1}
  \lambda  w = \lim_{\nu \to 0} {\nu}\ln \left(:\!\!\rme^{ {\nu}
[\Delta (w)-\Delta (0)] t^{2}\partial_{w}\partial_{w} } \!\!:\, \rme^{\frac{1}{\nu} \mathbb Z ( \lambda , w)}\right)\ .
\end{equation}
The dots around the first exponential function indicate that the derivatives act only on $\mathbb{Z}(\lambda,w)$, not on $\Delta(w)$. This can  be written as
\begin{equation}\label{tree-1b}
  \lambda  w = \lim_{\nu \to 0} {\nu}\ln \left(\rme^{ {\nu}
[\Delta (w)-\Delta (0)] t^{2}\partial_{u}\partial_{u} } \rme^{\frac{1}{\nu} \mathbb Z ( \lambda , u)}\right)\Big|_{u=w}\ .
\end{equation}
The limit of $\nu \to 0$ selects the  tree diagrams, and the ln selects a single connected component. Note that  $\rme^{ {\nu}
[\Delta (w)-\Delta (0)] t^{2}\partial_{u}\partial_{u} }$ is defined by its series expansion in $t$, thus the   limit $\nu \to 0$ is
done term by term in the expansion in powers of $t$.

To proceed, we observe that independent of $\nu$ and for all functions $f(u)$ which are infinitely differentiable\footnote{
The formula is valid for $\nu [\Delta(w)-\Delta(0)] >0$, i.e.\ the integral should be evaluated  for
$\nu<0$, and the limit to be taken is $\nu\to 0^-$.}
\begin{align}\label{trick}
&\rme^{ {\nu}
[\Delta (w)-\Delta (0)] t^{2}\partial_{u}\partial_{u} } f(u)
\Big|_{u=w} \\ \nn
& = \frac{1}{\sqrt{4\pi \nu[\Delta (w)-\Delta (0)]t^2}}  \int_{-\infty}^{\infty} \rmd u\, \rme^{-\frac{ (u-w)^2}{4\nu[\Delta (w)-\Delta (0)]t^2}} f(u) \ .
\end{align}
Inserting this relation into (\ref{tree-1b}) yields 
\begin{align}\label{tree-1c}
  \lambda  w = \lim_{\nu \to 0} {\nu} \ln \Bigg(&\frac{1}{\sqrt{4\pi \nu[\Delta (w)-\Delta (0)]t^2}} \\
  & \times \int_{-\infty}^\infty \rmd u\, \rme^{-\frac{ (u-w)^2}{4\nu[\Delta (w)-\Delta (0)]t^2}}  \rme^{\frac{1}{\nu} \mathbb Z ( \lambda , u)}\Bigg)\ . \nn
\end{align}
In the limit of $\nu\to 0$ to be 
taken here, the integral is dominated by its saddle point, and  we get
\begin{eqnarray}\label{78}
w \lambda & =& \mathbb Z (\lambda ,u) +\left[\Delta (0)-\Delta (w) \right] t^{2} \left[\partial_{u}
\mathbb Z (\lambda ,u) \right]^{2} \qquad \\
\label{79}
u &=& w +2 t^{2} \left[\Delta (w)-\Delta (0) \right] \partial_{u} \mathbb{Z} (\lambda ,u)\ .
\end{eqnarray} 
The new variable $u$ has to be eliminated between these two equations. 
Note that it is an independent variable not to be confused with $u(w)$. 

We  remark that passing from (\ref{tree-1b}) to the self-consistent set of equations (\ref{78})-(\ref{79}) is a quite common feature in tree-resummation problems. It also appears  in the large-$N$ resummation for the disorder itself, where the links are the 1-loop momentum integral, and which therefore are termed cactus-diagrams,  see \cite{LeDoussalWiese2003b,LeDoussalWiese2001}. 

Below we devise another method to compute $\mathbb Z (\lambda ,w)$ and
we have checked to high orders, $\sim \lambda^{100}$, that (\ref{78}) reproduces the solution Eq.\
(\ref{173}), e.g.\ (\ref{expandZt}) as well as the lowest-order Kolmogorov cumulants
(62)-(66) in \cite{LeDoussalWiese2008c}.

 One
needs the derivative $\frac{\rmd u}{\rmd w}$, obtained by deriving
(\ref{79}) w.r.t.\ $w$, 
\begin{equation}\label{80}
\frac{\rmd u}{\rmd w} = \frac{1 +2 t^{2} \Delta' (w) \partial_{u}
\mathbb Z
(\lambda ,u)}{1+2 t^2\left[\Delta (0) -\Delta (w)\right] \partial_{u}^{2} \mathbb Z (\lambda ,u)}\ .
\end{equation}
Deriving Eq.\ (\ref{78}) w.r.t.\ $w$, and using (\ref{80}) yields the
astonishingly rather similar equation, 
\begin{equation}\label{81}
\lambda = \partial_{u} \mathbb{Z} (\lambda ,u) +\Delta' (w) t^{2}  \left[\partial_{u}
\mathbb Z (\lambda ,u) \right]^{2}\ .
\end{equation}
We now make contact to the results given in Eq.\ (\ref{Uu1}). 
Since $u$ and $w$ are simply
variables, and we are more interested in $\mathbb{Z} (\lambda ,w)$
than in $\mathbb{Z} (\lambda ,u)$, one can exchange their names, to obtain a second set of
equations, 
\begin{eqnarray}\label{78b}
u \lambda & =& \mathbb Z (\lambda ,w) +\left[\Delta (0)-\Delta (u) \right] t^{2} \left[\partial_{w}
\mathbb Z (\lambda ,w) \right]^{2} \qquad  \\
\label{79b}
w-u &=& 2 t^{2} \left[\Delta (u)-\Delta (0) \right] \partial_{w} \mathbb{Z} (\lambda ,w)\\
\label{81b}
\lambda &=& \partial_{w} \mathbb{Z} (\lambda ,w) +\Delta' (u) t^{2}  \left[\partial_{w}
\mathbb Z (\lambda ,w) \right]^{2}\ .
\end{eqnarray}
Eq.\ (\ref{81b}) is redundant, or can be used instead of (\ref{78b}).
We have explicitly checked that up to order $t^{8}$, both expressions
are correct. 

Graphically, the interpretation of Eqs.~(\ref{78})--(\ref{81}) is
rather different from that of Eqs.\ (\ref{78b})--(\ref{81b}). To see this, recursively replace $u$ in Eq.\ (\ref{78}) by its value given by Eq.~(\ref{79}). This yields a perturbative expansion in $t$ of $\lambda w$, which can be read as a self-consistent equation for $\mathbb{Z}(\lambda,w)$. Graphically, it contains links made out of $\left[\Delta (w)-\Delta (0) \right]t^2$, which end in vertices made out of  $\mathbb{Z}(\lambda,w)$. If $n$ links enter into such a vertex $\mathbb{Z}(\lambda,w)$, it means to take $n$ derivatives. 

The picture is different when replacing recursively in Eq.\ (\ref{78b}) $u$ by its value given by Eq.~(\ref{79b}), thus  constructing again a perturbative expansion in $t$ for $\mathbb Z(\lambda, w)$. Apart from a single term $\mathbb Z(\lambda, w)$, all other terms are proportional to powers of $\partial_w \mathbb Z(\lambda, w)$, and no higher derivative of $\mathbb Z(\lambda, w)$ appears. The objects with more derivatives are $(n-2)$-nd derivatives of the disorder $[\Delta(w)-\Delta(0)]t^2$, which have $n$ outgoing lines which end either in a disorder, a $\partial_w \mathbb Z(\lambda, w)$ or $\lambda$. 
This will in more detail be discussed in \cite{LeDoussalWiese2010c}.
 
Identifying 
\begin{eqnarray}\label{x4}
 \partial_{w} \mathbb{Z} (\lambda ,w)&=& m^{2}U \\
u &=& 2 u_{1}
\end{eqnarray}
shows that Eq.\ (\ref{81b}) is equivalent to Eq.\ (\ref{Uu2}), and
(\ref{79b}) to (\ref{Uu1}). 
We still have to check expression (\ref{33}). We know that $\hat{\mathbb{Z}}
(\lambda ,w) = \mathbb{Z}
(\lambda ,w) - \lambda w = -{\cal S}_{\lambda}$.
Multiplying Eq.\ (\ref{79b}) with $\lambda$, and adding the result to (\ref{78b})  gives 
\begin{eqnarray}\label{x5}
 \mathbb Z (\lambda ,w) -\lambda w &=& \left[\Delta (u)-\Delta (0) \right] t^{2} \left[\partial_{w}
\mathbb Z (\lambda ,w) \right]^{2} \nn \\
&& -2 t^{2}\lambda  \left[\Delta (u)-\Delta (0) \right] \partial_{w}
\mathbb{Z} (\lambda ,w) \qquad 
\end{eqnarray}
The r.h.s.\ is nothing but $-\mathcal{S}_{\lambda}$, given in Eq.\
(\ref{33}).

\section{Beyond mean field: 1-loop calculation}
\label{s:1-loop}

\subsection{Simpler example: 1-point probability} 

\subsubsection{Perturbation around mean field}
\label{perturb}  

We start with the simpler case of a 1-point probability, e.g.\ as given 
by Eqs.~(\ref{ex1}) and (\ref{ex2}). As we explain in detail below, in the dimensional expansion
around mean field, the effective action can be written as 
\begin{equation}\label{pert1}
\Gamma[u] = {\cal S}[u] + \Gamma_1[u]\ .
\end{equation}
Here $S[u] \equiv S_{R}[u]$ is the improved action, and $\Gamma_1[u]$
is ``small'' in a sense to be 
specified below. Hence we can expect that $u^\lambda$ and
$u^{\lambda,\mathrm{tree}}$, the solutions of (\ref{ex2}) and
(\ref{mintree}),  are close to each others. Schematically we write 
\begin{eqnarray}\label{y6}
 {\cal S}'[u^\lambda]+ \Gamma'_1[u^\lambda] &=& \lambda \\
 {\cal S}'[u^{\lambda,\mathrm{tree}}]&=& \lambda \ .
\end{eqnarray}
Inserting into Eqs.\ (\ref{ex1}) and (\ref{ex2}) and expanding to
lowest order, i.e.\ in the differences $u^\lambda-u^{\lambda,\mathrm{tree}} \sim O(\Gamma_1)$
and  $\Gamma_1$, we find
\begin{eqnarray}
\left<\rme^{\int_x \lambda_x u_1(x)} \right>_S = \left<\rme^{\int_x \lambda_x u_1(x)} \right>^{\mathrm{tree}}_S \rme^{- \Gamma_1[u^{\lambda,{\mathrm{tree}}}] }\ .
\end{eqnarray}
Hence to compute this generating function to lowest order around the tree result we only need to evaluate $\Gamma_1$ at the tree saddle point.

\subsubsection{Effective action for the pinned interface}
 
\label{s:eafpi}
The (replica) effective action $\Gamma[u]$  associated to
the (bare) action (\ref{action0}) of the pinned interface takes the form
\begin{equation}
 \Gamma[u] = {\cal S}_{\mathrm{el}}[u] - \sum_{p=2}^\infty \sum_{a_1,\dots ,a_p} \frac{1}{T^p p!} S^{(p)}[u_{a_1},\dots ,u_{a_p}] \ .
\end{equation}
Here ${\cal S}_{\mathrm{el}}[u]=\frac{1}{T} \sum_a
H_{\mathrm{el}}[u_a]$ arises from the elastic and quadratic well
energy, defined in (\ref{def1}). The $p$-replica terms, $S^{(p)}$, are
the $p$-th cumulants of the renormalized disorder\footnote{Due to
statistical tilt symmetry,  the term $p=1$ is  a constant, dropped
here. }. 
The local part (i.e.\ $u_{a}
(x)=u_{a}$) of the second cumulant, $S^{(2)}[u_a,u_b]=L^d R(u_{ab})$,
$u_{ab}=u_a-u_{b}$ defines the  
renormalized disorder correlator $R(u)$. 

Let us now consider  $T=0$. As $m \to 0$, $R(u)$ flows to a
fixed-point function  $R=O(\epsilon)$ where 
$\epsilon=d_{\mathrm{uc}}-d$. In general $\Gamma[u]$ can be computed in an expansion in 
powers of $R$, i.e.\ of $\epsilon$. For $p=2$ one has \cite{LeDoussalWieseChauve2003,LeDoussal2008}:
\begin{align}\label{p2}
& S^{(2)}[u_a,u_b] = \int_x \sum_{ab} R\big(u_{ab}(x)\big) \\
& + \frac{1}{2} \int_{xy} (g_{xy}^2 - \delta_{xy} g_z^2) {\sf R}''\big(u_{ab}(x)\big) {\sf R}''\big(u_{ab}(y)\big) 
+ O(R^3) \nonumber\ ,
\end{align}
where here and below we denote ${\sf R}''(u):=R''(u)-R''(0)$. Purely
nonlocal parts are thus $O(\epsilon^2)$ and higher. For $p \geq 3$,
each $S^{(p)}$ is of  order $O(R^p)=O(\epsilon^p)$. They were computed
previously \cite{LeDoussalWiese2008c,ChauveLeDoussal2001}; the result can be summarized by Eq.~(\ref{pert1}) with
\begin{eqnarray}\label{g1}
 \Gamma_1[u] &=&  \frac{1}{2} \Tr \ln\big( g^{-1} \delta_{ab}  -
 W^{1}_{ab} \big)- \frac{1}{2}  
\Tr \ln\big( g^{-1} \delta_{ab}  - W^{0}_{ab} \big)  \nn  \\
&&   + \frac{I_2}{4} \,\Tr [(W^{1})^2 - (W^{0})^2] 
 \\
  W^{\kappa}_{ab,xy} &=& \frac{1}{T} \delta_{xy} [ \delta_{ab} \sum_c R''\big(u_{ac}(x)\big)  - \kappa R''\big(u_{ab}(x)\big) ]  
\end{eqnarray}
where $I_n = \int_k g_k^n$.
(\ref{g1}) has the usual expression of a 1-loop effective action,
$\frac{1}{2} \Tr\, \ln {\cal S}''$, apart from a subtraction of the
1-loop graphs leading to $p+1$ replica terms, proportional to $T$
(second term on first line) and of the $p=2$ part, already taken into
account in (\ref{p2}), thus expressing $\Gamma_{1}[u]$ as a
functional of the renormalized instead of the bare disorder. (Note
that as written $\Gamma_1$ also contains the bilocal $O(\epsilon^2)$ part of $p=2$ in Eq.~(\ref{p2})). 
Upon expanding the $\Tr\, \ln$, which acts both on replica and space
indices, the $S^{(p)}$ are recovered to $O(W^p)$; see e.g.\ formula (113) in \cite{LeDoussalWiese2008c}. 
Note that the two $O(W)$ terms cancel (using   $n=0$). 

\subsubsection{1-loop probability-distribution of displacements}
 Let us compute the 1-point generating function to lowest
order beyond mean field,¤
\begin{equation}
 \overline{ \left< \rme^{  \int_x \lambda_x u(x;w)} \right> } =  \overline{ \left< \rme^{  \int_x \lambda_x u(x;w)} \right>}^{\mathrm{tree}}
\rme^{-\Gamma_1[\{ u_a(x) \}]}\ .
\end{equation}
Here $u_a(x)$ satisfies the tree-level saddle-point  equation
\begin{equation}
\int_{x'} g_{xx'}^{ -1} [u_a(x')-w_{x'}]  -  \frac{1}{T} \sum_{c}
R'\big(u_{ac}(x)\big)  = T \lambda_x \delta_{a1} \ .
\end{equation}
Supposing that there are exactly two different fields $u_1$ and $u_{a}$ for $a>1$, this
gives 
\begin{eqnarray}\label{62}
\int_{x'} g_{xx'}^{ -1} [u_1(x')-w_{x'}]  +  \frac{1}{T}
R'\big(u_{12}(x)\big)  &=& T \lambda_x \qquad \\
\label{63}
\int_{x'} g_{xx'}^{ -1} [u_2(x')-w_{x'}]  -  \frac{1}{T}
R'\big(u_{21}(x)\big)  &=& 0
\end{eqnarray}
Note that in the first line we have dropped the term $\sim R' (0)=0$, and the sign
change comes from the factor of $(n-1)$ replica. On the other hand, from
the sum in the
second line, only the term $c=1$ survives, while all other terms are  $\sim R' (0)=0$. 

As in equation (\ref{def1bis}), we now look for a solution $u_a=u_1- T U$ for
$a \neq 1$. Eq.~(\ref{62}) and the difference between Eqs.~(\ref{62}) and (\ref{63})  become
\begin{eqnarray}\label{U1}
T \lambda_{x}&=&  \int_{x'} g_{xx'}^{-1} [u_1(x')-w_{x'}]  +
\frac{1}{T} R'\big(T U (x)\big) \qquad   \\
\label{U1bis}
U_{x}   &=&  \int_{x'}g_{xx'} \lambda_{x'} 
\end{eqnarray}
In the limit of $T\to 0$, the first equation is expanded as 
\begin{eqnarray}\label{x6}
u_1(x) &=& w_x - R''(0) \int_{x'} g_{xx'} U (x') +O (T)\nn\\
 &=& w_x - R''(0) \int_{y}\int_{z} g_{xy} g_{yz} \lambda_{z} +O (T)\ .\qquad 
\end{eqnarray}
This gives the tree contribution at $T=0$, 
\begin{equation}
 \overline{ \left< \rme^{ \int_x \lambda_{x} u (x;w)} \right>  }_{T=0}^{\mathrm{tree}} = \rme^{\int_x \lambda_x w_x - \frac{1}{2} R''(0) \int_{xx'y} \lambda_x g_{xy}g_{yx'} \lambda_{x'}  }  \ .
\end{equation}
It is  a  Gaussian distribution for the displacements at tree level (recall $R''(0)<0$). 

Let us now compute the  1-loop corrections to mean field. Here we restrict to uniform $\lambda_x=\lambda$ and  uniform $w_x=w$. Since we consider a 1-point function, $w=0$ can be chosen. The saddle point is  uniform $u_a(x)=u_a$, and 
\begin{eqnarray}\label{g2}
\frac{ \Gamma_1[u]}{L^{d}} &=&  \frac{1}{2} \int_k  \tr \ln\big( g_k^{-1} \delta_{ab}  + M^{1}_{ab} \big) \nonumber \\
&& - \frac{1}{2} 
 \int_k \tr \ln \big( g_k^{-1} \delta_{ab}  + M^{0}_{ab} \big) \nn  \\
&& + \frac{I_2}{4} \tr \left[(M^{1})^2 - (M^{0})^2\right]
 \\
 M^\kappa_{ab} &=& \frac{1}{T}  \Big\{  \delta_{ab} \Big[ R''(u_{ac})- R''(u_{a1}) \Big] + \kappa R''(u_{ab})  \Big\}\ \ \ \ \ \  \  
\end{eqnarray}
Here $\tr$ refers to replica indices, and we have used the saddle-point
properties denoting by $c$ any index with $c \neq 1$. 
More explicitly, $M^\kappa$ is a replica matrix, where replica $1$ is
singled out, and which is symmetric in the other $n-1$ replica $a \neq 1$. It
thus has (for any $n$) four distinct components (denoting  with  $a,b$  indices different from $1$):
\begin{eqnarray} \label{mkappa1}
  M^\kappa_{11} &=& \frac{1}{T} \left[R''(T U)  + (\kappa-1) R''(0)\right] \\
M^{\kappa}_{1}&=& M^\kappa_{1a}=M^\kappa_{a1}= \frac{1}{T} \kappa R''(T U) \\
 M^\kappa_{ab}& =& \delta_{ab} M^\kappa_c 
+ M^\kappa  \\
 M^\kappa_c &=& \frac{1}{T} [ R''(0) - R''(T U)  ]\\
M^\kappa &=& 
 \frac{1}{T} \kappa R''(0) \ 
\end{eqnarray}
The field $U$ is  $U=\lambda/m^2$, see Eq.\ (\ref{U1bis}). 
The diagonalization of this matrix is performed in Appendix
\ref{app:diag}. The eigenvalues are given in Eqs.\ (\ref{eigen1}) and
(\ref{eigen2}).  In
the limit $n\to 0$ they are:

 (i) a $( -2)$-dimensional space with
eigenvalue $\mu=M_c$; since $M_c^\kappa$ is independent of
$\kappa$, its contribution cancels between the two first lines of
(\ref{g2}).

(ii) a $2$-dimensional space with eigenvalues given in
(\ref{eigen2}). Since $\bar \mu^{\kappa } =M^\kappa_{11}+M^\kappa_c-M^\kappa=0$,
the eigenvalues in (\ref{eigen2}) are $\mu^{\kappa} = \pm \frac{1}{2} \sqrt
{A^{\kappa} B^{\kappa}}$. The latter vanishes for $\kappa =1$, while
for $\kappa =0$ one has $A^{\kappa}=B^{\kappa}$. Their contribution
can be regrouped leading to  
\begin{eqnarray}\label{g3}
\frac{ \Gamma_1[u]}{ L^{d}} &=&   -\frac{1}{2} \int_k \bigg[   \ln\left( 1 -  \frac{g^2_k}{T^2} \left[R''(T U)-R''(0) \right]^2 \right) \nonumber \\
&&  \hphantom{ \frac{1}{2} \int_k \Big[  }+ \frac{g^2_k}{T^2} \Big(R''(T U)-R''(0)\Big)^2 \bigg] 
\ .\qquad 
\end{eqnarray} 
The limit of $T\to 0$ can then be taken unambiguously, i.e.\
independent of the sign of $U$: 
\begin{equation}\label{g4}
\frac{ \Gamma_1[u]}{ L^{d}} =-  \frac{1}{2} \int_k \bigg [  \ln\Big( 1 - R'''(0^+)^2 g^2_k U^2 \Big)  + R'''(0^+)^2 g_k^2 U^2 \bigg ] 
\end{equation}
This gives the final result for the characteristic function of the probability distribution 
of the center of mass $u(w=0)$ of the interface, to lowest order in $\epsilon=d_{\mathrm{uc}}-d$: 
\begin{eqnarray}
&& \frac{1}{L^{d}} \ln \overline{ \left< \rme^{ \lambda L^d u(0)} \right>  } = - R''(0) \frac{\lambda^2}{2 m^4} \\
&& \quad  +  \frac{1}{2} \int_k \left[  \ln \left( 1 -
R'''(0^+)^2 \frac{g^2_k}{m^{4}} \lambda^2\right)   + R'''(0^+)^2
\frac{g_k^2}{m^{4}} \lambda^2  \right].
\nonumber 
\end{eqnarray}
This result is in agreement with Eq.~(G2) in \cite{LeDoussalWiese2008c}. At
depinning, the distribution is different, see Eq.\ (42) of
\cite{FedorenkoLeDoussalWiese2006}.

\subsection{2-point probabilities and avalanche-size distribution}
\subsubsection{General considerations}
Following similar arguments as in Section \ref{perturb}, the 2-point
generating function can be computed as  
\begin{align}\label{gam}
& \overline{ \rme^{ \int_x \lambda_x [u(x;w/2) - w - u(x;-w/2)]} }  \\
& = \rme^{\int_x \lambda_x [u^\lambda_1(x) - w - v^\lambda_1(x)] - \Gamma[u^\lambda,v^\lambda] }
\nonumber \\
& 
= \overline{ \rme^{ \int_x \lambda_x [u(x;w/2) - w - u(x;-w/2)]} }^{\mathrm{tree}}
\rme^{- \Gamma_1[u^{\lambda,\mathrm{tree}},-u^{\lambda,\mathrm{tree}}] } \nonumber 
\end{align}
In the second line, which is exact, $u^\lambda,v^\lambda$ denote the
saddle-point  solutions obtained  from $\Gamma[u,v]$. Here
$\Gamma[u,v]={\cal S}[u,v]+\Gamma_1[u,v]+O (\epsilon^{2})$ is the effective action   
of   ${\cal S}[u,v]$ given by (\ref{Slambdauv}), setting $\lambda=0$.
The third line is only correct to 1-loop order (i.e.\ lowest order in
$\epsilon = d_{\mathrm{uc}}-d$), and requires
the evaluation of $\Gamma_1$ at the tree-level saddle point; hence we
can set $u^{\lambda}=u^{\lambda,\mathrm{tree}}$, and
$v^{\lambda}= v^{\lambda,\mathrm{tree}}=-u^{\lambda,\mathrm{tree}}$.
Note that this symmetry property carries over to the effective action,
hence one can set from the outset
$\Gamma[u^\lambda,v^\lambda]\to \Gamma[u^\lambda,-u^\lambda]$. An
important property is that the dependence on $w$ of $\Gamma[u,v]$ is
the same as the one of ${\cal S}[u,v]$, i.e.\ only through the elastic
energy\footnote{This is because changing $w$  results in a shift $(w^1_a(x) \to
w^1_a(x)+w/2,w^2_a(x) \to w^2_a(x)-w/2)$ in the generating functional
$W[w^1,w^2]$ associated to $S[u,v]$.} in  (\ref{Slambdauv}). We can thus derive (\ref{gam})
w.r.t $w$, to obtain two alternative expressions:
\begin{eqnarray}\nn 
\lefteqn{ \partial_{w_y} \overline{ \rme^{ \int_x \lambda_x [u(x;w/2) - w - u(x;-w/2)] } }\Big|_{w \to 0}=} \\
& =&  - \lambda_y + \sum_a\int_{y'} \frac{g^{-1}_{yy'}}{T} u^\lambda_a(y') \label{75} \\
& =&  - \lambda_y + \sum_a\int_{y'} \frac{g^{-1}_{yy'}}{T} u^{\lambda,\mathrm{tree}}_a(y') 
- \partial_{w_y} \Gamma_1[u^{\lambda,\mathrm{tree}}_w ] \Big|_{w \to 0} \qquad  \label{useful}
\end{eqnarray}
The first one (\ref{75}) is exact in terms of the exact saddle point, hence only explicit
derivatives w.r.t.\ $w$ are needed. The second is expressed in terms of the tree saddle point and is true to 1-loop order. $\Gamma_1$ has no explicit
$w$ dependence, hence the last derivative acts only on the dependence on $w$ of the tree solution (emphasized in the notation).

\subsubsection{Computation of $\Gamma_1$} 
The general expression of $\Gamma_1[u,v]$ is given in Appendix \ref{s:appB}. 
Again we restrict to  uniform  $\lambda_x=\lambda$ and  uniform $w_x=w$,
thus we only need the expression for $\Gamma_1[u,-u]$ at the uniform tree saddle point. 
Dropping the superscripts $\lambda$ and $\mathrm{tree}$, this is 
\begin{eqnarray}\label{g1uv}
\frac{ \Gamma_1[u,-u]}{ L^{d}} &=&  \frac{1}{2} \Tr \ln\!\Big( g^{-1} \delta_{ab}  1\!\mathrm{l}  + {\mathbb{M}}^1_{ab}  \Big)\nn \\
&& - \frac{1}{2} 
\Tr \ln\Big( g^{-1} \delta_{ab} 1\!\mathrm{l} + {\mathbb{M}}^0_{ab}  \Big) \nn  \\
&&  + \frac{I_2}{4} \Tr \Big [({\mathbb{M}}^{1})^2 - ({\mathbb{M}}^{0})^2 \Big] 
\end{eqnarray}
with
\begin{eqnarray} \label{y18} 
{\mathbb{M}}^\kappa_{ab}&=& \left(\begin{array}{cc}
M^{\kappa}_{ab} & P^{\kappa}_{ab} \\
P^{\kappa}_{ab} & M^{\kappa}_{ab} \end{array} \right) 
\\
1\!\mathrm{l} &=& \left(\begin{array}{cc}
1 & 0 \\
0 & 1 \end{array} \right)\\
 M^\kappa_{ab} &=& \frac{1}{T}  \Big[  \delta_{ab}  ( - R''(u_{a1}) + R''(u_{ac}) - R''(u_{a}+u_1) \nonumber  \\
&& \hphantom{\frac{1}{T}  \Big[ }+ R''(u_{a}+u_c) )  + \kappa R''(u_{ab})  \Big]  \label{y7} \\
 P^\kappa_{ab} &=&   \frac{1}{T} \kappa R''(u_a + u_b) \ .\label{y19}
\end{eqnarray}
Here $\Tr$ refers to a trace over replica and $u,v$ indices (space), and $c$ to $c \neq 1$.
More explicitly,  the matrices $M^\kappa$ and $P^\kappa$ have for any $n$ again each  four distinct components (denoting $a,b$ indices different from $1$),
\begin{eqnarray}
  M^\kappa_{11} &=& \frac{1}{T} \Big[R''(T U)  - R''(2 u_1) + R''(2 u_1-T U) \nn\\ \label{y8}
&& \hphantom{ \frac{1}{T} \Big[ }+ (\kappa-1) R''(0)\Big] \\ \label{y9}
 M^\kappa_{1}&=& M^\kappa_{1a}=M^\kappa_{a1}= \frac{\kappa}{T}  R''(T U)\\ \label{y10}
  M^\kappa_{ab} &=& \delta_{ab} M^\kappa_c 
+ M^\kappa \\ \label{y11}
 M^\kappa_c &=& \frac{1}{T} \Big[ R''(0) - R''(T U) -  R''(2 u_1 - T U)\nn \\ \label{y12}
&&\hphantom{ \frac{1}{T} \Big[} + R''(2 u_1 - 2 T U)  \Big]\\ \label{y13}
 M^\kappa &=& \frac{ \kappa}{T} R''(0)  
\end{eqnarray}
and 
\begin{eqnarray}\label{y14}
  P_{11}^{\kappa } &=&  \frac{\kappa}{T}  R''(2 u_1) \\ \label{y15}
 P_{1}^{\kappa }&=& P_{1a}^{\kappa }=P_{a1}^{\kappa }=
 \frac{\kappa}{T}  R''(2 u_1 - T U) \qquad \\ \label{y16} 
 P_{ab}^{\kappa } &=& \delta_{ab} P_{c}^{\kappa }+ P^{\kappa }\\
P^{\kappa}_{c}&=&0\\
P^{\kappa}&=& \frac{\kappa}{T}  R''(2 u_1 - 2 T U) 
\end{eqnarray}
We recall that $U$ and $u_1$ are solutions of  (\ref{Uu1}). 

The matrix ${\mathbb{M}}^\kappa$ is diagonalized in Appendix \ref{s:appB} for $n=0$. It has 

(i) with multiplicity $-4$ the eigenvalue $M^\kappa_c$ (since $P^{\kappa}_c=0$). Since $M^\kappa_c$ in (\ref{y12}) does not depend on $\kappa$, the contribution of these eigenvalues to (\ref{g1uv}) cancels between $\kappa=1$ and $\kappa=0$. 

(ii) 2 eigenvalues each for $\sigma=\pm1$, of the form  $\mu=\frac{1}{2} (\bar \mu_\sigma \pm \sqrt{A_\sigma B_\sigma})$, where $\bar \mu_\sigma$, $A_\sigma$ and $B_\sigma$ will be calculated below.

Regrouping in Eq.~(\ref{g1uv}) we get
\begin{eqnarray}\label{g1uv-BIS}
\frac{ \Gamma_1[u,-u]}{ L^{d}} &=&  \frac{1}{2} \int_k \sum_{\sigma=\pm1} \bigg\{  \ln\!\Big( \big[g_k^{-1} + \frac{1}{2} \bar \mu^1_\sigma\big]^2 - \frac{1}{4}  A^1_\sigma B^1_\sigma\Big) \nonumber \\
&& \hphantom{ \frac{1}{2} \int_k \sum_{\sigma=\pm1} \bigg\{ }-  \ln\!\Big( \big[g_k^{-1} + \frac{1}{2} \bar \mu^0_\sigma\big]^2 - \frac{1}{4}  A^0_\sigma B^0_\sigma\Big) \nn\\
&&\hphantom{ \frac{1}{2} \int_k \sum_{\sigma=\pm1} \bigg\{ }- \mbox{quadratic part}  \bigg\}\ .
\end{eqnarray}
We find using formula (\ref{b9}), (\ref{b9n}) of  appendix  \ref{app:diag}
\begin{eqnarray}\nn
 \bar \mu^\kappa_\sigma &=& \frac{1}{T} \Big[R''(2 u_1-2 T U)-R''(2 u_1)\Big] (1- \sigma \kappa) \\
& =&2 U R'''(2 u_1) (\sigma \kappa-1) + O(T)
\end{eqnarray}
\begin{eqnarray}
 A^1_\sigma &=& (\sigma-1) U^2 R''''(2 u_1) T + O(T^2) \\
 B^1_\sigma &=& \frac{4}{T} \Big[R''(0) + \sigma R''(2 u_1)\Big] + O(T^{0}) 
\\ A^0_\sigma = B^0_\sigma &=& \frac{1}{T} \Big[ 2 R''(2 u_1-T U) -R''(2 u_1-2 T U) \nonumber \\
&& \hphantom{\frac{1}{T} \Big[}+2 R''(T U)-R''(2 u_1)-2 R''(0) \Big]\nn \\
& =& \frac{1}{T} \Big[2 R''(T U)-2 R''(0) + O(T^2)\Big] 
\end{eqnarray}
In Eq.\ (\ref{g1uv-BIS}) appears the product
 $A^0_\sigma B^0_\sigma$, which, contrary to each factor, has no ambiguity at $T=0$, 
\begin{equation}
A^0_\sigma B^0_\sigma = 4 R'''(0^+)^2 U^2 \ .
\end{equation}
Putting everything together we obtain
\begin{widetext}
\begin{eqnarray}\nn
 \frac{\Gamma_1[u,-u]}{L^{d} } &=&  \frac{1}{2} \int_k \bigg\{ 2 \ln g_k^{-1} + \ln\Big(\big[ g_k^{-1} - 2 U R'''(2 u_1)\big]^2 -  2 U^2 R''''(2 u_1) \big[R''(2 u_1)-R''(0)\big] \Big)  \\
&& \hphantom{ \frac{1}{2} \int_k \bigg\{} - 2 \ln\Big( \big[g_k^{-1} -  U R'''(2 u_1)\big]^2 - 
R'''(0^+)^2 U^2 \Big)    - \mbox{quadratic part} \bigg\}\ , \label{g1uvfinal} 
\end{eqnarray}
\end{widetext}
where subtraction of terms quadratic in $U$, or equivalently in $g_k$, is indicated. 
Inserting into formula (\ref{gam}) this gives the correction to the tree expression for the
2-point generating function for arbitrary $w$, the distance between the points. It is
expressed in terms of $U$ and $u_1$ which, we recall, are solutions of  (\ref{Uu1}).
Since $u_1 \to 0^+$ as $w \to 0^+$ we check that $\Gamma_1[u,-u]$ indeed vanishes as $w \to 0^+$
as it should from (\ref{gam}). The linear term in $w$ contains the information about avalanches  as we discuss now.

\subsubsection{Limit of $w \to 0^+$ and generating function of avalanche moments}
We now want to use formula (\ref{useful}), i.e.
\begin{eqnarray} 
 \hat Z &=& L^{-d} \partial_{w} \overline{ \rme^{ L^d \lambda (u(w/2) - w - u(-w/2)) } }\Big|_{w \to 0^+}\nn \\
& =&  - \lambda + m^2 U
- L^{-d} \partial_{w} \Gamma_1[u,-u ] \Big|_{w \to 0^+} \label{wewant}\qquad
\end{eqnarray}
In Eq.\ (\ref{g1uvfinal}) the dependence on $w$ is only contained in $U$ and $u_1$. Since $\Gamma_1[u,-u]$ vanishes for any $U$ as $w \to 0^+$, it is of the form $\Gamma_1[u,-u] = w f(U) + O(w^2)$; hence the dependence of $U$ on $w$ 
is not needed and  we can consider $U$ as its $w=0^+$ limit, i.e the
solution of $m^2 U  - R'''(0^+) U^2  = \lambda$. To obtain (\ref{wewant}) we thus replace $u_1=y w/2$, expand to linear order in $w$ 
using the $w=0^+$ limit given in (\ref{y2}), $y  = \big[{1-\frac{2 R'''(0^+)}{m^4} m^2 U}\big]^{-1}$. 
We find $\hat Z=Z - \lambda$, with 
\begin{align}\label{98}
&Z = Z_{\mathrm{tree}} + y R''''(0^+) R'''(0^+) \times \\
&\times \int_k \frac{U^2}{[g_k^{-1} {-} 2 U R'''(0^+)]^2} +  \frac{2 U^2 g_k}{g_k^{-1} - 2 U R'''(0^+)} - 3 g_k^2 U^2  \nn
\end{align}
with $Z_{\mathrm{tree}}=m^2 U$. Specifying to $g_k^{-1} =
k^{2}+m^{2}$, and rescaling $k\to k m$ yields 
\begin{align}\nn 
Z =&\, Z_{\mathrm{tree}} + \frac{ R''''(0^+) m^{d-4} Z_{\mathrm{tree}}}{1-2 S_{m} Z_{\mathrm{tree}}}\times \\
&\times \int_k\bigg[ \frac{Z_{\mathrm{tree}} S_{m}}{( k^{2}+1 - 2
Z_{\mathrm{tree}} S_{m})^2} \nn \\
& \hphantom{\times \int_k\bigg[}+  \frac{1}{  k^{2}+1 - 2
Z_{\mathrm{tree}} S_{m} }-  \frac{1}{  k^{2}+1} \nn \\
&  \hphantom{\times \int_k\bigg[} - 3 \frac{Z_{\mathrm{tree}} S_{m}}{(
1+k^{2})^{2}} \bigg ] +O (\epsilon^{2})\ ,
\label{99}
\end{align}
where the terms appear in the same order as in (\ref{98}). 
We have abbreviated the characteristic scale of avalanches, 
\begin{equation}\label{b1}
S_{m}= \frac{R''' (0^{+})}{m^{4}}
\end{equation}
already introduced in (\ref{Sm}). $R'''' (0^{+}) \sim \epsilon$ is the small
expansion parameter, and as indicated  subleading terms
are of order $\epsilon^{2}$.
Equation (\ref{99})  has the form 
\begin{equation}\label{b2}
( Z-Z_{\mathrm{tree}} )(1-2 S_{m} Z_{\mathrm{tree}}) = \epsilon \, \delta Z (Z_{\mathrm{tree}}) +O (\epsilon^{2})\ ,
\end{equation}
and since $Z-Z_{\mathrm{tree}} \sim \epsilon$, it can be rewritten as 
\begin{equation}\label{b3}
( Z-Z_{\mathrm{tree}} )\big[1- S_{m}(Z+ Z_{\mathrm{tree}} ) \big] = \epsilon \, \delta Z (Z)+O (\epsilon^{2})\ .
\end{equation}
Rearranging gives 
\begin{eqnarray}\label{b4}
Z&=& S_{m} Z^{2} + Z_{\mathrm{tree}} (1-S_{m} Z_{\mathrm{tree}})+  \epsilon\,
\delta Z (Z)+O (\epsilon^{2}) \nonumber \\
&=& S_{m} Z^{2} +\lambda +   \epsilon\,
\delta Z (Z)+O (\epsilon^{2}) 
\end{eqnarray}
Explicitly, this is 
\begin{align}\nn 
Z =& \,\lambda + S_{m}Z^{2}+ { R''''(0^+) m^{d-4} Z} \times \\
&\times \int_k\bigg[ \frac{Z S_{m}}{( k^{2}+1 - 2
Z S_{m})^2} \nn \\
& \hphantom{\times \int_k\bigg[}+  \frac{1}{  k^{2}+1 - 2
Z S_{m} }-  \frac{1}{  k^{2}+1  } \nn \\
&  \hphantom{\times \int_k\bigg[} - 3 \frac{Z S_{m}}{(
1+k^{2})^{2}} \bigg ] +O (\epsilon^{2})\ .
\label{99bis}
\end{align}
We see that $Z S_{m}$ always appear together. It is therefore useful
to introduce the dimensionless function $\tilde Z$ of the
dimensionless argument $\lambda S_{m}$, 
\begin{equation}\label{b5}
\tilde Z (\lambda S_{m}) := Z ( \lambda) S_{m}\ .
\end{equation}
Inserting into the above equation yields
\begin{align}\nn 
\tilde Z =& \,\lambda  +\tilde Z^{2}+ {\epsilon \tilde I_{2} R''''(0^+) m^{d-4} } \times \\
&\times \frac{1}{\epsilon \tilde I_{2}}\int_k\bigg[ \frac{\tilde Z^{2} }{( k^{2}+1 - 2
\tilde Z)^2} \nn \\
& \hphantom{\times\frac{1}{\epsilon \tilde I_{2}} \int_k\bigg[}+  \frac{\tilde Z}{  k^{2}+1 - 2
\tilde Z }-  \frac{\tilde Z}{  k^{2}+1  } \nn \\
&  \hphantom{\times\frac{1}{\epsilon \tilde I_{2}} \int_k\bigg[} - 3 \frac{\tilde Z^2 }{(
1+k^{2})^{2}} \bigg ] +O (\epsilon^{2})
\label{99c}\ ,
\end{align}
where the combination
\begin{equation}\label{b6}
\alpha :={\epsilon \tilde I_{2}} {R'''' (0^{+}) m^{d-4}}  \equiv \tilde R'''' (0^{+})
\end{equation}
is the fourth derivative of the rescaled renormalized disorder, and $\tilde I_{2}$, defined in equation (\ref{I2}), is the (dimensionless) 1-loop integral used to
eliminate the normalization of $\int_{k}$. The result (\ref{99c}) is
equivalent to Eq.\ (151) of
\cite{LeDoussalWiese2008c}, noting that the force-force correlator
used in  Eq.~(143) of \cite{LeDoussalWiese2008c} is $\Delta'' (0)=-R'''' (0^{+})$.

The generalization to a more general elastic kernel is straightforward,
and can be obtained replacing $k^2+1 \to \tilde g_k^{-1}$ as detailed in Appendix E of
\cite{LeDoussalWiese2008c}, and for the contact-line experiment in \cite{LeDoussalWiese2009a}.

\subsection{Avalanche-size distribution}\label{b7}
Let us recall the results for the normalized probability-distribution
function $p (s)$, defined in Eq.\ (\ref{p(S)}), as obtained in
\cite{LeDoussalMiddletonWiese2008,LeDoussalWiese2008c} for
standard elasticity:  For $d \geq 4$,
the tree or MF result is relevant. It reads
\begin{equation}
p_{\mathrm{MF}} (s) =\frac{1}{2 \sqrt{\pi}}  \rme^{-s/4}\ .
\label{meanfield}
\end{equation}
For dimension $d$ smaller than  $4$, loop corrections are
relevant. The 1-loop result, obtained by inverse-Laplace
transforming (\ref{99c}) is
\begin{equation}\label{final}
p(s) = \frac{A}{2 \sqrt{\pi}}   \exp\!\left(C \sqrt{s} -
\frac{B}{4} s^\delta\right)\ ,
\end{equation}
 with  exponents
\begin{eqnarray}\label{a64}
&& \tau = \frac{3}{2} + \frac{3}{8} \alpha =  \frac{3}{2} - \frac{1}{8} (1 - \zeta_1) \epsilon 
\\
&& \delta= 1 - \frac{\alpha}{4} = 1 + \frac{1}{12} (1 - \zeta_1) \epsilon\,
\end{eqnarray}
where $\alpha= - \frac{1}{3} (1 - \zeta_1) \epsilon$ and $\zeta_1=1/3$
for the RF class, relevant to the present study. The constants $A$,
$B$ and $C$  depend   on  $\epsilon$, and must satisfy  the
normalization conditions 
\begin{eqnarray}\label{b8}
\int_{0}^{\infty}\rmd s\, s p (s) &=& 1\\
\int_{0}^{\infty}\rmd s\, s^{2} p (s) &=& 2\ .
\end{eqnarray}
At first order in $\epsilon$ they are   
\begin{eqnarray}\label{ABC}A &=& 1+
\frac{1}{8} (2-3 \gamma_{\mathrm{E}} ) \alpha\\
B &=& 1-\alpha(1+\frac{\gamma_{\mathrm{E}}}{4})\\
C&=&- \frac{1}{2} \sqrt{\pi} \alpha\ ,
\end{eqnarray}
where $\gamma_{\mathrm{E}}=0.577216\dots $ is Euler's constant.

\section{Uncorrelated avalanches: the Brownian Force Model (BFM)}
\label{sec:bfm} 
\subsection{The BFM model}
In this section we study the Brownian-force model (BFM), which corresponds to
a Gaussian bare disorder with a force correlator in Eq.~(\ref{bare}) of
\beq \label{bareabbm} 
- R_0''(u ) = - R''_0(0) + \sigma |u| \  , \quad R'''_0(u) = \sigma ~ {\rm sign}(u)
\eeq
For a point, i.e.\ in $d=0$, $V'(u)$ performs a Brownian motion
in $u$.
The potential $V(u)$ is thus given by a so-called random acceleration
process \cite{Burkhardt1993,MajumdarRossoZoia2010,WieseMajumdarRosso2010}. In the present framework we assume that the distribution of $V'(u)$ 
has statistical translational invariance, hence the 
model needs a regularization. It can for instance be
defined in a periodic box $V(u+W)=V(u)$ with $W \to \infty$, the 
increments $V'(u_1)-V'(u_2)$ being those of the Brownian motion,
\begin{equation}
\overline{[V'(u_1) - V'(u_2)]^2} = \sigma |u_1-u_2|\ .
\end{equation} 
The zero mode then has very large fluctuations, i.e.\ $R''_0(0) = O(W)$. 
The generalization to an interface is straightforward with  $V'(u,x)$ being a
set (indexed by $x$) of mutually uncorrelated Brownian motions along $u$. Note that
a dynamical version of this model was studied in the context
of non-equilibrium depinning \cite{LeDoussalWiese2008a,LeDoussalWiese2011a}.
In $d=0$, it is known as the ABBM model (see \cite{LeDoussalWiese2008a} for a review).

A remarkable property of this model defined in the continuum,
is that it appears to be an exact fixed point of the FRG in any dimension $d$, i.e.
the renormalized disorder correlator $R(u)$ (for its definition see Section \ref{s:eafpi} and
Sections II and III of Ref.\ \cite{LeDoussalWiese2008c}) remains of the same form
as (\ref{bareabbm}).  More precisely
\begin{equation} \label{abbm}
R'''(u) = \sigma \, {\rm sign}(u)   \quad , \quad \tilde R'''(u) = \tilde \sigma \,{\rm sign}(u)\ ,
\end{equation} 
where the rescaled disorder correlator was defined in (\ref{rescR}). Its flow, i.e.
its dependence on $m$, is given by the FRG equation
\begin{equation}\label{144}
 - m \partial_m \tilde R'''(u) = (\epsilon - \zeta) \tilde R'''(u) + \beta[R]'''(u) \ .
\end{equation} 
The $\beta$-function, taken for $u>0$, contains only higher derivatives which
vanish for (\ref{abbm}), and this {\it to any loop order}. This property is detailed in 
Appendix \ref{a:FRG4BFM}, together with a stability analysis, which shows that this fixed point is  attractive. More precisely, it is at least  linearly attractive up to 2-loop order.  The
roughness exponent for the BFM can be read off from Eq.~(\ref{144}) to be $\zeta=\epsilon=4-d$. Hence
$\sigma = A_d \tilde \sigma$ with $A_0=1/4$.

At this stage, this remarkable property is not rigorously established for arbitrary $d$. In fact, some of the statements have to be qualified, see Appendix \ref {a:FRG4BFM}. 
It should be considered as a (quite solid) conjecture. In $d=0$, however, there exists some 
theorems, discussed below, which strengthen the case.

In $d=0$, this model has been studied in the context of the 1D Burgers equation \cite{Bertoin1998,CarraroDuchon1998,Valageas2009}. Let us recall
the connection. It is convenient to denote space by $w$ and consider the time dependent velocity field 
$v \equiv v_t(w)$ satisfying the Burgers equation
\begin{eqnarray}
\partial_t v + \frac{1}{2} \partial_w v^2 = \nu \partial_w^2 v
\end{eqnarray} 
in the inviscid limit $\nu=0^+$. It is solved via the Cole-Hopf
transformation \cite{Wiese1998a}
\begin{eqnarray} \label{co}
v_t(w) = \frac{1}{t} \big[w-u(w)\big] = \hat V'(w)  \ ,
\end{eqnarray} 
where $u(w)$ realizes the minimum of
\begin{eqnarray}
 \hat V(w) = \min_{u} \left[ \frac{1}{2 t} (u-w)^2 + V(u) \right]\ .
\end{eqnarray}
Hence this is exactly the  disordered model in $d=0$ with the (Burgers) ``time'' $t=1/m^2$,
taken at temperature $T=\nu/2=0^+$ (minimization condition), identical to the inviscid limit. 
At initial time $t=0$, $\hat V=V$, hence the initial velocity field is $v_{t=0}(w)=V'(w)$. 
The Burgers velocity correlator thus equals the renormalized disorder correlator,
\begin{equation}
t^2\; \overline{v_t(w_1) v_t(w_2)} =  -  R''(w_1-w_2)\ ,
\end{equation}
with $R=R_0$ at $t=0$ (i.e.\ $m=\infty$). 

Until now, these statements  were completely general. The
BFM corresponds to a choice of a random initial velocity  $v_{t=0}(w)$ with the same
increments as the Brownian motion. As a consequence, $\tilde R'''(u)$ is for all times given by (\ref{abbm}).

\subsection{Shocks in the BFM and Levy processes}
\label{s:Levy}
If we admit that there are no loop-corrections for the BFM, then
we can conjecture that the (improved) tree level (i.e.\ mean field)
result is {\it exact} for the BFM in any dimension $d$. From
the fact that $R'''(u) = \sigma$ (for $u>0$), i.e.\ all higher
derivatives vanish, the results of Section \ref{s:IIG} then show that, for $w>0$,
\begin{equation} \label{aga}
\overline{ \rme^{ L^d \lambda [u(w/2) - w - u(-w/2)]}} = \rme^{ L^d w \hat Z(\lambda)  }\ .
\end{equation}
This should hold in any $d$ for the two-point correlation of the center-of-mass 
displacements. In (\ref{aga}) $\hat Z(\lambda)$ takes the tree expression 
$\hat Z(\lambda)=Z_{\mathrm{tree}}(\lambda)-\lambda$
from (\ref{16})
\begin{equation} \label{mf2}
\hat Z(\lambda)= \frac{1}{2 S_m} \left(1- 2 S_m \lambda - \sqrt{1- 4 S_m \lambda}\right)\ ,
\end{equation}
with $S_m=\sigma/m^4=\sigma t^2$. 
In Appendix \ref{app:bfm} we show, from our saddle-point  method,
that the same holds for an {\it arbitrary number} of ordered points $w_1<w_2<\ldots<w_p$,
\begin{equation}
\label{Zh2} 
\overline{ \rme^{ L^d \sum_{i=1}^p \lambda_i [u(w_i) - w_i]}} = \rme^{ L^d \sum_{i=1}^{p-1} (w_{i+1}-w_i)
 \hat Z(\mu_i)  }
\end{equation}
with $\sum_{i=1}^p \lambda_i=0$   and $\mu_i=-\sum_{j=1}^i \lambda_j$. It admits a more general formulation
\begin{equation}
\label{Zhcontinuous}
\overline{ \rme^{ - L^d \int \rmd w\, \mu'(w) [u(w) - w]}} = \rme^{ L^d \int \rmd w\, \hat Z(\mu(w))  }
\end{equation}
for any function $\mu(w)$ which vanish at $w=\pm \infty$, derived  in Appendix \ref{a:Carraro-Duchon}.
Inserting $\mu(w) = - \sum_i \lambda_i \theta(w-w_i)$ one recovers (\ref{Zh2}).

Let us now make contact with a remarkable set of results obtained by
Carraro-Duchon and Bertoin for the Burgers equation, i.e.\ the case $d=0$ \cite{Bertoin1998,CarraroDuchon1998}.  

We recall the definition of a (homogeneous) {\it Levy process}. It is a real random function
$X(w)$, continuous on the right with a limit on the left, i.e.\ it can have jumps. It has homogeneous and independent
increments, i.e.\ $\{X(w_{i+1}) - X(w_i) \}_{i=1,\ldots ,p}$ are independent random variables
for any ordered set $w_1<w_2<\ldots <w_p$ and any $p$; and for all $w<w'$ the law of $X(w')-X(w)$ is the same
as the law of $X(w'-w)-X(0)$. Its characteristic function satisfies, for $w>0$, and
 $\omega \in i \mathbb{R}$,
\begin{equation} \label{2ptbis}
\overline{\rme^{ \omega [X(w)-X(0)] } } = \rme^{w \phi(\omega)} \ .
\end{equation}  
A Levy process is thus fully determined by its {\em Levy exponent} $\phi(\omega)$, with $\phi(0)=0$. More generally,
\begin{equation}\label{f:Carraro-Duchon1}
\overline{\rme^{ - \int \rmd w\, \omega'(w) X(w) } } = \rme^{\int \rmd w\, \phi \left(\omega(w)\right)} 
\end{equation} 
for any function $\omega(w)$ which vanishes at $w=\pm \infty$ \footnote{In Bertoin \cite{Bertoin1998} $\phi$ is called $\psi$ and $\omega$ is called $q$.}.
(\ref{2ptbis}) is recovered using $\omega(v)=\omega \theta(w-v)\theta(v)$. 
The Levy-Khintchine theorem  \cite{BertoinBook}
then establishes that $X(w)$ is a sum of a Brownian motion (with drift) and
an independent jump process, with measure $n({\sf s}) \rmd {\sf s}$. (We use sans-serif $\sf s$ in order not do confuse with $s=S/S_m$ used earlier.) Here we  need the case
of (i) only positive or zero jumps (resp.\ only negative jumps); (ii) finite first and second moments $\int \max({\sf s},{\sf s}^2) n({\sf s}) \rmd {\sf s} < \infty$.
In that case
\begin{eqnarray} \label{phi}
\phi(\omega) = b \omega + \int_{{\sf s}>0}(\rme^{- \omega {\sf s}} -1) \, n({\sf s}) \,\rmd {\sf s}  \ .
\end{eqnarray}
(The same formula holds with ${\sf s} \to -{\sf s}$ for only negative jumps). In
(\ref{phi}) $\omega$ can be taken in a domain of convergence
which includes $\mathrm{Re}(\omega) \geq 0$ (but usually is larger). 

A remarkable theorem by Carraro and Duchon \cite{CarraroDuchon1998} establishes that if
the velocity field $X_t(w)=v_t(w)$ of the inviscid Burgers equation is
a Levy process (with only negative jumps) at initial time 
(with $\phi'(0) \geq 0$), then (i) it remains a Levy process
with only negative jumps for all times; (ii) its associated Levy exponent $\phi_t(\omega)$ satisfies 
itself a Burgers equation
\begin{eqnarray}\label{f:Carraro-Duchon2}
\partial_t \phi + \phi \partial_{\omega} \phi = 0 \ .
\end{eqnarray} 
We recall in Appendix \ref{a:Carraro-Duchon} a simple-minded derivation of this formula. 
Its solution for $\omega>0$ is obtained by inverting
\begin{eqnarray}
\phi_t(\omega + t \phi_0(\omega)) = \phi_0(\omega)\ ,
\end{eqnarray}
i.e.\ $\phi_t(\omega)=\phi_0(h_t(\omega))$ where 
$h_t(\omega)$ is the inverse function of $\omega \to \omega+t \phi(\omega)$.
This was applied to the case of the initial Brownian velocity
\begin{eqnarray}
\phi_0(\omega) = \frac{a^2}{2} \omega^2 \ ,
\end{eqnarray}
leading to \cite{CarraroDuchon1998,Bertoin1998}
\begin{eqnarray}\nn
 \phi_t(\omega) &=& \frac{1 + a^2 \omega t - \sqrt{1 + 2 a^2 \omega t}}{a^2 t^2} \\
 &=&\frac{\omega}{t} + \int_{{\sf s}<0} (\rme^{\omega {\sf s}} -1)\, n({\sf s})\, \rmd {\sf s} \\
 n({\sf s})  &=& \frac{1}{a \sqrt{2 \pi t^3} |{\sf s}|^{3/2}} \rme^{- \frac{|{\sf s}|}{2 a^2 t}} \label{duchon3}\ .
\end{eqnarray}
This is the same law for the shock-size distribution as the mean-field result (\ref{mf2}) for the interface!

We can now identify the results from our present method with
those in $d=0$. Since Eq.\ (\ref{co}) gives $v_t(w) = [w - u(w)]/t$, in
$d=0$ the process $u(w)-w$ in the BFM is a Levy process 
with only positive jumps. This is consistent with the above, Eq.\ (\ref{mf2}), noting 
\begin{equation}
\hat Z(\lambda) = \phi_t(\omega = - t \lambda) 
\end{equation} 
where we recall $t=1/m^2$. The result (\ref{duchon3}) then gives the
$P(S)$ of Eqs.\ (\ref{p(S)}), (\ref{pmf}) with ${\sf s}=- m^2 S$ and $a^2 = 2 \sigma$.

To conclude, we conjecture that the BFM model for the interface
in any $d$ has center-of-mass displacements given by a Levy process with positive jumps,
i.e.\ perfectly uncorrelated shocks. In $d=0$ this was proven in \cite{CarraroDuchon1998,Bertoin1998}.
Since we argue that for interfaces for more general disorder (i.e.\ not restricted to
the BFM model but with shorter-ranged correlations) the mean-field theory becomes
exact for $d \geq d_{\mathrm{uc}}$, we conclude that at (and above) the upper critical dimension
the BFM becomes a good description (with $\zeta=0$) and the
center of mass of the interface  undergoes a Levy process. 
The $\epsilon$ expansion then allows to compute deviations from
the independent-avalanche properties.

\section{Generalization: Tree-level differential equation for an arbitrary disorder correlator}
\label{sec:generalduchon} 

We have seen in the previous section that in the case where $|R'''(u)|$ is a
constant,  (i) the generating functions for the joint probabilities of
$u(w)-w$ at an arbitrary number of points is easily computed, from the one at 2 points, (ii)
the (Levy) exponent of the 2-point generating function itself satisfies the Burgers equation, as shown by
Carraro-Duchon \cite{CarraroDuchon1998}.

Here we show an even more striking result: We find a generalization of the differential equation,
satisfied at tree level, to an {\it arbitrary} number of points, and for {\it any} disorder correlator $R(u)$. 
This equation encodes  the complete mean-field results developed in this paper.
Here we show how it arises. The question of its solution, and further applications, 
will be examined elsewhere, but for illustration  we discuss an explicit solution
for {\it periodic disorder} at the end of this section.

\subsection{Observable}

The observable of interest is
\begin{equation}
e^{\hat{\mathbb Z}_t [\lambda] } := \overline{ \rme^{\int_w \lambda(w) [u(w)-w]} } \ ,
\end{equation}
a slight generalization of (\ref{defZmath}) since it is now a functional of $\lambda(w)$ (hence the notation with a
square bracket) and contains information about multiple-point correlations. 
As in Section \ref{s:graph} we work in dimension $d=0$ or equivalently (up to the volume factor of $L^d$)
we study the center-of-mass displacement in any $d$. We compute it in the (improved) tree   approximation,
i.e.\ it is the sum of all connected tree graphs (for details of the graphical rules see section \ref{s:Graphical proof} below):
\begin{eqnarray} \label{tree1}
\hat{\mathbb Z}_t [\lambda] &: = & \overline{ \rme^{\int_w \lambda(w) [u(w)-w]} }^\mathrm{c,tree}  \nn\\
& =&\sum_{\cal G} \fig{0.7}{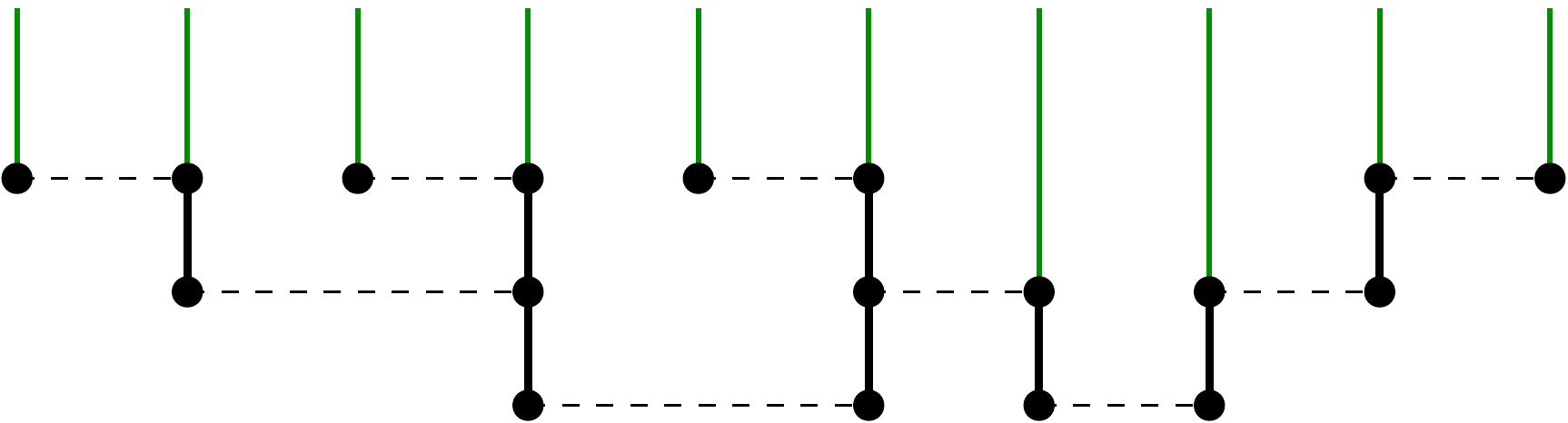} \qquad
\end{eqnarray}
It is 
obtained by the expansion of $ \rme^{\int_w \lambda(w) [u(w)-w]}$,
 where the external lines on the top link to the external $u(w_i)-w_i$ fields at
various $w_i$, following the graphical rules
defined in  section V.C of \cite{LeDoussalWiese2008c}. It will be convenient to use the notation,
\begin{equation}
\Omega(w) = - \lambda(w) t\ , \quad \quad t = 1/m^2 
\end{equation}
and loosely denote by the same symbol 
\begin{equation}
\hat{\mathbb Z}_t [\Omega] := \hat{\mathbb Z}_t [\lambda=-\Omega/t] \ .
\end{equation}
\begin{table}
\begin{tabular}{|c|}
\hline
$\displaystyle t=\frac1{m^2} \rule[-4mm]{0mm}{10mm}$\\
\hline
$\displaystyle \lambda(w) = -\mu'(w) =-\frac{\Omega(w)}t = \frac{\omega'(w)}t \rule[-4mm]{0mm}{10mm}$\\
\hline 
$e^{\hat{\mathbb Z}_t [\lambda] } := \overline{ \rme^{\int_w \lambda(w) [u(w)-w]} } \rule[-4mm]{0mm}{10mm}$
\\
\hline
$
~~\hat{\mathbb Z}_t [\Omega] = \hat{\mathbb Z}_t [\lambda=-\Omega/t]=  \hat{{\mathbb Y}}_t[\Omega,{\sf u}]\big|_{{\sf u}(w)=w}  \rule[-4mm]{0mm}{10mm}~~$
\\
\hline
$ \displaystyle {\mathbb Z}(\lambda,w)= Z(\lambda) w + O(w^2)\rule[-4mm]{0mm}{10mm}$ \\
\hline 
$ \displaystyle \hat {\mathbb Z}(\lambda,w)= {\mathbb Z}(\lambda,w)  - \lambda w \rule[-4mm]{0mm}{10mm}$ \\
\hline 
\end{tabular}
\caption{Conventions used for the various sources and generating functions.}
\label{t:conv}
\end{table}It contains information about all $n$-point cumulants $\hat C^{(n)}$ of the displacement, or  equivalently in $d=0$
of the Burgers velocity (see table \ref{t:conv} for the conventions used). $ \hat{\mathbb Z}_t [\Omega] $ can be expanded as
\begin{equation}\label{expa}
 \hat{\mathbb Z}_t [\Omega] = \sum_{p=2}^\infty \frac{1}{p!} \int_{w_1,\ldots, w_p}\!\! \Omega(w_1)\ldots \Omega(w_p) 
\hat C^{(p)}_t(w_1,\ldots,w_p) 
\end{equation}
where the cumulants
\begin{equation}
(-t)^p  \hat C^{(p)}_t(w_1,\ldots,w_p) = \overline{[u(w_1)-w_1]\ldots [u(w_p)-w_p]}^c 
\end{equation} were defined in \cite{LeDoussalWiese2008c}. There we have seen how to calculate
them at tree level as a sum over all connected tree graphs $\cal
G$ as in (\ref{tree1}) and obtained them explicitly for $p\le 4$. 

\subsection{Differential equation}

Here we show that  $\hat{\mathbb Z}_t [\Omega]$ can be obtained very elegantly
from {\it a suitable (functional) generalization of the Carraro-Duchon equation
which naturally sums up all tree graphs in the field theory}. The idea is to write an evolution
equation in the variable $t$; hence we have emphasized the dependence on
this variable.

To achieve this, one needs to generalize  $\hat{\mathbb Z}_t [\Omega]$
into a functional  $\hat {\mathbb Y}_t [\Omega,{\sf u}]$ of two variables ${\sf u}$ and $w$, 
defined as
\begin{align}\label{defY}
& \hat {\mathbb Y}_t [\Omega,{\sf u}] = \sum_{n=2}^\infty \frac{1}{n!}
 \\
&~~~ \times \int_{w_1,\ldots, w_n} \Omega(w_1)\ldots \Omega(w_n) 
\hat C^{(n)}_t({\sf u}(w_1),\ldots, {\sf u}(w_n)). \nn
\end{align}
Hence it depends on a background field ${\sf u}(w)$, not to be confused with $u(w)$, the center of
mass of the manifold in a given disorder realization, the fluctuating field which is averaged over. 
Then the following property holds:

\medskip

$\hat {\mathbb Y}_t[\Omega,u]$ is the solution of the flow equation
\begin{equation}\label{Du-gen1}
\partial_t \hat {\mathbb Y}_t[\Omega,{\sf u}] = - \int_w \frac{\delta}{\delta {\sf u}(w)} \hat {\mathbb Y}_t[\Omega,{\sf u}] \frac{\delta}{\delta \Omega(w)} \hat {\mathbb Y}_t[\Omega,{\sf u}]
\end{equation}
with initial condition 
\begin{equation}\label{initgen}
 \hat {\mathbb Y}_{t=0}[\Omega,{\sf u}] = \frac{1}{2} \int_{w,w'} \Omega (w) \Omega (w')
 \Delta \big({\sf u}(w)-{\sf u}(w')\big) \ .
\end{equation}

\subsection{Graphical proof} 
\label{s:Graphical proof}
By definition one has:
\begin{eqnarray}\label{199}
\hat {\mathbb Y}_t[\Omega,{\sf u}]&\hat = & \overline{ \rme^{-\int_w \Omega(w)t^{-1} [u(w)-w]} }^\mathrm{c,tree}  \nn\\
& =&\sum_{\cal G} \fig{0.7}{general-tree} \qquad
\end{eqnarray}
By the notation $\hat{=}$ we mean to use the graphical rules
extending the ones defined in \cite{LeDoussalWiese2008c} Sec.~V.C.
as follows:
\begin{enumerate}
\item Draw all connected tree diagrams obtained by the expansion of
$ \rme^{-\int_w \Omega(w) t^{-1}[u(w)-w]}$.               
\item Each external point is a contracted variable $-\int_w \Omega(w) t^{-1}  [u(w)-w]$, i.e.\ will contribute a factor of $-\int_w \Omega(w)/t$, hence it does not depend on the background field ${\sf u}$.
\item \label{r:3} Each dashed line is a disorder correlator, $R\big({\sf u}(w_1)-{\sf u}(w_2)\big)$, with $n_1$ derivatives taken w.r.t.\ ${\sf u} ( w_1)$ and $n_2$ derivatives taken w.r.t.\ ${\sf u} ( w_2)$, where $n_1$ and $n_2$ are the number of lines entering the left and right vertex respectively.  
\item Each solid line is a correlation function at zero momentum, $g_{q=0}=1/m^2 =t$. All points connected with such a line have the same argument $w_i$. 
In the drawing of Eq.~(\ref{199}), we have distinguished external propagators, i.e.\
lines which end in a $\Omega(w)$ (in green/grey/thin) from internal ones (bold, black). The reason is that the factor of $t$ on an {\em external} line cancels with the factor of $1/t$ which comes with each $\Omega(w)$. Thus only {\em internal} lines carry a factor of $t$. 
\item Once $\hat{{\mathbb Y}} [\Omega,{\sf u}]$ has been evaluated, one sets ${\sf u}(w)\to
w$ to get $\hat{\mathbb Z}_t [\Omega]$,  
\begin{equation}\label{x7}
\hat{\mathbb Z}_t [\Omega] = \hat{{\mathbb Y}}[\Omega,{\sf u}]\big|_{{\sf u}(w)=w}  \ .
\end{equation}
In order to allow for a recursion relation, we perform this last step
only at the end.  
\end{enumerate} 
We now show that there exists a recursion relation  for derivatives w.r.t.\
{\em  internal lines}: 

Consider $\partial_t \hat{{\mathbb Y}} [\Omega,{\sf u}]$. Since each {\em internal} line carries a factor of $t$, graphically this means a sum over all possibilities $\cal M$ to mark an internal line (here dotted, red),
\begin{align}
&\partial_t \hat {\mathbb Y}_t[\Omega,{\sf u}] \nn\\
&\quad =\sum_{\cal M} \sum_{\cal G} \fig{0.7}{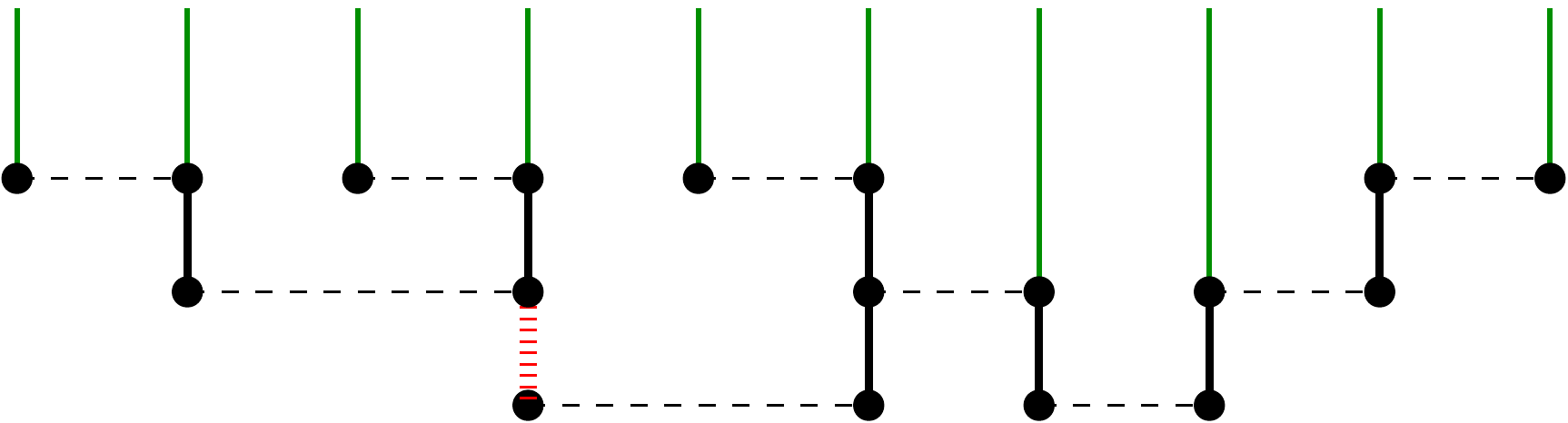} 
\end{align}
One  realizes that above and below the marked propagator appear
functional derivatives of $\hat{Y} [\Omega,{\sf u}]$ itself:
$\frac{\delta}{\delta {\sf u}(w)} \hat{Y} _t[\Omega,{\sf u}] $ at the top, and
$-\frac{\delta}{\delta \Omega(w)} \hat Y _t[\Omega,{\sf u}]$ at the bottom. This implies 
 equation (\ref{Du-gen1}). The initial condition (\ref{initgen}) is then 
made to recover the exact second cumulant $\hat C^{(2)}$, i.e.
the quadratic term in the graphical expansion.

\subsection{Consequences and particular cases}
The above mean-field differential equation is remarkable in
several respects. First it allows to compute explicitly the $n$-point function
$\hat C^{(n)}$ by  integration of (\ref{Du-gen1}) in a small-$t$ expansion. One immediately checks that 
 the terms of order $\Omega^{3}$ and $\Omega^{4}$ coincide with the 
 expressions (59)--(61)   of
\cite{LeDoussalWiese2008c}. 

We now analyze some special cases, for which the above equations  simplify. 
Suppose we want to compute the $p$-point expectation
\bea
e^{\hat {\mathbb Z}_{t}(\{\omega_{i},w_{i} \})}
:= \overline{ \rme^{-\frac1t \sum_{i=1}^p  \omega_{i}  [u(w_{i})-w_{i}]} } \eea
within the tree approximation. We can use the above formalism with the choice
 \begin{equation} \label{delt}
\Omega(w) :=\sum_{i=1}^p \omega_i \delta(w-w_i)\ .
\end{equation}
The two functions $\Omega (w)$ and ${\sf u}(w)$ have been
replaced by the two sets of discrete variables $\omega_{i}$ and
$w_{i}$. The variation w.r.t.\ ${\sf u} (w)$ gets replaced by the derivative
w.r.t.\ $w_{i}$, s.t.\ one can write the recursion relation directly
for $\hat {\mathbb{Z}}_{t} $, 
\begin{align}
& \hat {\mathbb{Z}}_{t}(\{\omega_{i},w_{i} \}) =  \overline{ \rme^{-\frac1t \sum_{i=1}^p  \omega_{i}  [u(w_{i})-w_{i}]} }^\mathrm{c,tree}  \nn\\
& =\sum _{\cal G}\fig{0.7}{general-tree} 
\end{align}
Rule \ref{r:3}  now  gives $R(w_1-w_2)$ instead of
$R\big({\sf u}(w_1)-{\sf u}(w_2)\big)$. 
The functional differential Eq.\ (\ref{Du-gen1}) simplifies to an ordinary
 differential equation, 
\begin{equation}\label{169}
\partial_t \hat {\mathbb{Z}}_t (\{\omega_{i},w_{i} \}) = - \sum_{i=1}^p \frac{\partial}{\partial w_{i}} \hat {\mathbb{Z}}_t(\{\omega_{i},w_{i} \}) \frac{\partial}{\partial \omega_{i}} \hat {\mathbb{Z}}_t(\{\omega_{i},w_{i} \})\ .
\end{equation}
The initial condition to this equation is
\begin{equation}\label{x8}
 \hat {\mathbb{Z}}_{t=0} (\{\omega_{i},w_{i} \}) =\frac{1}{2} \sum_{i,j=1}^p
 \omega_{i}\omega_{j} \Delta (w_{i}-w_j) \ .
\end{equation}
Solving Eq.\ (\ref{169}) iteratively in powers of $t$ reproduces again
the
$n$-point functions of \cite{LeDoussalWiese2008c}, Eqs.\ (59)--(61). 
Indeed one also has, from (\ref{expa}) and (\ref{delt}):
\bea
&&  \hat {\mathbb{Z}}_t (\{\omega_{i},w_{i} \}) \\
 && = \sum_{n=2}^\infty \frac{1}{n!} 
\sum_{i_1,\ldots ,i_n=1}^p \omega_{i_1} \ldots \omega_{i_n} \hat C_t^{(n)}(w_{i_1},\ldots ,w_{i_n}) \nn
\eea

Formula (\ref{169}) simplifies even more for the $p=2$ generating function
expressed as a function of the position-difference, 
\begin{equation}\label{x9}
\rme^{\hat{{\mathbb{Z}}}_{t}(\omega ,w)}  :=  \overline{ \rme^{-\frac{\omega}{t}  [u
(w)-u (0)-w]}} \ .
\end{equation}
Equation (\ref{169}) and the initial condition  become 
\begin{eqnarray}\label{173}
\partial_{t} \hat{{\mathbb{Z}}}_{t} (\omega,u) &=& - \frac{\partial}{\partial w}
\hat{{\mathbb{Z}}}_{t} (\omega,w) \frac{\partial}{\partial \omega} \hat{{\mathbb{Z}}}_{t}
(\omega,w)  \qquad \\ \label{174}
\hat{{\mathbb{Z}}}_{t=0} (\omega,w) &=& \omega^{2} \left[\Delta (0)-\Delta (w)  \right]\ .
\end{eqnarray}
As an example, we give the solution up to order $t^{2}$, 
\begin{eqnarray} \label{expandZt}
 \hat {\mathbb{Z}}_t(\omega,w) &=& \omega^2 \Big[\Delta(0)-\Delta(w) \Big] \\
&& + 2 \omega^3 t \Delta'(w)  \Big[ \Delta(0)-\Delta(w) \Big] \nn \\
&& 
+ \omega^4 t^2 \Big[ 5 (\Delta(0)-\Delta(w)) \Delta'(w)^2 \nn\\
&& \qquad\ \ \  - 2 \Big(\Delta(0)-\Delta(w)\Big)^2 \Delta''(w) \Big] + O(t^3) \nn
\end{eqnarray}
This recursion easily reproduces the 6 first connected Kolmogorov cumulants, explicitly
calculated in \cite{LeDoussalWiese2008c}, Eqs.\ (62)-(66).

There is an interesting property related to the expansion in $w$: If one writes
\beq
\hat {\mathbb{Z}}_t(\omega,w) = \sum_{n=1}^\infty w^n z_n(\omega,t)\ , 
\eeq
then the equations for the $z_p(\omega,t)$, $p \leq n$, close, i.e.
\bea
&& \partial_t z_1 + z_1 \partial_\omega z_1 = 0\ , \\
&& \partial_t z_2 + 2 z_2 \partial_\omega z_1 + z_1 \partial_\omega z_2 = 0 \ ,\label{z2} \\
&&  \partial_t z_3 + 3 z_3 \partial_\omega z_1 + 2 z_2 \partial_\omega z_2 + z_1 \partial_\omega z_3 = 0\ .\qquad 
\eea 
More generally
\beq\label{216}
\partial_t z_n + \sum_{q=1}^n q z_q \partial_\omega z_{n-q+1} = 0 \ ,
\eeq
with initial conditions $z_n(\omega,t=0)= - \frac{1}{n!} \omega^2 \Delta^{(n)}(0^+)$. 

One particular solution of these equations is $z_n(\omega,t)=0$ for $n \geq 2$. It
corresponds to the BFM  (where $\Delta^{(n)}(0^+) = 0$ for $n \geq 2$), discussed in the previous section, and 
thus to the equation (\ref{f:Carraro-Duchon2}) originally derived by Carraro and Duchon
\cite{CarraroDuchon1998}. It describes a Levy process with exponent
$z_1(\omega,t)=\phi_t(\omega)$. The initial condition is
\beq\label{178}
z_1(\omega,t=0) = \phi_{t=0} (\omega) =- \omega^{2} \Delta' (0^{+}) = \omega^{2} \sigma\ . 
\eeq 
$z_1$ is uniquely determined by its initial condition, hence we can use the result (\ref{duchon3})
\beq
z_1(\omega,t) =  \frac{1 + 2 \sigma \omega t - \sqrt{1 + 4 \sigma \omega t}}{2 \sigma t^2} \ .
\eeq
Inserting into (\ref{z2}) we find the general solution for $z_2$,
\beq
z_2(\omega,t) = \frac{F\left( \frac{1 + \sqrt{1+ 4 \sigma \omega t}}{4 \omega \sigma} \right) }{1 + 4 \omega \sigma t }\ .
\eeq
This can be seen by introducing $a=\ln \omega$ and $b=\ln(1+\sqrt{1+4 \sigma t \omega})$ in which
variables one gets $(\partial_a+\partial_b) z_2 = \frac{2}{e^{-b}-1} z_2$. The function $F(x)$ is determined
by the initial condition as
\bea
F(x) = - \frac{1}{{ 8} x^2} \frac{\Delta''(0^+)}{\Delta'(0^+)^2} \ .
\eea 
Hence
\beq
z_2(\omega,t) = - \frac{2 \Delta''(0^+) \omega^2}{ (1 + \sqrt{1+ 4 \sigma \omega t})^2 (1 + 4 \omega \sigma t )}\ .
\eeq
Now one can check that $z_1$ and $z_2$ reproduce  the terms $O(w)$ and $O(w^2)$ in (\ref{expand}) 
obtained there by a completely different method. 
For the general case one can determine the $z_n$ recursively. Their systematic study is left for the future.

\subsection{Connection with Exact RG equations}

One easily sees that the generalized Carraro-Duchon equation (\ref{Du-gen1}) together with the definition 
(\ref{defY}) is equivalent to the following RG equation for the cumulants
\begin{align} \label{carraroduchonCum}
& \partial_t \hat C_t^{(n)}(w_1,\ldots, w_n)  \\
& = - \sum_{p,q,p+q=n+1} \frac{n!}{(p-1)! (q-1)!} \times \nn \\
&~~~~ \times  [  \hat C_t^{(p)}(w_1,w_2,\ldots, w_p)) \partial_{w_1} \hat C_t^{(q)}(w_1,w_{p+1},\ldots, w_n) ]\ . \nn
\end{align}
Here $[\ldots ]$  means symmetrization over the $n$ variables $w_1,\ldots,w_n$. The summation over $p,q$ is for $p,q \geq 2$
 in case of STS and $p,q \geq 1$  in the absence of STS. One can show that if $\hat C^{(1)}=0$ at $t=0$,
 it remains so. In that case the equation for $\hat C^{(n)}$ involves only $\hat C^{(n-1)}$. In  appendix 
 \ref{app:erg} we recall the Exact RG (ERG) equations in $d=0$, and show that neglecting one term in these
 equations (which corresponds to loop corrections) we indeed recover (\ref{carraroduchonCum}) which hence appears as a
 tree approximation
 
\subsection{Including loops}
It is shown in  Appendix \ref{s:LpaBfm} that $\hat{\mathbb Z}_t[\Omega]$ satisfies a 
more general evolution equation
\beq\label{newF7bis}
\partial_t  \hat{\mathbb Z}_t =   \frac{1}{2} 
   \int_{w} \Omega'(w) \left[  \frac{\delta^2  \hat{\mathbb Z}_t}{\delta \Omega(w)^2} 
 +
 \frac{\delta  \hat{\mathbb Z}_t}{\delta \Omega(w)} 
   \frac{\delta  \hat{\mathbb Z}_t}{\delta \Omega(w)} \right]\ .
\eeq
This equation is exact (i.e.\ valid beyond the tree approximation)
and equivalent to the ERG equations given in Appendix \ref{app:erg}.
Neglecting the first term corresponds to the tree approximation,
and is equivalent to the recursion of  moments (\ref{carraroduchonCum}). As shown in Appendix \ref{s:LpaBfm}, Eq.~(\ref{newF7bis}) can also be obtained by replacing the tree level equation (\ref{Du-gen1})
by the equivalently exact equation 
\bea \label{Du-gen-new}
 \partial_t \hat {\mathbb Y}_t[\Omega,{\sf u}] &=& 
- 
\int_w \lim_{w'\to w} [\frac{\delta}{\delta \Omega(w')} \frac{\delta}{\delta {\sf u}(w)} \hat {\mathbb Y}_t[\Omega,{\sf u}] ] \nn \\
&&  
- \int_w \frac{\delta}{\delta {\sf u}(w)} \hat {\mathbb Y}_t[\Omega,{\sf u}] \frac{\delta}{\delta \Omega(w)} \hat {\mathbb Y}_t[\Omega,{\sf u}]\ ,\qquad 
\eea
The first term generates all loop corrections. 

We now consider the BFM, with statistical translation invariance. In appendix \ref{s:LpaBfm} we show that then a solution  for $\hat{\mathbb Z}_t$ can be obtained from
\bea \label{aga2bis}
\hat{\mathbb Z}_t &=&  f_t(\omega_\infty) + \int_{w_1} \phi_t(\omega(w_1), \omega_{\infty})\\
\omega(w) &=& \int_w^{\infty} \rmd w'\, \Omega(w') \  , \quad  \omega_{\infty} = \int_w \Omega(w) \qquad  \\
\label{trara1bis}
 \partial_t \phi_t(x,y) &=& - \phi_t(x,y) \partial_1 \phi_t(x,y)  \\
 \partial_t f_t(y) &=& \frac{1}{2} \Big[\partial_1 \phi_t(0,y) + \partial_1 \phi_t(y,y) \Big] \ ,\qquad \label{trara2bis}
\eea
where $\partial_1$ denotes the partial derivative w.r.t.\ the first argument.
The initial condition for the BFM is
\bea\label{init1bis}
 \phi_{t=0}(x,y) &=& \sigma x^2 - \sigma x  y\ ,  \\
 f_{t=0}(y) &=& \frac{1}{2} \Delta(0) y^2 \ .\label{init2bis}
\eea 
The solution of the system (\ref{trara1bis})--(\ref{trara2bis}) with this  initial condition is
\bea 
 \phi_{t}(x,y) &=& \frac{1}{2 \sigma t^2} \bigg[1 + 2 \sigma t \left(x - \frac{y}{2}\right) \\
&&\qquad - \sqrt{ 1 + 4 \sigma t \left(x - \frac{y}{2}\right) + \sigma^2 t^2 y^2 } \,\bigg] \nn \\
 f_t(y) &=& \frac{1}{2} \ln( 1 - t^2 s^2 y^2) + \frac{1}{2} \Delta_0(0) y^2 \ .
\eea 
The $\ln$ term corresponds to 1-loop corrections, while for this model  higher loop contributions identically vanish, as
discussed in appendices \ref{s:LpaBfm} and \ref{sec:loops}.

\subsection{Periodic case}
\label{sec:periodic} 

In the periodic case in any dimension it is conjectured that $R''(u) -
R''(0)  = R'''(0^+) u(1-u).$ Noting $ \sigma:=R''' (0^{+})$, we have
$r_{4}:=R'''' (0^{+})=-2\sigma$. 

Here we compute (in $d=0$ for simplicity but extension is straightforward)
the most general 2-point generating function using an arbitrary value for
$\sigma$. The calculation is performed in Appendix
\ref{a:periodic}.
The general result for any function $\lambda(w)=-\mu'(w)$ on the
interval $[0^-,1^-]$ (this is sufficient
since $u(w)-w$ is periodic) with $\mu(0)=\mu(1)$ is
\begin{align}
& \overline{\left<\rme^{ \int_w \lambda(w) [u(w) - w] } \right>}^{\mathrm{tree}} = \rme^{ -S_\lambda} \\
&  -{\cal S}_\lambda = - \int_0^1 \rmd w\,[\mu(w) + m^2 V(w)] + \sigma  A^2 \ .
\end{align} 
\beq
V(w) = \frac{m^2 -2 \sigma   A}{2 \sigma} \left(\sqrt{1 - 
\frac{4\sigma}{(m^2 -2 \sigma A)^2} \mu(w) } -1\right)\ .
\eeq
 $A$ is given by the self-consistent equation
\begin{eqnarray}
A =  \int_0^1 \rmd w\, \frac{m^2 V(w)}{m^2  -2\sigma  A + 2 \sigma V(w)}\ .
\end{eqnarray}
From this  the 2-point function is obtained by taking $\mu(w')= \lambda
\theta(0<w'<w)$: 
\begin{align}
& \overline{\left<\rme^{ \lambda [u(w) - w -u (0)] } \right>}^{\mathrm{tree}} = \rme^{ -S_\lambda} \\
& -{\cal S}_\lambda = - w (\lambda +  m^2 V)     +\sigma  A^2 \label{223}
\end{align}
$V$ and $A$ both depend on the length of the interval $w$, and the
function $V (w')$ introduced above is $V (w') = \theta(0<w'<w) V$. The
self-consistent equations are 
\begin{eqnarray}
&& V  m^{2} -2 \sigma  A V + \sigma  V^2 = - \lambda \label{224}\\
&& A ( m^{2} -2 \sigma A + 2\sigma  V) = {w V}  m^{2}  \label{225}
\end{eqnarray}
to be solved for the branch such that $V = \frac{-\lambda}{m^2}+O(\lambda^2)$ and
$A = \frac{-\lambda w}{m^2}+O(\lambda^2)$.
Up to order $t^{10}$, or equivalently $\lambda^{6}$ or $\sigma^{5}$ this gives,
denoting  $t=1/m^2$:  
\begin{eqnarray}\label{x13}
 -{\cal S}_\lambda&=& - (\lambda t)^2 \sigma  (w-1) w+2 (\lambda t)^3 \sigma ^2 t (w-1) w (2 w-1)\nn \\
&& - (\lambda t)^4 \sigma ^3 t^2 (w-1) w (24 (w-1) w+5)\nn \\
&& +2 (\lambda t)^5 \sigma ^4 t^3 (w-1)
   w (2 w-1) (44 (w-1) w+7)\nn \\
&& -2 (\lambda t)^6 \sigma ^5 t^4 (w-1) w \times \nn\\
&& \qquad \times (52
   (w-1) w (14 (w-1) w+5)+21) \nn \\
&& +O (t^{12},\lambda^7)
\end{eqnarray}
This is in agreement both with our previous expansions in (\ref{expand}) (up to order $w^3$) and 
with (\ref{expandZt}).

\section{Conclusion}\label{y17}

In this paper we have presented novel and efficient algebraic tools to study  multi-point
correlations of the displacement field of an elastic
manifold of internal dimension $d$ in a random potential, upon variation of an external parameter. In $d=0$ these identify with the 
correlations of the Burgers velocity field with random initial conditions
(playing the role of the disorder).  Such correlations are of 
interest in the field of turbulence. In both cases, they yield the 
statistics of avalanches, i.e.\ shocks in Burgers.

The first method uses replica. The saddle-point equations obtained
at $T=0$ resum all tree diagrams and yield among others the avalanche-size distribution in the mean-field limit, i.e.\ for $d \geq d_{\mathrm{uc}}$. We have then extended this method
to compute the 1-loop corrections. It allowed us to derive the
1-loop avalanche-size distribution more systematically than in
our previous work  \cite{LeDoussalWiese2008c}, providing an independent check of the latter.
This method has a natural extension to the dynamics, which allows
to compute the distribution of velocities in an avalanche near the depinning
transition \cite{LeDoussalWiese2011a,DobrinevskiLeDoussalWiesetobe,LeDoussalWiese2010b}. Apart from the avalanche-size
distribution, other distributions, which were obtained by a resummation of diagrams,
as the width of an interface 
\cite{RossoKrauthLeDoussalVannimenusWiese2003,LeDoussalWiese2003a}, or
the distribution of critical forces at depinning 
\cite{FedorenkoLeDoussalWiese2006}, should now be obtainable in a purely
algebraic way. 

The second method arises from the study of the Brownian force model (BFM). 
That model has the unique property that ``its mean-field treatment is exact'', i.e.
summation of tree diagram yields (almost) the exact result. We have argued that
this property holds in any $d$. We also proved the stability, i.e.\ attractive character of this model under RG 
to one loop, but we believe it to be valid more generally. In $d=0$ this model identifies 
with the Burgers equation with a (stationary) Brownian initial condition. We 
recalled results from the mathematical literature: at all times the velocity field
remains a Levy process, implying that the shocks are uncorrelated. 
Furthermore it was shown that the Levy exponent of this process obeys itself a  Burgers
 evolution equation in time, the Carraro-Duchon equation. We then pointed out
a more general connection between the Carraro-Duchon equation and the mean-field
theory of elastic manifolds, not restricted to the BFM, which allowed us to:
(i) show that avalanches in elastic manifolds at and above their upper critical dimension 
are described by a Levy process; 
(ii) derive a Generalized Carraro-Duchon functional
equation, which is in essence the exact RG equation satisfied by the mean-field theory, i.e.\ the sum of
all tree graphs. This allows in particular to recover very efficiently most of the results of the first method
at the level of the mean-field theory. Extensions
including loop corrections were presented, but their study was left for the future.
They should in principle lead to another, maybe more powerful method to
study loop corrections. 

Both methods presented here have recently been extended to a manifold with a $N$-component
displacement field, and the results are presented in \cite{LeDoussalWiese2010c,LeDoussalRossoWiese2011}.

\begin{acknowledgments} We are very grateful to V.\ Vargas for pointing out Ref.\ \cite{CarraroDuchon1998} and for a useful discussion on Levy processes. We thank J.L.\ Jacobsen for an inspiring discussion on tree resummations. 
\end{acknowledgments}

\appendix

\section{General non-uniform $w_x$}
\label{appgen}
In the general case of the 2-point function (\ref{2pt}) at points $w_x= \pm \frac{w}{2} f(x)$ 
we need to analyze Eq.\ (\ref{a3}). Expanding for small $w$, we find
\begin{equation}\label{y4}
\int_{x'} g_{xx'}^{-1}\left[y (x')-f (x') \right] = 2 R''' (0^{+}) \left|y
(x) \right| U (x) \ .
\end{equation} 
The other equation to be satisfied is 
\begin{equation}\label{41bis}
\int_{x'}g^{-1}_{xx'} U(x')   - R''' (0^{+})\, {\rm sgn}\big(y(x)\big)\, U(x)^2  =\lambda_{x}\ .
\end{equation}
These equations can first be studied as an expansion in small $\lambda_x$:
\begin{eqnarray}\label{AA}
 y(x) &=& f(x) + 2 R'''(0^+) \int_{x'x''} g_{xx'}  |f(x')| g_{x'x''} \lambda_{x''} 
\nn\\ &&+ \ldots    \\
U(x) &=& g_{xx'} \lambda_{x'} + R'''(0^+) g_{xx'}  {\rm sgn}(f(x')) (g_{x'x''} \lambda_{x''})^2  \nn \\
&& + \ldots \ .\label{aa}
\end{eqnarray}
The second-order part of $U(x)$ allows to retrieve the second moment of 
local avalanche sizes, from (\ref{local}),  
\begin{eqnarray}
\rho_0^f \langle S_x S_{x'} \rangle = 2 R'''(0^+) g_{xx''} g_{x'x''} |f(x'')|\ .
\end{eqnarray}
These equations were studied in \cite{LeDoussalWiese2008c} in the case $f(x)=1$
and $\lambda_x=\lambda \delta(x)$. In that case $y(x)>0$ does not change sign
and the equation (\ref{41bis}) can be studied separately. More generally however, 
we see from (\ref{AA}) that if $f(x)$ changes sign, $y(x)$ will also change sign (at least
for small enough $\lambda_x$), but not necessarily  at the  same location.  
A general analysis of these equations demands a more thorough study.

\section{Diagonalization of replica matrices}
\label{app:diag}
\subsection{1-point formulas}\label{x14}

Consider a replica matrix ${ M}$ as in (\ref{mkappa1}) specified by the four
components ${ M}_{11}$, ${ M}_{1a}={ M}_{a1}=:{M}_1$ for $a\neq 1$, 
${ M}_{aa}=:M_c+M$ and ${ M}_{ab}=:M$, where $a \neq b$ are two arbitrary
replica indices distinct from 1. The eigenspaces 
can be split into two groups:

(i) a 2-dimensional subspace of vectors of the form
\begin{equation}
V = \left(\begin{array}{c}
 v_1 \\
 v ~ \vec \omega
 \end{array} \right) \quad , \quad \vec \omega = \left(\begin{array}{c}
 1 \\
 \vdots \\
1 
 \end{array} \right)\ .
\end{equation}
There the action of ${ M}$ reduces to a simple $2\times 2$ matrix:
\begin{eqnarray}
 {  M} V  &=&    \left(\begin{array}{c}
 v'_1 \\
 v' ~ \vec \omega
 \end{array} \right) \\
 \left(\begin{array}{c}
 v'_1 \\
 v'
 \end{array} \right) &=&
 \left(\begin{array}{cc}
{ M}_{11}& (n-1) M_{1} \\
M_1 & ~M_c + (n-1) M \end{array} \right) \left(\begin{array}{c}
 v_1 \\
 v 
 \end{array} \right)\ .\qquad  \label{A3}
\end{eqnarray}
(ii) $n-2$ eigenvectors associated to the eigenvalue $\mu=M_c$:
\begin{equation}
V_p = \left(\begin{array}{c}
 0 \\
  \vec \omega^{(p)}
 \end{array} \right) \quad , \quad \vec \omega^{(p)} = \left(\begin{array}{c}
 \omega^{(p)}_1 \\
 \ldots  \\
 \omega^{(p)}_{n-1} 
 \end{array} \right)\ ,
\end{equation}
with $\omega^{(p)}_{j}  = \rme^{ p j \frac{2 i \pi}{n-1}}$, ($j=1,\ldots,n-1$; $p=1,\ldots,n-2$),
the $( n-1)$-vector constructed from the $( n-1)$-th root of unity, with $\sum_{j=1}^{n-1} \omega^{(p)}_{j} =0$. 

To summarize, the eigenvalues and multiplicities are for $n\to 0$
\begin{align}\label{eigen1}
 \mu&=M_c \ , \  &&{\rm multiplicity} ~ -2  \\ \label{eigen2}
 \mu &= \frac{1}{2} (\bar \mu \pm \sqrt{A B}) \ , \ &&
\mbox{multiplicity 1 for each sign}  
\end{align}
with
\begin{eqnarray}
&& \bar \mu  = { M}_{11} + M_c -M  \\
&& A= { M}_{11} + M - M_c -2 M_1   \\
&& B = { M}_{11} + M - M_c +2 M_1\ .   
\end{eqnarray}

\subsection{2-point formulas}
The same analysis can be repeated for the $2n \times 2n$ symmetric
matrix $\mathbb{M}$, with 
\begin{eqnarray}
{\mathbb{M}}^\kappa_{ab}= \left(\begin{array}{cc}
{ M}^{\kappa}_{ab} & P^{\kappa}_{ab} \\
P^{\kappa}_{ab} & { M}^{\kappa}_{ab} \end{array} \right) \ .
\end{eqnarray}
Since the $2\times 2$ structure is obviously diagonalized by the symmetric $ \left(\begin{array}{c}
 1 \\
 1
 \end{array} \right)$ and antisymmetric $ \left(\begin{array}{c}
 1 \\
- 1
 \end{array} \right)$ combination, the task reduces to finding the
 eigenvalues of ${ P}^{\kappa }\pm { M}^{\kappa }$; hence the
 first $-4$ eigenvalues are for  $n\to 0$ obtained from
 (\ref{eigen1}), 
\begin{eqnarray} \label{eigen3}
 \mu&=&M_c+P_c \quad , \quad {\rm multiplicity} ~ - 2 \ \\
 \label{eigen4}
 \mu&=&M_c-P_c \quad , \quad {\rm multiplicity} ~ - 2 \ .
\end{eqnarray}
The remaining four eigenvalues are according to (\ref{A3}) the  eigenvalues of the two following $2\times 2$
matrices
\begin{eqnarray}
{\cal U}_{+}&=&   \left(\begin{array}{cc}
M_{11}+ P_{11} & - (M_{1}+P_1) \\
M_1+P_1 & M_c - M +P_c - P \end{array} \right) \ ,  \\
{\cal U}_{-}&=& \left(\begin{array}{cc}
M_{11}- P_{11} & - (M_{1}-P_1) \\
M_1-P_1 & M_c - M -P_c + P \end{array} \right) \ .\qquad  
\end{eqnarray}
These  four eigenvalues are 
\begin{equation}\label{b9}
\mu_{\sigma,\pm}  = \frac{1}{2} (\bar \mu_\sigma \pm \sqrt{A_\sigma B_\sigma})\ ,
\end{equation}
with $\sigma=\pm1$, and 
\begin{eqnarray} \label{b9n}
 \bar \mu_\sigma &=& M_{11} + M_c -M + \sigma (P_{11} + P_c -P) \\
 A_\sigma&=& M_{11} + M - M_c -2 M_1 + \sigma (P_{11} + P - P_c -2 P_1) \nonumber \\
 B_\sigma &=& M_{11} + M - M_c +2 M_1 + \sigma (P_{11} + P - P_c +2 P_1) \ .\nonumber 
\end{eqnarray}

\section{$\Gamma_1[u,v]$}
\label{s:appB}
The general expression of $\Gamma_1[u,v]$ for 2-point observables is an extension of (\ref{g1}):\begin{eqnarray}\label{g1uvAppB}
 \Gamma_1[u,v] &=&  \frac{1}{2} \Tr \ln\left( g^{-1} \delta_{ab}  \mbox{1\!l} - {\mathbb{W}}^1_{ab}  \right) \\
&& - \frac{1}{2} 
\Tr \ln( g^{-1} \delta_{ab} \mbox{1\!l} - {\mathbb{W}}^0_{ab} ) \nn\\&& + \frac{I_2}{4} \Tr \left [({\mathbb{W}}^{1})^2 - (\mathbb{W}^{0})^2 \right] 
\nonumber \\
{\mathbb{W}}^\kappa_{ab}&=& \left(\begin{array}{cc}
W^{\kappa,uu}_{ab} & W^{\kappa,uv}_{ab} \\
W^{\kappa,vu}_{ab} & W^{\kappa,vv}_{ab} \end{array} \right) 
\end{eqnarray}
with
\begin{eqnarray}
  W^{\kappa,uu}_{ab,xy} &=& \frac{1}{T} \delta_{xy} \bigg[ \delta_{ab} \sum_c R''\big(u_{ac}(x) \big)  - \kappa R''\big(u_{ab}(x) \big) \nn \\
&& \hphantom{ \frac{1}{T} \delta_{xy} \bigg[}
+ \delta_{ab} \sum_c R''\big(u_{a}(x)-v_c(x) \big) \bigg] \\ W^{\kappa,vv}_{ab,xy} &=& \frac{1}{T} \delta_{xy} \bigg[ \delta_{ab} \sum_c R''\big(v_{ac}(x) \big)  - \kappa R''\big(v_{ab}(x) \big) \nn \\
&&\hphantom{ \frac{1}{T} \delta_{xy} \bigg[}
+ \delta_{ab} \sum_c R''\big(v_{a}(x)-u_c(x) \big) \bigg] \\
 W^{\kappa,uv}_{ab,xy} &=& -  \frac{1}{T} \delta_{xy} R''\big(u_{a}(x)-v_b(x) \big) = W^{\kappa,vu}_{ab,xy} 
\end{eqnarray}

\section{Diagrammatic representation of  1-loop corrections}
\label{s:diagrammatic}
Let us recall the graphical interpretation of the (improved) tree-level self-consistency equation.  
As already discussed in the text $Z_{\mathrm{MF}} = \lambda +S_{m}
Z_{\mathrm{MF}}^{2}$ is graphically written as 
\begin{equation}\label{C3}
Z_{\mathrm{MF}}=\   \parbox{0.8\figwidth}{\fig{.8}{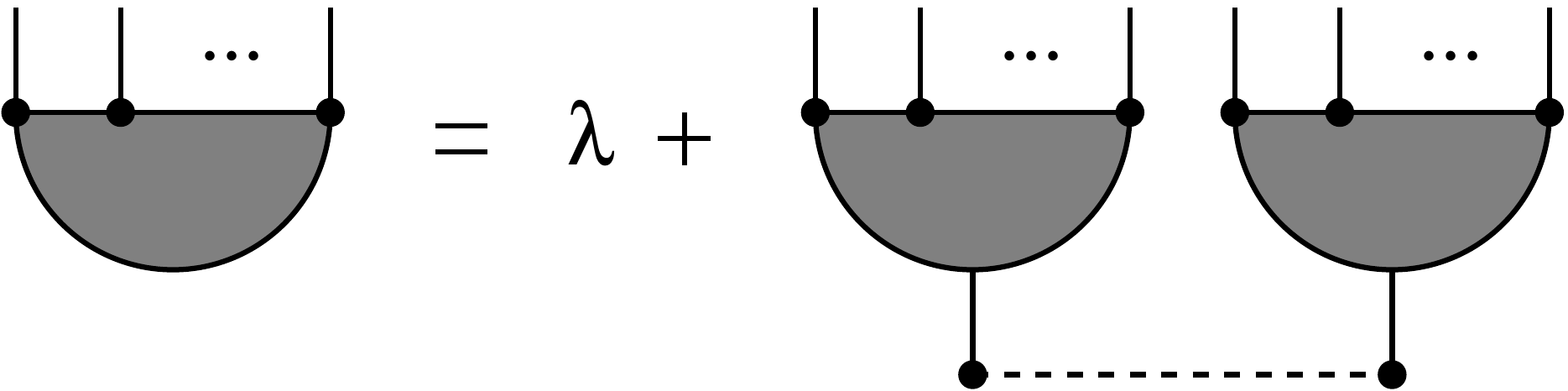}}
\end{equation}
As indicated, the blob denotes $Z$ itself. Note that we work in rescaled variables, where $\Delta'(0^+) =-1$, and in order to lighten the notation, we count the lower vertex as $1$ instead of $\Delta'(0^+)=-1$, which explains the change in sign w.r.t.\ Eq.~(\ref{74}).

Let us now consider loop corrections. More details can be found in
\cite{LeDoussalWiese2008c}. A graphical interpretation of
the dressed propagator $1/(k^2 + m^2 - 2 S_{m} Z)$, appearing e.g.\ in
(\ref{99bis})  is 
\begin{equation}\label{x15}
\diagram{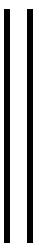} := \diagram{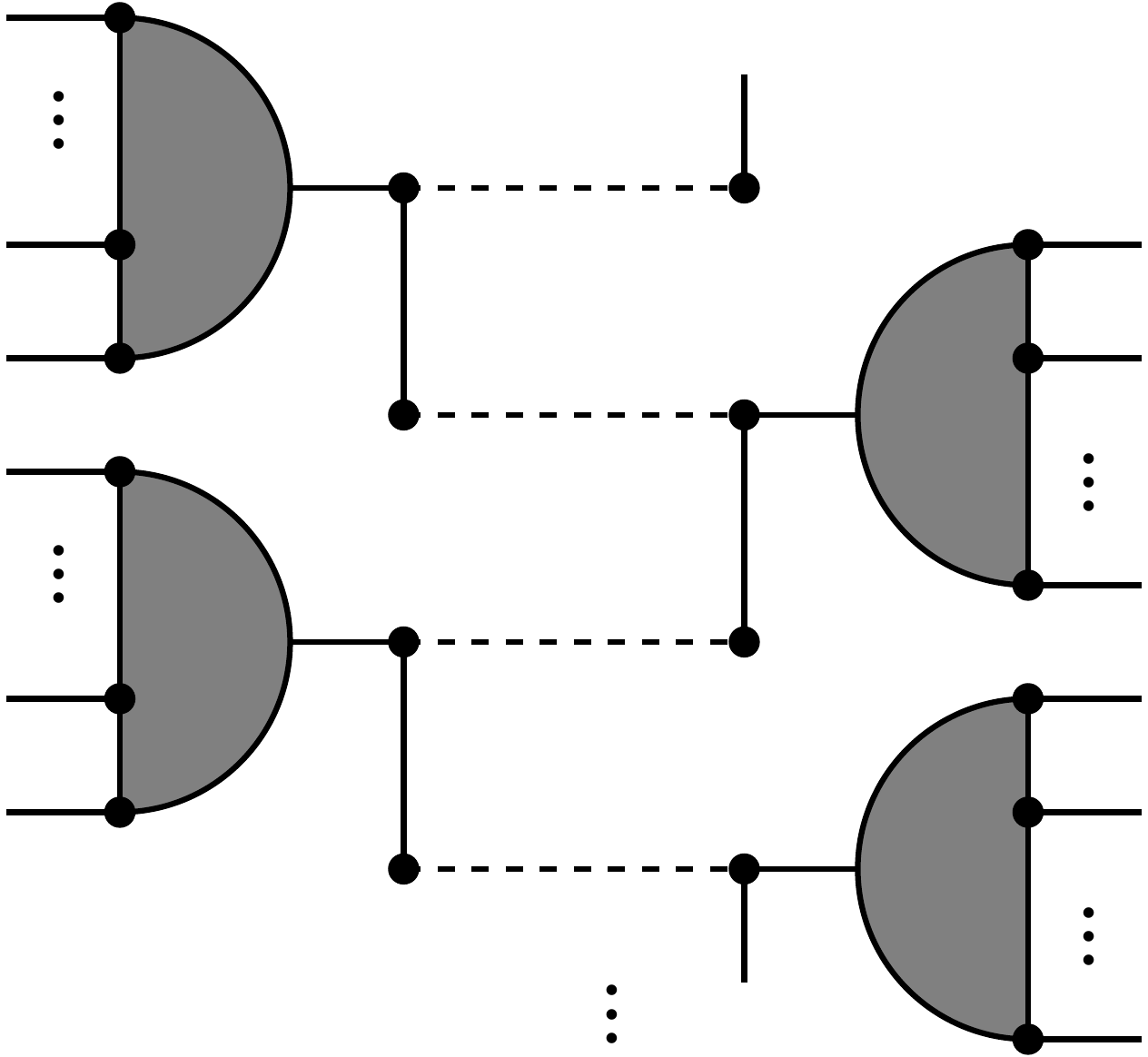}
\end{equation}
The notation on the r.h.s.\ of the equation is as follows:  The left vertex of each
disorder is at $0$, the right one at $w$. This is a graphical
representation  of the
{\em antiferromagnetic} rule described in \cite{LeDoussalWiese2008c}. 
The outgoing lines  all end in a factor of $Z$ which is explicitly drawn. 

There appear two classes of diagrams, corresponding to the first and
second line (of the integral) in Eq.\ (\ref{99bis}). 
The first class, denoted ${\cal C}_{1}$ in  \cite{LeDoussalWiese2008c}, can be written as:
\begin{align}\label{A6}
\displaystyle  {\cal C}_{1} &= \displaystyle \diagram{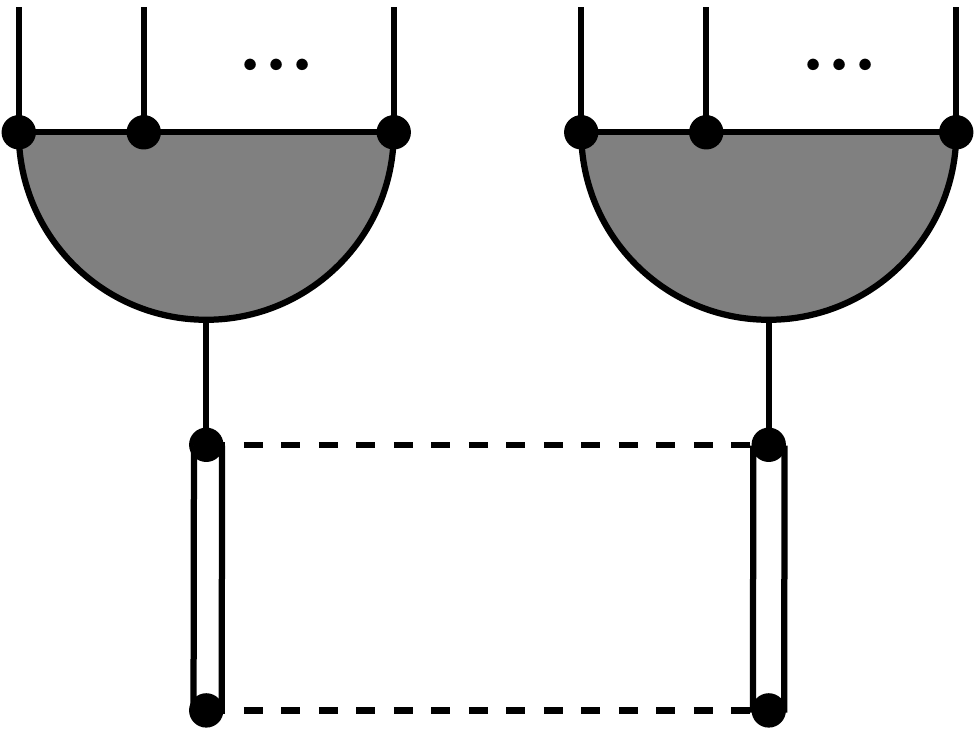}+  \diagram{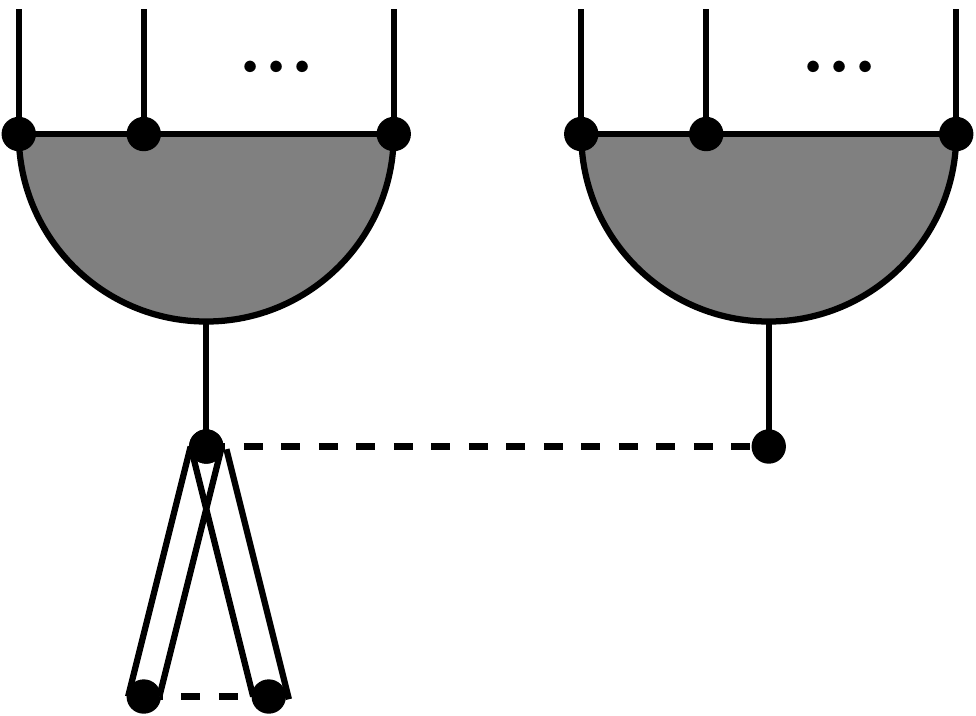}\nonumber \\
& =\displaystyle
\Delta'' (0)[\Delta (w)-\Delta (0)] \int_{k}\frac{Z^{2}}{(k^{2}+m^{2}-2 Z)^{2}} +{\cal O}(w^{2})\ .
\end{align}
The diagrams on the second line (of the integral) of Eq.~(\ref{99c}) are termed class
${\cal C}_{2}$ in \cite{LeDoussalWiese2008c}. They  look like a
correction to the critical force and can be represented as follows 
\begin{align}\label{A7}
\displaystyle  {\cal C}_{2} &= \displaystyle \diagram{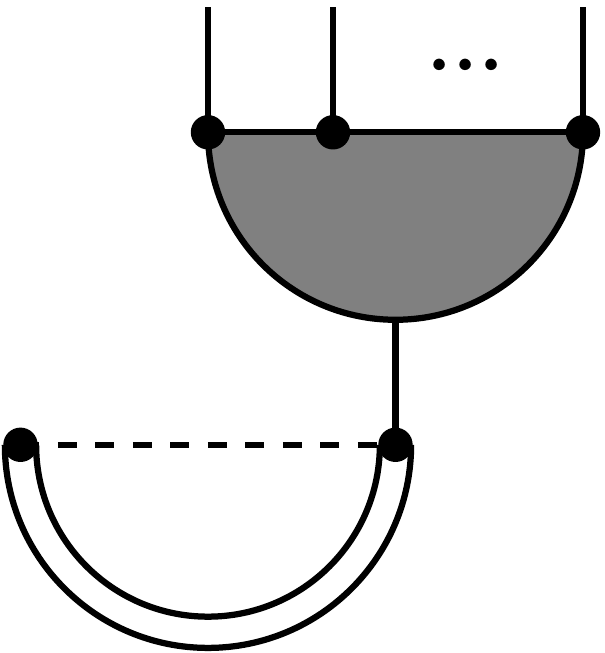} \nonumber \\
& =\displaystyle
[\Delta' (w)-\Delta' (0^{+})] \int_{k}\frac{Z}{(k^{2}+m^{2}-2 Z)}-\frac{Z}{k^{2}+m^{2}}\ .
\end{align}
Indeed it should be viewed as a loop of $\Delta'$, of which exactly
one is expanded in $w$, leading to a loop with one marked
vertex. This is the vertex drawn above. Note that the double line
needs {\em at least one} $\Delta'(w)$ otherwise it cannot start at 0
and go to $w$ as indicated. This leads to the last term in (\ref{A7})
being subtracted.

If one wants the expression in terms of the renormalized disorder, one  has to subtract the contribution proportional to $\int_k \frac 1{(k^2+m^2)^2}$, giving an additional term $-[\Delta' (w)-\Delta' (0^{+})] \int_k \frac {2 Z}{(k^2+m^2)^2}$.

\section{Brownian force model: Many point correlations}
\label{app:bfm}

In this Appendix we derive the correlation function of the center-of-mass 
displacement for the BFM model for an arbitrary number of points, thereby
giving another derivation of its Levy process character discussed in
the text. We provide both a discrete derivation (of the $p$-point correlation)
and a continuum one (functional average). To simplify notations, we set $d=0$,  which amounts
to omitting  the factor of $L^d$,  restored in the main text. Of course
this is achieved within  tree-level, since we have argued that this be exact for the BFM.

\subsection{Discrete calculation}
We start with the tree-level equations (\ref{multi}), derived for
arbitrary $R(u)$ and specify them to $R'''(u)=\sigma \, \mbox{sign}(u)$. 
In this section, for notational simplicity we set $m^2=1$,
$R'''(0^+)=\sigma =1$ and denote $\Delta(0):=-R''(0)$. We must solve the following system of equations 
for $U_i$ and $u_{1i}$:
\begin{eqnarray}
&& u_{1i} - w_i + \sum_{j} |u_{1i}-u_{1j}| U_j  = \Delta(0) \sum_j U_j  \nonumber \\
&& U_i  + \sum_{j \neq i} {\rm sgn}(u_{1i}-u_{1j}) U_i U_j  =  \lambda_i\ .  \label{system}
\end{eqnarray}
Insert the result in
\begin{equation}
  \overline{ \rme^{  \sum_i \lambda_i [u(w_i)- w_i] }}^{\mathrm{tree}}   = \rme^{\sum_i (\lambda_i - \frac{1}{2} U_i) (u_{1i} - w_i) } \ .
\end{equation} 
We choose the  $w_i$ ordered as $w_1<w_2<\ldots < w_n$. The second equation above implies $\sum_i U_i = \sum_i \lambda_i$. Hence we can shift
$u_{1i} \to u_{1i} + \Delta(0) \sum_i \lambda_i$ and eliminate $\Delta(0)$ without changing the equations. Thus the dependence on $\Delta(0)$ is trivial, and we now compute the rest setting $\Delta(0)= 0$:
\begin{align}
 & \overline{ \rme^{  \sum_i \lambda_i (u(w_i)- w_i) }}^{\mathrm{tree}}  \\
  &=\rme^{\frac{1}{2} \Delta(0) (\sum_i \lambda_i)^2 }
 \rme^{ - \frac{1}{2} \sum_{ij} (\lambda_i U_j + \lambda_j U_i - U_i U_j) |u_{1i} - u_{1j}| }\Big|_{\Delta(0)=0} \nn
\end{align}
The solution of the first equation of (\ref{system}) for $n$ points 
satisfies
\begin{eqnarray}
u_{i+1} - u_i = \frac{w_{i+1} - w_i}{1 + \sum_{j=1}^i U_j -  \sum_{j=i+1}^n U_j } \ , \quad 1 \leq i \leq n-1.\nn
\end{eqnarray}
The second set of equations in (\ref{system}) can be rewritten as
\begin{equation}
\lambda_i = U_i \bigg(1 + \sum_{j=1}^{i-1} U_j - \sum_{j=i+1}^{n} U_j \bigg) \quad , \quad 1 \leq i \leq n \ .
\end{equation}
Its solution is
\begin{align}
& 2 U_1 = - 1 + \sum_{j=1}^n \lambda_j + \sqrt{\Delta_1} \\
& 2 U_i = \sqrt{\Delta_i} - \sqrt{\Delta_{i-1}} \quad , \quad 2 \geq i \geq n-1 \\
& 2 U_n = 1 + \sum_{j=1}^n \lambda_j - \sqrt{\Delta_{n-1}}
\end{align}
with 
\beq
\Delta_i 
= 1 + \Big(\sum_{j=1}^n \lambda_j\Big)^2 + 2 \Big( \sum_{j=1}^i \lambda_j  - \sum_{j=i+1}^n \lambda_j  \Big) \ .
\eeq
Hence we find
\beq
u_{i+1} - u_i = \frac{w_{i+1} - w_i}{\sqrt{\Delta_i}} \ .
\eeq
This allows to rewrite
\begin{eqnarray}
&& \!\!\! - \frac{1}{2} \sum_{ij} (\lambda_i U_j + \lambda_j U_i - U_i U_j) |u_{1i} - u_{1j}| \\
&& =
 - \sum_{i=1}^{n-1} u_{i+1,i} \Big[\sum_{j=1}^i \lambda_j \sum_{j=i+1}^n U_j + \sum_{j=1}^i U_j \sum_{j=i+1}^n (\lambda_j -U_j) \Big] \nn \\
 && =   - \sum_{i=1}^{n-1} \frac{w_{i+1,i}}{4 \sqrt{\Delta_i}} \Big[  (\sum_{j=1}^n \lambda_j)^2  + (1- \sqrt{\Delta_i})^2 \nn \\
&& \qquad  \qquad  \qquad  \qquad +
 2 (1- \sqrt{\Delta_i}) ( \sum_{j=1}^i \lambda_j - \sum_{j=i+1}^n \lambda_j ) \Big] \nn
\end{eqnarray}
Using now twice the definition of $\Delta_i$ it can be rewritten and simplified to 
\begin{align} \label{gene2}
&  \overline{ \rme^{  \sum_{i=1}^n \lambda_i [u(w_i)- w_i] }}^{\mathrm{tree}} = \rme^{\frac{1}{2} \Delta(0) (\sum_i \lambda_i)^2 }  \\
&\qquad  \times \rme^{\frac{1}{2} \sum_{i=1}^{n-1} w_{i+1,i} (1 + \sum_{j=1}^i \lambda_j - \sum_{j=i+1}^n \lambda_j  - \sqrt{\Delta_i} ) } \nn
\end{align}
This proves formula (\ref{Zh2}) in the main text.

Consider now $X_i=u(w_i)- w_i$ and choose
\beq
\lambda_1 = - \mu_1 + \frac{1}{2}  \mu ,~   \lambda_2 = \mu_1 - \mu_2 , \ldots   ,~  \lambda_n =\mu_{n-1} + \frac{1}{2}  \mu \ .
\eeq
We then find
\begin{eqnarray}
&&  \overline{ \rme^{  \sum_{i=1}^{n-1} \mu_i (X_{i+1}-X_i) + \frac{1}{2} \mu (X_1+X_n) }}^{\mathrm{tree}}  \nn \\
 &&
 =  
 \rme^{\frac{1}{2} \Delta(0) \mu^2 + \frac{1}{2} \sum_{i=1}^{n-1} w_{i+1,i} (1 -2 \mu_i  - \sqrt{1 + \mu^2 - 4 \mu_i   }) }\ ,\qquad
\end{eqnarray}
 i.e.\ the variables $X_{i+1}-X_i$ are still independent for fixed $\mu$, but are not independent
 of $X_1+X_n$. However, if one considers the rescaled variable $(X_1+X_n)/\sqrt{\Delta(0)}$,  
 then for large $\Delta(0)$ one recovers statistical independence.  A similar result holds 
 with $(X_1+X_2+\ldots+X_n)/\sqrt{\Delta(0)}$.

\subsection{Continuous version}

Let us consider  equations (\ref{multicont}), (\ref{Scont}) in the main text, and specify to the 
BFM model, with $\sigma = R'''(0^+)$. One must solve
\begin{eqnarray}
&& m^2 ( u_1(w)-w ) +  \sigma \int_{w'} |u_1(w)-u_1(w')| U(w')\nn  \\
&& \ = \Delta(0) \int_{w'} U(w')  \\
&& m^2 U(w) + \sigma \int_{w'} {\rm sign}(u_1(w)-u_1(w')) U(w) U(w') \nn \\
&& \ = \lambda(w) \ ,
\end{eqnarray}
and insert into
\beq
 -S_\lambda = \int_{w} \Big[\lambda(w) - \frac{m^2}{2} U(w)\Big] [u_1(w)-w] \ .
\eeq
Here and below $\int_w= \int_{-\infty}^{\infty} \rmd w$.
We now restrict to test functions such that $\int_w \lambda(w)=0$, hence $\int_w U(w)=0$.  We define
\begin{eqnarray}
U(w) = V'(w)
\quad , \quad \lambda(w) = - \mu'(w) \ .
\end{eqnarray} 
Let us assume that $\mu(w)$ vanishes sufficiently fast at $w=\pm \infty$, hence the
same holds for $V(w)$ and no boundary term arises in any integration by part. 
The first equation becomes
\begin{eqnarray}
 && m^2 [ u_1(w)-w ] \\
 &&+  \sigma \int_{w'} {\rm sign}\big(u_1(w)-u_1(w')\big) u_1'(w') V(w') = 0\ . \qquad \nn
\end{eqnarray}
Now assume that
\beq \label{mon}
{\rm sign}\big(u_1(w)-u_1(w')\big) = {\rm sign}(w-w')\ ,
\eeq and take a derivative
w.r.t.\ $w$, leading to 
\beq
 u_1'(w) = \frac{m^2}{m^2 + 2 \sigma V(w)}  \ .
\eeq
Hence monotonicity holds indeed as long as $-1 < \frac{2 \sigma}{m^2} V(w) $, which we now assume, and discuss below. 
The second equation gives
\be
V'(w) \left[m^2 + \sigma \int_{w'} {\rm sign}\big(u_1(w)-u_1(w')\big)  V'(w') \right] = \lambda(w) \ .
\ee
Using monotonicity (\ref{mon}) and integration by part yields
\beq
 V'(w) \left[m^2 + 2 \sigma  V(w) \right] = \lambda(w) = - \mu'(w)\ .
\eeq
Hence
\beq
m^2 V(w) + \sigma V(w)^2 = - \mu(w) \ , 
\eeq
which can be solved as
\beq
V(w) = \frac{m^2}{2 \sigma} \left[ \sqrt{1 - 4 \frac{\sigma}{m^4} \mu(w)} - 1 \right] \ .
\eeq
Note that $m^2 V(w) = - Z\big(\mu(w)\big)$ with $Z$ given in (\ref{16}). Integrating by parts we find
\begin{eqnarray}
 -S_\lambda  &=&
\int_{w} \left[ - \mu'(w) - \frac{m^2}{2} V'(w)\right] \left[u(w)-w \right] \nn \\
& = &
2 \sigma \int_{w} \left[- \mu(w) - \frac{m^2}{2} V(w)\right] \frac{V(w)}{m^2 + 2 \sigma V(w)}  \nn \\
& = &\sigma \int_{w} V(w)^2=  \int_{w} \hat Z\big(\mu(w)\big) \ .
\end{eqnarray}
Here $ \hat Z(\mu) = 
\frac{m^4}{\sigma} {\hat {\tilde Z}}( \frac{\sigma}{m^4} \mu)$ and 
\beq
 {\hat {\tilde Z}}(\mu) = \frac{1}{2} \left(1 - 2 \mu - \sqrt{1 - 4 \mu}  \right) \ .
\eeq
which shows formula (\ref{Zhcontinuous}) in the text. Note that the above monotonicity condition
for $u_1(w)$ is equivalent to $\frac{4 \sigma}{m^4} \mu(w) < 1$, the usual analyticity domain
where the generating function is convergent.

\subsection{Cumulants of $u(w)-w$ (Burgers velocity)}

It is equivalently interesting to obtain, for the BFM model, the expression for the cumulants
of the renormalized pinning force. These were defined in
\cite{LeDoussalWiese2008c}, Section III B, as
\begin{align} 
& m^{2 n} \overline{h_{1}h_{2}\dots h_{n} } ^{c} = L^{-(n-1) d} (-1)^n 
\hat C^{(n)} (w_{1},\dots ,w_{n}) \\
& h_{i}:= u(w_{i})-w_{i} \ .
\end{align} 
We will choose $w_1<\ldots <w_n$. For simplicity, we use dimensionless
units setting $m=1$, $R'''(0^+)=1$, and $d=0$, all factors being easily recovered. The lowest moments can be computed
using the tree-level formula (61) in \cite{LeDoussalWiese2008c}, which should be
the exact result for the BFM model according to our conjecture, giving
the following simple expressions:
\begin{eqnarray}\label{x16}
\hat C^{(2)} (w_{1},w_{2}) &=& w_1-w_2 - R''(0) \\
- \hat C^{(3)} (w_{1},w_{2},w_{3}) &=&2\left(w_1-2 w_2+w_3 \right) \\
\hat C^{(4)} (w_{1},w_{2},w_{3},w_{4})&=& 3!\left( w_1-3 w_2+3 w_3-w_4\right) \qquad \ 
\end{eqnarray}
Higher cumulants have been re-calculated and we find that the 
general result can be written as:
\begin{equation}\label{x17}
(-1)^n \hat C^{(n)} (w_{1},\dots, w_{n}) = { (n-1)!} \sum_{i=1}^{n}\left({n-1
\atop i-1} \right) (-1)^{i+1} w_{i}\ .
\end{equation}
Equivalently 
\begin{equation}\label{finalC}
(-1)^n C_{n} (w_{1},\dots ,w_{n}) = (n-1)!\, a (1-a)^{n-1} \Big|_{a^{i} \to w_{i}}\ ,
\end{equation}
where the rule $a^i\to w_i$ means to expand in powers of $a$, and to replace the $i$-th power of $a$ by $w_i$.
This formula has been checked against the Kolmogorov-cumulants
$K^{n} (w):=\left< ( h_{2}-h_{1})^{n} \right>^{c} = a_{n}
(w_{2}-w_{1})$ obtained in (96) of \cite{LeDoussalWiese2008c}.

We have also checked that this result is consistent
with the result for the $n$-point generating function (\ref{gene2}).

\section{Derivation of the Carraro-Duchon formula}
\label{a:Carraro-Duchon}

In this Appendix we give a physicist's derivation of 
Eq.\ (\ref {f:Carraro-Duchon2}) entering (\ref{f:Carraro-Duchon1}). For a mathematical derivation see \cite{CarraroDuchon1998}. 
We use the notation $\int_w = \int_{-\infty}^{\infty} \rmd w$, and recall that our conventions are summarized in table \ref{t:conv}.

Consider the Burgers velocity field to be a Levy process at time $t$. Then the Levy-Khintchine theorem \cite{BertoinBook} implies that 
\begin{eqnarray}
&& \overline{\rme^{ \int_{w} v_t(w) \Omega(w) } } = \rme^{\int_{w} \phi_t(\omega(w)) } \\
&&  \Omega(w)= - \omega'(w) \quad , \quad \omega(w) = \int_{w}^{\infty} \rmd w_1 \,\Omega(w_1) \nn
\end{eqnarray}
for any function $\omega(w)$ such that $\omega(\pm \infty)=0$. Below we also assume that $\Omega(w)$ vanishes (sufficiently
fast) at infinity. 

Let us assume that it remains of this form at all times, and check that it is correct provided that
$\phi_t$ satisfies some differential equation. To show this, first take $\partial_t$ on both sides and use the Burgers equation $\partial_t v_t(w) + \frac{1}{2} \partial_w [v_t(w)^2]=0$.
This leads to
\begin{eqnarray} \label{cd1}
&& \int_{w} \partial_t \phi_t\big(\omega(w)\big) \rme^{\int_{w'}  \phi_t(\omega(w')) } \\&&
= 
 \overline{ \int_w \Omega(w) \partial_t v_t(w) \rme^{ \int_{w'} v_t(w') \Omega(w') } }\nn
 \\
 && = 
 \frac{-1}{2} \overline{ \int_{w} \Omega(w) \partial_w v_t(w)^2 \rme^{ \int_{w'} v_t(w') \Omega(w') } } \nn
 \\
 && = \frac{1}{2} \overline{ \int_{w} \Omega'(w)  v_t(w)^2 \rme^{ \int_{w'} v_t(w') \Omega(w') } }\nn \\
&&=
 \frac{1}{2} \int_{w} \Omega'(w)  \frac{\delta^2}{\delta \Omega(w)^2} \overline{  \rme^{ \int_{w'} v_t(w') \Omega(w') } }  \nn \\
 && =  \frac{1}{2} \int_{w} \Omega'(w)  \frac{\delta^2}{\delta \Omega(w)^2}  \rme^{\int_{w'}  \phi_t(\omega(w')) }\nn
 \\
 &&
 =  \frac{1}{2} \int\limits_{w} \int\limits_{w_1} \Omega'(w)  \frac{\delta}{\delta \Omega(w)}   \theta(w{-}w_1) \phi_t'\big(\omega(w_1)\big) \rme^{\int_{w'} \phi_t(\omega(w')) } \ .\nn
 \end{eqnarray}
We have used that $\frac{\delta}{\delta \Omega(w)} \omega(w') = \int_{w'}^{\infty} \rmd w_1 \delta(w-w_1) = \theta(w-w')$; then
we obtain
\begin{eqnarray}
 \int_w \partial_t \phi_t(\omega(w)) &=& \frac{1}{2} \int_w \int_{w_1} \int_{w_2} \Omega'(w) \theta(w-w_1) \theta(w-w_2) \nn\\
&&\qquad\qquad\qquad~~\times\phi_t'\big(\omega(w_1)\big) \phi_t'\big(\omega(w_2)\big) \nn\\
&&  + \frac{1}{2} \int_{w} \int_{w_1} \Omega'(w) \theta(w-w_1) \phi_t''\big(\omega(w_1)\big) \nn \\
&&  \label{cd2}
\end{eqnarray}
where we have divided by $\rme^{\int_{w'} \phi_t(\omega(w')) }$.
Integration by parts leads to
\begin{eqnarray}
&& \int_w \partial_t \phi_t(\omega(w))  \nn\\
&&=
  \int_w \int_{w_1}  \omega'(w) \theta(w-w_1)  \phi_t'\big(\omega(w_1)\big) \phi_t'\big(\omega(w)\big) \nn\\
  &&\qquad + \frac{1}{2} \int_w  \omega'(w) \phi_t''\big(\omega(w)\big) \nn\\
&& =  \int_w \int_{w_1}  \theta(w-w_1)  \phi_t'(\omega(w_1)) \frac{\rmd}{\rmd w} \phi_t\big(\omega(w)\big) \nn\\
&&\qquad + \frac{1}{2} \Big[\phi_t'\big(\omega(w)\big)\Big]_{-\infty}^{\infty} \nn\\
&&= - \int_w \phi_t'(\omega(w)) \phi_t(\omega(w)) \ ,
\end{eqnarray}
since $\omega(w)$ vanishes at infinity (and $\phi_t$ vanishes in zero). Since this is true for any function $\omega(w)$, it implies 
\begin{eqnarray} \label{cnew}
\partial_t \phi_t(\omega) + \phi_t(\omega) \partial_\omega \phi_t(\omega) = 0\ .
\end{eqnarray} 
This is nothing but Eq.\ (\ref{f:Carraro-Duchon2}).

\section{Beyond the Carraro-Duchon formula: including loops} 

\subsection{Evolution equation}

We now give the more general evolution equation, 
valid beyond Levy processes. We define, as in the text, 
\beq 
e^{ \hat{\mathbb Z}_t }
 := \overline{ e^{ \int_w \Omega( w)  v_t( w) }} \ ,
\eeq
where  $\hat{\mathbb Z}_t$ is 
a priori an arbitrary functional of $\Omega(w)$,
and we only assume that $\Omega(w)$ vanishes at infinity (i.e.\ $\int_w \Omega(w)$ is
not necessarily zero). Similar manipulations as above in ({\ref{cd1}), (\ref{cd2}) using Burgers' equation yield the exact evolution equation
\beq\label{F7}
\partial_t  \hat{\mathbb Z}_t =   \frac{1}{2} 
   \int_{w} \Omega'(w) \left[  \frac{\delta^2  \hat{\mathbb Z}_t}{\delta \Omega(w)^2} 
 +
 \frac{\delta  \hat{\mathbb Z}_t}{\delta \Omega(w)} 
   \frac{\delta  \hat{\mathbb Z}_t}{\delta \Omega(w)} \right]\ .
\eeq
Inserting (\ref{expa}) and expanding in $\Omega$,  this equation provides yet another derivation of the
exact RG equations (\ref{carraroduchonCum}) for the cumulants $\hat C^{(n)}$. (See  Appendix
\ref{app:erg} for a derivation using replica). 
  
To make contact with our equation (\ref{Du-gen1}), we note that $\Omega(w)$ and $u(w)$ always appear together, and thus 
\begin{equation}
\frac{\delta}{\delta u(w)} \hat {\mathbb Y}_t [\Omega,{\sf u}] = \frac{\Omega(w)}{u'(w)} \frac{\partial}{\partial w} 
\frac{\delta}{\delta \Omega(w)} \hat {\mathbb Y}_t [\Omega,{\sf u}] \ .
\end{equation}
Setting $ \hat{{\mathbb Y}}[\Omega,{\sf u}]\big|_{{\sf u}(w)=w}\to \hat{\mathbb Z}_t [\Omega] $ (possible since we no longer derive w.r.t.\ $u(w)$), inserting this into (\ref{Du-gen1}) and integrating by part yields the second term in (\ref{F7}).

The first term in (\ref{F7}) corresponds to loop corrections and to the first term in the more general
equation (\ref{Du-gen-new}). One can see that these are equivalent as follows:
\begin{align}\label{F8}
&- 
\int_w \lim_{w'\to w} \left[\frac{\delta}{\delta \Omega(w')}  \frac{\delta}{\delta {\sf u}(w)} \hat {\mathbb Y}_t[\Omega,{\sf u}] \right]\\
& = - \int_w \lim_{w'\to w} \left[\frac{\delta}{\delta \Omega(w')} 
 \frac{\Omega(w)}{u'(w)} \frac{\partial}{\partial w} 
\frac{\delta}{\delta \Omega(w)} \hat {\mathbb Y}_t [\Omega,{\sf u}] \right] \nn
\end{align}
Replacing ${\sf u}(w)\to w$ (possible since we no longer derive w.r.t.\ $u(w)$) yields
\begin{align}\label{F9}\nn
&-\int_w \lim_{w'\to w} \left[  \frac{\delta}{\delta \Omega(w')} 
 {\Omega(w)} \frac{\partial}{\partial w} 
\frac{\delta}{\delta \Omega(w)} \hat {\mathbb Z}_t [\Omega] \right] \\
& =-\frac12 \int_{w}  \lim_{w'\to w} \Omega(w) \left[
\left( \frac{\partial}{\partial w} {+} \frac{\partial}{\partial w'} \right)
\frac{\delta}{\delta \Omega(w)}\frac{\delta}{\delta \Omega(w')}  \hat {\mathbb Z}_t [\Omega]\right] \nn\\
& =-\frac12 \int_{w}\Omega(w) \frac{\partial }{\partial w} 
\frac{\delta^2}{\delta \Omega(w)^2}  \hat {\mathbb Z}_t [\Omega] \nn
\nn\\
& =\frac12 \int_{w}\Omega'(w) 
\frac{\delta^2}{\delta \Omega(w)^2}  \hat {\mathbb Z}_t [\Omega] \ .
\end{align}
Thus (\ref{F8}) is the 1-loop correction (in $d=0$) to be added to (\ref{Du-gen1}). 

\subsection{Levy processes and Brownian force model}\label{s:LpaBfm}
We now study some particular solutions of the evolution equation (\ref{F7}).
The first one corresponds to the Levy processes discussed above. Suppose
one restricts to the case where $\int_w \Omega(w)=0$, and where 
$\hat{\mathbb Z}_t$ is a  function of $\omega(w)=\int_w^\infty \rmd w'\, \Omega(w')$,
\beq \label{agabis}
\hat{\mathbb Z}_t = \int_w \phi_t(\omega(w))\ .
\eeq
Using that $\frac{\delta}{\delta \Omega(w)} = \int \rmd w_1\, \theta(w-w_1) \frac{\delta}{\delta \omega(w)}$, 
one recovers equation (\ref{cnew}).
Equation (\ref{F7}) could thus  be used to study deviations from Levy processes. 

For a Levy process, we note that the first term in (\ref{F7}) vanishes when 
$\int_w \Omega(w)=0$, i.e.\ there are no loop corrections to averages of
velocity differences, and the tree approximation is exact for such observables.

An interesting generalization, within Levy processes, is to allow for $\int_w \Omega(w) \neq 0$, i.e.
study observables which involve the full velocity and not simply velocity differences. In particular, we want to know the full solution for the BFM. In the
case of discrete $p$-point correlations of the BFM this was done in Appendix \ref{app:bfm}. A  generalization
of the Carraro-Duchon approach allows to treat  that case for continuum observables such
as $\hat{\mathbb Z}_t[\Omega]$. An interesting output is that we will recover quite simply the
full loop corrections for this model obtained via ERG in Appendix \ref{app:erg}. 

We  define as before $\omega(w)=\int_w^\infty \rmd w'\, \Omega(w')$, but now
$\omega(-\infty)=\omega_\infty:=\int_{-\infty}^\infty \rmd w'\, \Omega(w')$ may be non-zero,  
while $\Omega(w)$ still vanishes at
infinity. We  show that (\ref{aga}) is replaced by
\beq \label{aga2}
\hat{\mathbb Z}_t =  f_t(\omega_\infty) + \int_{w_1} \phi_t(\omega(w_1), \omega_{\infty})\ .
\eeq
For the integral over $w_1$ to be convergent at $\pm \infty$ we need the function $\phi_t(x,y)$ to
satisfy
\bea \label{newcond}
\phi_t(0,y) = 0 \  , \quad \phi_t(y,y) = 0\ ,
\eea 
which we assume from now on,  and which our solution (\ref{G22}) given below satisfies. Upon differentiation this also implies $\partial_2\phi_t(0,y) =0$ 
and $\partial_1 \phi_t(y,y)  + \partial_2 \phi_t(y,y) =0$, which we use below\footnote{Note that assuming the disorder to be statistically translational invariant (STS)
can be expressed as a Ward identity
\bea \label{rel}
\int_w \Omega'(w) \frac{\delta \hat{\mathbb Z}_t}{\delta \Omega(w)} = 0 \ .
\eea 
obtained by performing the change of variables
$w \to w+ a$ in all $w$ integrals appearing in $\hat{\mathbb Z}_t$, e.g.\ in its definition (\ref{expa}).
It implies that if $\hat{\mathbb Z}_t$ is a solution of (\ref{F7})  then  $\hat{\mathbb Z}_t + F(\int_{-\infty}^{\infty} \Omega(w))$,  where $F(y)$ is an arbitrary function, is also a solution, with a different initial condition. This invariance corresponds to adding the so-called Larkin random force in the language of interface pinning. One easily checks that
STS is satisfied by our ansatz. Inserting (\ref{aga2}) into (\ref{rel}) one finds that it vanishes after integration by part 
because of (\ref{newcond}).}}.

Let us prove that (\ref{aga2}) is indeed a solution of (\ref{F7}):
\begin{align}
& \frac{\delta \hat{\mathbb Z}_t}{\delta \Omega(w)}  =  f'_t(\omega_\infty) \\
& + \int_{w_1} \Big[ \theta(w-w_1) \partial_1  \phi_t(\omega(w_1), \omega_{\infty}) + \partial_2 \phi_t(\omega(w_1), \omega_{\infty}) \Big] \nn \\
&  \frac{\delta^2 \hat{\mathbb Z}_t}{\delta \Omega(w)^2}  =  f''_t(\omega_\infty) + \int_{w_1} \Big[\theta(w-w_1) \times \nn\\
& ~~~ \qquad \qquad \times \Big( \partial^2_1  \phi_t(\omega(w_1), \omega_{\infty}) + 2 \partial_1 \partial_2 \phi_t(\omega(w_1), \omega_{\infty}) \Big) \nn\\
&~~~  \qquad  \qquad + \partial^2_2 \phi_t(\omega(w_1), \omega_{\infty}) \Big] 
\end{align}
Let us first compute the loop contributions, using that $\int_w \Omega'(w)=0$, and $\frac{\rm d}{\rmd w} [  A(\omega(w), \omega_{\infty}) ] = -  \Omega(w) \partial_1  A(\omega(w), \omega_{\infty})$:
\begin{align}\nn
& \frac{1}{2}  \int_{w} \Omega'(w)   \frac{\delta^2  \hat{\mathbb Z}_t}{\delta \Omega(w)^2} 
\\
&= - \frac{1}{2}  \nn
   \int_{w} \Omega(w)   \Big[   \partial^2_1  \phi_t(\omega(w), \omega_{\infty}) + 2 \partial_1 \partial_2 \phi_t(\omega(w), \omega_{\infty}) \Big]  \\
   & =  \frac{1}{2}  \Big[ \partial_1  \phi_t(0, \omega_{\infty}) + 2  \partial_2  \phi_t(0, \omega_{\infty}) \nn\\
 &\qquad ~
   - \partial_1  \phi_t(\omega_{\infty}, \omega_{\infty}) - 2  \partial_2  \phi_t(\omega_{\infty}, \omega_{\infty}) \Big]\nn  \\
   &= \frac{1}{2} \Big[ \partial_1  \phi_t(0, \omega_{\infty})  +  \partial_1  \phi_t(\omega_{\infty}, \omega_{\infty}) \Big]
\end{align}
We now compute the tree contribution $\frac{1}{2}  \int_{w} \Omega'(w)   \frac{\delta \hat{\mathbb Z}_t}{\delta \Omega(w)}  \frac{\delta \hat{\mathbb Z}_t}{\delta \Omega(w)} = A + B$: 
\bea
 A &=&\Big[ f'_t(\omega_\infty)+ \int_{w_1} \partial_2 \phi_t(\omega(w_1), \omega_{\infty})   \Big]\nn\\ && \times \int_{w,w_2} \Omega'(w)  \theta(w-w_2) \partial_1  \phi_t(\omega(w_2), \omega_{\infty}) \nn\\
&=& \Big[ f'_t(\omega_\infty)+ \int_{w_1} \partial_2 \phi_t(\omega(w_1), \omega_{\infty})   \Big] \nn\\ && \times \Big[  \phi_t(0, \omega_{\infty})  -  \phi_t(\omega_{\infty}, \omega_{\infty}) \Big] = 0\ ,
\\
B 
    &=&  \frac{1}{2} 
   \int_{w,w_1,w_2} \Omega'(w)    \theta(w-w_1) \theta(w-w_2)   \nn\\
   && \qquad\qquad\qquad \times\partial_1  \phi_t(\omega(w_1), \omega_{\infty}) \partial_1  \phi_t(\omega(w_2), \omega_{\infty}) \nn \\
   & =& - \int_{w,w_1} \Omega(w)   \theta(w-w_1) \partial_1 \phi_t(\omega(w_1), \omega_{\infty}) \nn\\
   && \qquad\qquad\qquad\times \partial_1 \phi_t(\omega(w), \omega_{\infty}) 
 \nn   \\
   & = & - \int_{w}  \phi_t(\omega(w), \omega_{\infty}) \partial_1  \phi_t(\omega(w), \omega_{\infty})\ .
\eea 
Hence we find that if the  unknown functions $\phi_t(x,y)$ and $f_t(y)$ satisfy the following two equations, then
(\ref{F7}) is satisfied:
\bea \label{trara1}
 \partial_t \phi_t(x,y) &=& - \phi_t(x,y) \partial_1 \phi_t(x,y)  \\
 \partial_t f_t(y) &=& \frac{1}{2} \Big[\partial_1 \phi_t(0,y) + \partial_1 \phi_t(y,y) \Big] \ .\qquad \label{trara2}
\eea
Consider now the BFM. The initial condition is
\bea\label{init1}
 \phi_{t=0}(x,y) &=& \sigma x^2 - \sigma x  y  \\
 f_{t=0}(y) &=& \frac{1}{2} \Delta(0) y^2 \ .\label{init2}
\eea 
This can be seen by rewriting
\begin{eqnarray}
 Z_0[\Omega]&=& \frac{1}{2} \int_{w_1,w_2} \Omega(w_1) \Omega(w_2) \Delta(w_1-w_2) 
\nn\\
&=& \frac{1}{2} \Delta(0) ( \int \Omega)^2  - \sigma  \int_{w_1>w_2} \Omega(w_1) \Omega(w_2) (w_1-w_2) \nn \\
& =&  \frac{1}{2} \Delta(0) ( \int \Omega)^2  - \sigma  \int_{w_1>w_2} \omega(w_1) \Omega(w_2) \nn\\
& &+ \sigma \big[  \omega(w_1) \Omega(w_2) (w_1-w_2) \big]^{w_1=\infty}_{w_1=w_2} \ .
\end{eqnarray}   
Using that $\omega(\infty)=0$, the boundary term is zero and one obtains (\ref{init1}), (\ref{init2}) . 

The solution of the system (\ref{trara1})--(\ref{trara2}) with this  initial condition is
\bea \label{G22}
 \phi_{t}(x,y) &=& \frac{1}{2 \sigma t^2} \bigg[1 + 2 \sigma t \left(x - \frac{y}{2}\right) \\
&&\qquad - \sqrt{ 1 + 4 \sigma t \left(x - \frac{y}{2}\right) + \sigma^2 t^2 y^2 } \;\bigg] \nn \\
 f_t(y) &=& \frac{1}{2} \ln( 1 - t^2 s^2 y^2) + \frac{1}{2} \Delta_0(0) y^2 \ .
\eea
One checks that $ \phi_{t}(x,y) $ satisfies  the conditions (\ref{newcond}), and that this recovers  the loop corrections obtained by the
ERG method in (\ref{finalnew})--(\ref{H13}).


\section{ERG equations and their mean-field version} \label{app:erg} 
It is instructive to compare the equations (\ref{carraroduchonCum}) with the known exact RG equations;
we restrict here to $d=0$. In Ref.\ \cite{LeDoussal2008} section IV.A.2 (arXiv version) the ERG equations for the cumulant of the
renormalized potential 
$\hat S^{(n)}(w_1,\ldots w_n)= (-1)^n \overline{\hat V(w_1) \ldots  \hat V(w_n)}^c$ were obtained. It was shown that
the function
\beq
\hat U(w_a) = \sum_n \frac{1}{n! T^n} \sum_{a_1,\ldots a_n} \hat S^{(n)}(w_{a_1},\ldots w_{a_n}) 
\eeq
satisfies
\beq
2 \partial_t \hat U = T \sum_a \partial^2_{w_a} \hat U + \sum_a (\partial_{w_a} \hat U)^2\ .
\eeq
Hence the $\hat S^{(n)}$ satisfy at $T=0$:
\bea
\lefteqn{ 2 \partial_t \hat S^{(n)}(w_1,\ldots w_n) } \\
& =& n [ \hat S_{110\ldots 0}^{(n+1)}(w_1,w_1,\ldots w_n) ]\nn\\&& + \sum_{p,q,p+q=n+1} \frac{n!}{(p-1)! (q-1)!}\times\nn\\
&&\times [ \hat S_{10\ldots 0}^{(p)}(w_1,\ldots w_p)  \hat S_{10\ldots 0}^{(q)}(w_1,w_{p+1}\ldots w_n)  ] \nn\ ,
\eea
where we have used that
\bea
\lefteqn{  \sum_{a_1,\ldots, a_p} \partial^2_{w_a} \hat S^{(p)}(w_{a_1},\ldots, w_{a_p})} \\
&=& p \sum_{a_1,\ldots, a_p} \delta_{aa_1} 
\hat S_{20\ldots 0}^{(p)}(w_a,w_{a_2},\ldots, w_{a_p}) \nn\\
&+&  p (p-1) \sum_{a_1,\ldots, a_p} \delta_{aa_1} \delta_{aa_2}
\hat S_{110\ldots 0}^{(p)}(w_a,w_a,w_{a_3},\ldots, w_{a_p})  . \nn
\eea
We now use that $\hat C^{(n)}(w_1,\ldots, w_n) = (-1)^n \partial_{w_1}\ldots \partial_{w_n} \hat S^{(n)}(w_1,\ldots, w_n)$ to obtain
\bea
\lefteqn{ \partial_t \hat C^{(n)}(w_1,\ldots, w_n) }\\
&=&
-  \frac{n}{2} [ \partial_{w_1} \hat C^{(n+1)}(w_1,w_1,\ldots, w_n) ] \nn \\
&& 
- \frac{1}{2} \sum_{p,q,p+q=n+1} \frac{n!}{(p-1)! (q-1)!}\nn \\
&& \times  [  \partial_{w_1} ( \hat C^{(p)}(w_1,\ldots, w_p) \hat C^{(q)}(w_1,w_{p+1},\ldots, w_n) )  ] \nn
\eea
This formula  works when the $\hat C$ are continuous functions of their arguments\footnote{Here we study only  the case with  STS (translation invariance). The non-STS case, e.g.\ a two-sided Brownian force landscape starting at zero violates this condition at the origin,  and
requires a special treatment, which is left for the future.}. Hence we see here that if one takes out the first term one recovers exactly the mean-field RG equation
(\ref{carraroduchonCum}) of the main text. Hence this provides a further derivation of this equation. 

In the STS case the lowest-order ERG equations (including loops) are:
\newcommand{\w}{\nn\\&}
\begin{align}
& \partial_t \hat C^{(2)}(w_1,w_2) = \\
& -\frac{1}{2} \partial_{w_1}  \hat C^{(3)}(w_1,w_1,w_2)-\frac{1}{2} \partial_{w_2} 
   \hat C^{(3)}(w_1,w_2,w_2) \nn\\
& \partial_t \hat C^{(3)}(w_1,w_2,w_3) =  
-\frac{1}{2} \partial_{w_1}  \hat C^{(4)}(w_1,w_1,w_2,w_3)
\w -\frac{1}{2} \partial_{w_2}    \hat C^{(4)}(w_1,w_2,w_2,w_3) 
\w-\frac{1}{2} \partial_{w_3}  \hat C^{(4)}(w_1,w_2,w_3,w_3)  
\w-\partial_{w_1}    \hat C^{(2)}(w_1,w_2) \hat C^{(2)}(w_1,w_3)
\w-\partial_{w_3}  \hat C^{(2)}(w_2,w_3)   \hat C^{(2)}(w_1,w_3)
\w-\partial_{w_2}  \hat C^{(2)}(w_1,w_2) \hat C^{(2)}(w_2,w_3)\ .
\end{align}
An exact solution {\it  including loop corrections} exists for all $\hat C^{(n)}$  
in the STS-Brownian case (stationary BFM discussed in Section \ref{sec:bfm}). It reads:
\begin{align}\label{finalnew}
& \hat C^{(1)} (w_{1}) = 0 \\
& \hat C^{(2)} (w_{1},w_{2}) = \sigma (w_1-w_2) - \sigma^2 t^2 \\
& \hat C^{(3)} (w_{1},w_{2},w_{3}) = - 2 t \sigma^2 \left(w_1-2 w_2+w_3 \right) \\
& \hat C^{(4)} = 3! t^2 \sigma^3 \left( w_1-3 w_2+3 w_3-w_4\right)  -6 \sigma^4 t^4  \\
& \hat C^{(5)} = - 4! t^3 \sigma^4 ( w_1 - 4 w_2 + 6 w_3 - 4 w_4 + w_5 ) \\
& \hat C^{(6)} = 120 t^4 \sigma^5 (w_1 - 5 w_2 + 10 w_3 - 10 w_4 + 5 w_5 - w_6) \nn \\
& \quad \qquad - c_6 \sigma^6 t^6  \label{H13}
\end{align}
One has thus a constant part $- c_n \sigma^n t^n$ with $c_n=(n-1)!$ for $n$ even and
$c_n=0$ for $n$ odd. If one sets $c_n$ to zero one recovers the expressions in (\ref{finalC}). 
The fact that this simple exact solution exists in that case is a consequence of the
property that $\Gamma={\cal S}$ in the sense discussed in Section  \ref{sec:loops}. 

\section{FRG properties of the BFM model}
\label{a:FRG4BFM}

\subsection{Stability of the BFM fixed point}
Let us express the FRG equation using the rescaled force correlator $\tilde \Delta(u)=- \tilde R''(u)$
 defined in (\ref{rescR}). To one loop (first 2 lines)
and 2 loops (third line) the FRG flow for $\tilde \Delta'(u)$, derived in 
\cite{ChauveLeDoussalWiese2000a,LeDoussalWieseChauve2002,LeDoussalWieseChauve2003} is 
 \begin{eqnarray} \label{flowdeltaprime1} 
 - \frac{m \partial}{\partial m} \tilde \Delta'(u) &=& (\epsilon -  \zeta) \tilde \Delta'(u) + \zeta u \tilde \Delta''(u) \\
 &&- 3 \tilde \Delta'(u) \tilde \Delta''(u) - \tilde \Delta'''(u) \Big[\tilde \Delta(u)-\tilde \Delta(0)\Big]  \nn \\
 &&+  \frac{1}{2} \partial_u^3 \Big[ \big( \tilde \Delta(u) - \tilde \Delta(0)\big) \big(\tilde \Delta'(u)^2- \tilde \lambda \Delta'(0)^2\big) \Big] \nn
 \end{eqnarray} 
with  $\tilde \lambda=1$ for the statics (the BFM model studied here) and $\tilde \lambda=-1$  for depinning (the ABBM model
generalized to an interface). In both cases, there is a fixed point corresponding to $\zeta=\epsilon$ with
\begin{equation}\label{G2}
\tilde  \Delta'(u) = - \tilde \sigma ~ {\rm sign}(u)  \quad , \quad \tilde  \Delta(0) - \tilde \Delta(u) =  \tilde \sigma |u|  
\end{equation}
We note that 1- and 2-loop corrections identically vanish for the flow
of $\tilde \Delta'(u)$ at this fixed point -- even for its generalization to a $N$-component field \cite{LeDoussalWiese2005a}.
This is in fact more general, and we claim it to be  true to all
loop orders. 
Indeed, it is easy to see that higher loops bring more derivatives, hence vanishing contributions. For the statics, it can be checked  to four loops in $d=0$ \cite{LeDoussal2008}. Hence
we conjecture that this property holds for any $d$. 

Note that some parts of the effective action are flowing. For instance, one has, in the BFM model,
\beq
- m \partial_m \tilde \Delta(0)= - \epsilon \tilde \Delta(0) - \tilde \sigma^2\ .
\eeq
Hence $\tilde \Delta(0)=C m^\epsilon - \frac{\tilde \sigma^2}{\epsilon}$, i.e.\ $\Delta(0)=C - \frac{\tilde \sigma^2 m^{-\epsilon}}{\epsilon}$, and 
 the constant $C=\Delta_0(0)$ has to be chosen sufficiently large. This is consistent with the fact that the model, 
defined with statistical translational invariance, must be defined with a regularization, e.g.\ a periodic box of
size much larger than any other scale, as discussed in the text. Another regularization would be to
choose a Brownian force with origin at $u=0$, i.e.\ $V'(0)=0$ but this leads to a different FRG equation
which we leave for future investigations.

Let us now show that the above fixed point is {\it attractive}, at least for a class of perturbations (defined precisely below) which
are at most of the same range than the BFM model. It thus defines 
a universality class in any $d$. The stability analysis is performed 
to first order in $\epsilon=4-d$, within the 1-loop FRG equation.
We look
for perturbations of the form
\begin{eqnarray}
\tilde \Delta'(u) = - \tilde \sigma + g(u) \ .
\end{eqnarray}
Physically acceptable solutions for $g(u)$ must vanish at infinity and
have a regular Taylor expansion in powers of $|u|$ at $u=0$. One then obtains to linear order in $g(u)$ 
\begin{equation}
-m\partial_m g(u) = (\epsilon - \zeta) g (u) + \zeta u g'(u) + 3  \tilde \sigma g'(u) + g''(u) \tilde \sigma |u| 
\end{equation} 
Using $\zeta=\epsilon$ we can rescale $a \to \epsilon a$, $\tilde \sigma \to \epsilon \tilde \sigma$ and
$u \to \tilde \sigma u$.  Thus one must solve for $u>0$ 
\begin{equation}
-m \partial_m g (u) = (u+3) g'(u) + u g''(u) = - a g(u) \ .
\label{G7}
\end{equation}
As indicated, we search for eigenvalues $a$, where $a>0$ are stable and $a<0$ unstable modes. 
We find a general solution, noting $L_n^a(x)$ the generalized Laguerre-$L$ polynomial and $U(a,b,z)$ the confluent hypergeometric function
\beq \label{G8}
g(u) = C_1 \rme^{-u}  U(3 - a, 3, u) +   C_2 \rme^{-u}  L_{a-3}^{2}( u)\ .
\eeq
The first solution $\rme^{-u}  U(3 - a, 3, u)$ behaves as $u^{-2}$ at small $u$ (not physically acceptable), 
except when $a-3=0,1,\ldots $ is a positive integer, in which case the two solutions are linearly dependent.
Hence we set $C_1=0$.
The remaining solution is the second independent function, indexed by $a$, 
\begin{equation}\label{H8}
g_a(u) := \rme^{-u}  L_{a-3}^{2}( u)\equiv  L_{- a}^2(- u)\ .
\end{equation}
which has a regular Taylor-expansion at $u=0$ for all $a$. It is thus a physically acceptable solution. 
For non-integer $a$, 
\begin{equation}\label{asymp}
g_a(u) \sim u^{-a} \qquad \mbox{for }u \to \infty\ .
\end{equation}
Thus for $a>0$ this is a long-ranged (attractive) perturbation of (\ref{G2}).  The cases of integer $a>0$ must be treated separately, because then $L_{- a}^2(- u)$ either vanishes identically, or is  a short-ranged solution (see below). Taking the derivative of Eq.\ (\ref{G7}) w.r.t.\ $a$ with the  solution (\ref{H8}) in mind yields 
\begin{equation}
-m \partial_m \left[ \partial_a g_a (u)\right] = - a \left[ \partial_a g_a (u)\right] - g_a(u) \ .
\end{equation}
For $a=1,2$, $g_a(u)$ vanishes, and $\partial_a g_a (u)$ is  a long-range correlated eigenfunction of the RG flow. For $a=3,4,5, \ldots$  we can restrict our analysis to the 2-dimensional space spanned by $g_a(u)$ and $\partial_a g_a (u)$. It has a Jordan block structure and the 
general solution of the flow equation is 
\begin{equation}
\tilde g_a(u) = c_1(m) \left[ \partial_a g (u)\right]\Big|_{m=m_0} + c_2(m) g (u)\Big|_{m=m_0}
\end{equation}
with 
\begin{eqnarray}
c_1(m) &=& \left(\frac{m}{m_0}\right)^{\!\!a} c_1(m_0) \\
c_2(m) &=& \left(\frac{m}{m_0}\right)^{\!\!a} \left[ c_2(m_0) +\ln\left(\frac{m}{m_0}\right) c_1(m_0)\right]\ .\qquad\ \ 
\end{eqnarray}
Note that for  $a=3,4,5, \ldots$ $\partial_a g_a(u) \sim u^{-a}$ for $u \to \infty$, whereas $g_a(u)$ is short-ranged, as we discuss now:
\begin{eqnarray}
g_{a=3}(u) &=& \rme^{-u} \\
g_{a=4}(u) &=& \rme^{-u} (3 - u) \\
g_{a=5}(u) &=& \rme^{-u} (6 - u)(2- u) \ .
\end{eqnarray} 
The functions $g_a(u)$ for $a=1,2$ vanish. For negative $a$, there are polynomial solutions, consistent with the asymptotic behavior (\ref {asymp}), 
\begin{eqnarray}
g_{a=-1}(u) &=& 3+u  \\
g_{a=-2}(u) &=& \frac12 (2 + u) (6 + u) \\
g_{a=-3}(u) &=&\frac{u^3}{6}+\frac{5 u^2}{2}+10 u+10\ .
\end{eqnarray}
These solutions are unstable and physically unacceptable, since they grow stronger than $|u|$ at
large $u$. They correspond to models with even longer-ranged correlations than the
BFM.

For instance the leading short-ranged eigenmode $a=3$ reads
\begin{eqnarray}
&&\!\!\! \Delta(0)- \Delta(u) = \epsilon |u| + b (1-\rme^{-|u|} )\nn\\
&&= \int \frac{\rmd q}{2 \pi} \left( \frac{2\epsilon}{q^2} +  \frac{2b}{q^2+1} \right) [1-\cos(q u)] 
\end{eqnarray}
hence it has a positive Fourier transform (as long as $b>-\epsilon$), and thus corresponds to a physical disorder direction.

The question remains whether we have found {\em the complete spectrum for all physically allowed perturbations}. We argue that this is indeed the case: First of all, we have found a complete basis for  short-ranged perturbations, the functions $g_a(u)$ for integer $a$. Functions decaying as a {\em power-law} can be expanded in the basis $g_a(u)$ for non-integer $a$, or $\partial_a g_a(u)$ for integer $a$. As perturbations depending on $m$, they decay to 0, either as a power-law (non-integer $a$), or with additional logarithmic corrections (integer $a$). 
In conclusion, the BFM fixed point is stable w.r.t.\ perturbations 
of $R'''(u)$ which decay at least as a power law at infinity.

\subsection{More on loop expansion} \label{sec:loops} 

To obtain a deeper understanding of the properties of the BFM we must look at its replicated effective
action functional $\Gamma[u]$, with $u\equiv \{ u_a(x) \}_{a=1,\ldots, n;x \in {\mathbb {R}}^d}$. The statement that the (improved) tree level is exact means that 
for any replica field $u_a(x)$, 
\begin{eqnarray}  \label{prop}
\Gamma[u] = {\cal S}_R[u]\ .
\end{eqnarray}  The flow of the {\em exact} $R(u)$ was discussed above, with the claim
that $R'''(u)=\sigma \,{\rm sign}(u)$ does not flow (i.e.\ is independent of $m$), while 
$R''(0)$ flows, but its flow is unimportant. (It is part of the regularization
required to define the model). 

Let us examine the  meaning and validity of the property (\ref{prop}). The {\it exact} RG flow (as a function of $m$ in any $d$) of
the $p$-replica part of $\Gamma[u]$, $S^{(p)}(u_{12,\ldots,1p})$ was written in \cite{LeDoussal2008}, see Eqs.\ (384), (385) of the arXiv
version. We want to study the force correlator (corresponding in $d=0$ to the Burgers velocity) hence
look at $\partial_{u_1}\ldots\partial_{u_p} S^{(p)}(u_{12,\ldots, 1p}) \equiv \partial^p S^{(p)}$. By analogy with the second moment,
where we found that $\partial^3 S^{(2)}=R'''$ does not flow (away from coinciding arguments), 
we want to examine $\partial^{p+1} S^{(p)}$. If we assume that all $\partial^4 R$ and higher derivatives vanish,
it is clear from these equations that the feeding term for this quantity {\it vanishes}. The natural
conclusion is thus that $\partial^{p+1} S^{(p)}$ is zero for all $p \geq 3$ for the BFM model. 
Hence (\ref{prop}) holds {\it in the sense of derivatives} i.e.\ up to terms $O(u^p)$ in the
$p$-replica terms. These terms are called random force terms since they can be set to
zero by a STS transformation $u^a(x) \to u^a(x) + g_{xx'} f(x')$ where $f(x)$ is a (Larkin) random
force, coupling linearly to the displacement field. In $d=0$ this is
sufficient to ensure that the tree calculation is exact for the BFM model. Recently
we also showed this property  for the dynamics in $d=0$ \cite{LeDoussalWiese2011a}.

In $d>0$ one should worry about $x$ (i.e.\ space) dependent fields, i.e.\ non-local
parts of the functionals $S^{(p)}$. From Eq.\ (\ref{p2}), (see also (461) of \cite{LeDoussal2008}, arXiv
version) where the complete local+nonlocal two-replica functional $p=2$ is obtained
from the exact RG equations to order $R^2$, one finds for the BFM model
\begin{equation}  \label{rpp}
R''_{xy}[u_{ab}] = \sigma^2 g_{xy}^2 {\rm sign} \big(u_{ab}(x)\big) {\rm sign} \big(u_{ab}(y)\big) \ .
\end{equation} 
If we take a third derivative it vanishes away from the singular points. Thus naively there
is no nonlocal part for $\partial^3 S^{(2)}$ and the same will be true for
higher $p$. This leaves open the question of how the derivatives act
on the singular points (where two replica fields coincide for some values of
their argument $x$). We will not attempt to answer this question here, but leave it for future research. 
However,  we emphasize that the
fact that the tree level is exact for ``sufficiently reasonable", i.e.\ uniform or nearly uniform, field configurations is sufficient for computing e.g.\ center of mass observables using tree-level formulas. This can be seen
from (\ref{rpp}) since if $u_{ab}(x)$ does not change sign, i.e.\ the replica are in partially ordered configurations,
taking a third derivative again gives zero. Hence we can safely assume
that (\ref{prop}) holds in any $d$ for: (i) the needed derivatives of the $p$-replica part; (ii)
partially ordered configurations.  This is sufficient to argue that tree calculations
are exact for the BFM model in most applications.

\section{Toy model: Markovian and Poisson process for avalanches}\label{poisson}
\label{app:poisson}

In this Appendix we describe two simple toy models: (i) avalanche positions
being a Markov process, (ii)~avalanche positions and sizes being a 
totally uncorrelated process (Poisson process). 

First consider a Markovian model where the location $w_n$ of avalanche $n$ 
depends only on the previous one, with the ``waiting time", or interval
between avalanche $\ell=w_{n+1,n}=w_{n+1}-w_n$ 
distributed according to a distribution $Q(\ell)$ with $\int_{0}^\infty \rmd \ell\, Q(\ell)=1$. 
Given that a first avalanche occurs in $w_1$, the probability that the $n$ subsequent ones 
occur in $\prod_{i=2}^n [w_i , w_i +d w_i]$ is thus $Q(w_{n,n-1})\ldots Q(w_{2,1}) \rmd w_2\ldots \rmd w_n$,
also normalized to unity. Also assume statistical translation invariance 
with a uniform density of avalanches, noted $\rho_0$, hence a given avalanche can occur
anywhere with the same probability. 
 
For this model one shows that the probability that the interval $[0,w]$ with $w>0$
contains $n$ avalanches and that their positions are $0< w_1<w_2<\ldots <w_n< w$, is given for $n \geq 1$ by
\begin{eqnarray}  \label{pwn}
 p^{(n)}_w(w_1,\ldots, w_n) &=& \frac{1}{\left<\ell\right>} \int \rmd w_0 \rmd w_{n+1} \theta(-w_0) \\
 && \qquad \times \theta(w_{n+1}-w) \prod_{i=0}^{n} Q(w_{i+1,i}) \nn\\
 &=&  \frac{1}{{\left<\ell \right>}} \tilde Q(w_1) \prod_{i=1}^{n-1} Q(w_{i+1,i}) \tilde Q(w-w_n) \ ,\nn \end{eqnarray} 
with $\tilde Q(w)=\int_w^\infty \rmd\ell Q(\ell)$. To prove this one notes that the probability that the origin belongs to an interval of size $w_{1,0}$ is
$\frac{w_{1,0} Q(w_{1,0})}{{\left<\ell \right>}}$, hence the probability that the first positive shock occurs in $[w_1,w_1+\rmd w_1]$,
$w_1>0$, is $\frac{1}{{\left<\ell \right>}} \int \rmd w_0 \frac{\theta(w_{1,0}-w_1)}{w_{1,0}} \frac{w_{1,0} Q(w_{1,0})}{{\left<\ell \right>}}
=\frac{1}{{\left<\ell \right>}} \tilde Q(w_1)$. Inserting the factors
$1=\prod_{i=1}^\infty [\theta(w-w_i)+\theta(w_i-w)]$ and $\prod_{j=1}^\infty Q(w_{j+1,j})$ 
and expanding one gets (\ref{pwn}). The probabilities that there are $n$ avalanches
in $[0,w]$ are thus
\bea
 p^{(n)}&=&\int \rmd w_1\, \ldots \rmd w_n\, p^{(n)}_w(w_1,\ldots ,w_n) \\
  p^{(0)}_w&=& \frac{1}{{\left<\ell \right>}} \int_w^\infty \rmd w_1 \int_{w_1}^\infty \rmd w_{10} Q(w_{10}) \nn\\
  &=&  \frac{1}{{\left<\ell \right>}} \int_w^\infty \rmd w_1 (w-w_1)Q(w_{1}) \ .\label{pn}
\eea
These expression are easier written Laplace transformed, and one finds for $p_n(s) = \int_0^\infty \rmd w\,\rme^{-w s} p^{(n)}_w$
\begin{eqnarray} 
p_n(s) &=& \frac{1}{{\left<\ell \right>} s^2} (1- Q(s))^2 Q(s)^{n-1}\quad  \mbox{ for } n \geq 1\qquad  \\
 p_0(s)&=&\frac{1}{{\left<\ell \right>} s^2} (Q(s)-1+s {\left<\ell \right>}) 
\end{eqnarray}
with $Q(s) = \int_0^\infty \rmd w\,\rme^{-s w} Q(w)$ and ${\left<\ell \right>}=-Q'(0)$, which satisfy the normalization $\sum_{n=0}^\infty p_n(s)=1/s$, i.e.
$\sum_{n=0}^\infty p^{(n)}=1$. 

The case where avalanches are fully independent events, i.e.\ a Poisson process of density $\rho_0$, corresponds to 
$Q(w) = \rho_0 \rme^{- \rho_0 w}$ and ${\left<\ell \right>}=1/\rho_0$. Then $Q(s)=\frac{\rho_0}{s+\rho_0}$ 
and one finds $p_n(s)=\frac{\rho_0^n}{(s+\rho_0)^{n+1}}$ and, not surprisingly
\begin{eqnarray} 
p^{(n)}=\frac{1}{n!} (\rho_0 w)^n \rme^{- \rho_0 w} 
\end{eqnarray} 
for $n \geq 0$, i.e.\ the Poisson distribution for the number of shocks in the interval. More precisely one finds
\begin{equation} 
p^{(n)}_w(w_1,\ldots ,w_n) = \rho_0^n \rme^{- \rho_0 w} \theta(0<w_1<\ldots <w_n<w)\ ,
\end{equation} 
i.e.\ a uniform distribution for the positions of the shocks.

Let us now add information about avalanche sizes. Assume the process contains only positive jumps 
$u(w)-u(0)=\sum_\alpha S_\alpha \theta(0<w_\alpha < w)$, then
\begin{align} 
\lefteqn{\overline{\rme^{\lambda [u(w)-u(0)]} }}\\
& = \sum_{n=0}^\infty \int_{w_i,S_i} p_w^{(n)}(w_1,\ldots ,w_n;S_1,\ldots ,S_n) \rme^{\lambda (S_1+\ldots +S_n)}\ .\nn
\end{align}
If we further assume that the avalanche sizes are independent events uncorrelated with their location,
\begin{align} 
&p_w^{(n)}(w_1,\ldots ,w_n;S_1,\ldots ,S_n) \nn \\
&\qquad = p_w^{(n)}(w_1,\ldots ,w_n) \prod_{i=1}^n P(S_i)\ ,
\end{align}
where $P(S)$ is a normalized probability distribution, one finds
\begin{eqnarray} 
\overline{\rme^{\lambda [u(w)-u(0)]} } &=& \sum_{n=0}^\infty p_w^{(n)} \left< \rme^{\lambda S} \right>^n \\
\left< \rme^{\lambda S} \right> : &=&\int \rmd S\, P(S) \rme^{\lambda S}\ .
\end{eqnarray}
The case of a general $Q(w)$ can be solved in Laplace,
\begin{eqnarray} \label{I12}
&& \int_0^\infty \rmd w\,\rme^{-s w} \overline{\rme^{\lambda [u(w)-u(0)]} }\nn\\
&& = \frac{1}{s} + \frac{1}{{\left<\ell \right>} s^2} \frac{\left[1-Q(s)\right]\left[ \left<\rme^{\lambda S}\right>-1\right]}{1-
\left<\rme^{\lambda S}\right> Q(s)} \ .
\end{eqnarray}
For the Poisson process, we find that (\ref{I12}) simplifies into $(s + \rho_0 (1- \left<\rme^{\lambda S}\right>))^{-1}$, hence
\begin{eqnarray} 
\overline{\rme^{\lambda [u(w)-u(0)]} }&=& \rme^{\rho_0 w \int \rmd S\, P(S) (\rme^{\lambda S}-1) } \nn\\
&=& \rme^{w \int \rmd S\, \rho(S) (\rme^{\lambda S}-1) } = \rme^{w Z(\lambda)}\qquad 
\end{eqnarray}
in agreement with our general result (we have set $d=0$). 
Although here we have assumed $\rho_0$ finite (which is the case, e.g.\ in numerical
simulations \cite{RossoLeDoussalWiese2009a}) the above formula remains valid
when the total density of shocks is infinite as long as the density for a given size
$\rho(S)$ (also noted $n(s)$ in the text) is finite and $\int \rmd S\, S \rho(S)$ is finite. 
One then recovers the Levy process formula for the case of only positive jumps. 
For a proper mathematical formulation see \cite{Bertoin1998, CarraroDuchon1998}. 

We observe that in the Poisson case (for $d=0$)  $Z(\lambda) \to - \rho_0 L^{-d} $  at
large negative $\lambda$. This is dominated by the probability that there is no avalanche in the interval $w$. 
This limit can also be written as $Z(- \infty)=- 1/\langle S \rangle$. Hence the mean-field formula
(\ref{16}) is valid only for $\lambda > - 1/S_{\mathrm{min}}$, where $S_{\mathrm{min}}$ is a typical small-scale cutoff
for the avalanche size, as discussed in  Sec.~V.E of \cite{LeDoussalWiese2008c}.

\section{Periodic case} \label{a:periodic}

We now solve the equations (\ref{multicont}) and (\ref{Scont}) for the case where $R(u)$ is periodic.
It is sufficient to choose $\lambda(w)= - \mu'(w)$
on the interval $[0,1]$ with $\mu(0)=\mu(1)$. Indeed since 
in the periodic case we know that $u_1(w)-w$ is periodic of period $1$, 
we can  write:
\begin{align}
& \overline{\left<\rme^{ \int_w \lambda(w) [u_1(w) - w)] } \right>} = \overline{\left<\rme^{ \int_0^1 \rmd w\,[u_1(w) - w)] \tilde \lambda(w)} \right>}\qquad \\
&  \sum_n \lambda(w+n) = \tilde \lambda(w)
\end{align}
with $\int_0^1 \rmd w\,\tilde \lambda(w)=0$. Using $u_1(w+n)=u_1(w)+n$ one finds for $w \in [0,1]$
that the equations (\ref{multicont}) and (\ref{Scont}) hold where all integrals are over $w' \in [0,1]$
and $U(w) \to \tilde U(w)$. We define again for $w \in [0,1]$ 
\begin{eqnarray}
\tilde \lambda(w) = - \mu'(w) \quad , \quad \tilde U(w) = V'(w)\ .
\end{eqnarray} 
Note that since $\tilde \lambda(0)= \tilde \lambda(1)$ and $\int_0^1 \rmd w\,\tilde \lambda(w)=0$ one has
$\mu(0)=\mu(1)$ and similarly $V(0)=V(1)$. The second equation in (\ref{multicont}) gives after integration by part
\begin{eqnarray}
&&\!\!\!V'(w) \left[ m^2 + \int_0^1 \rmd w' R''''\big(u_1(w)-u_1(w')\big) u_1'(w')  V(w')\right] \nn\\
&& = - \mu'(w)\ .
\end{eqnarray} 
The boundary term $[R'''(u_1(w)-u_1(w')) V(w')]_0^1$
vanishes only if the integral goes from $0^-$ to $1^-$, i.e.\ contains the delta function at $0$, a convention which we use here.
Noting that for the periodic case \beq
\label{J5}
R''''(u)=R''''(0) + \sum_n 2 R'''(0^+) \delta(u-n)
\eeq
we obtain for $w \in [0,1]$:
\begin{eqnarray}
&& \Big[ m^2 + R''''(0)  \int_0^1 \rmd w'  u_1'(w')  V(w') + 2 R'''(0^+) V(w)\Big]\times \nn\\
&& \qquad \times  V'(w) = - \mu'(w)\ .
\end{eqnarray}
Integration yields
\begin{eqnarray}
&&\!\!\!V(w) \Big[ m^2 + R''''(0)  \int_0^1 \rmd w'  u_1'(w')  V(w')\Big] \nn\\
&& + R'''(0^+) V(w)^2 = - \mu(w)\ .
\end{eqnarray}
 (Note that a possible integration constant has been dropped since it does not enter any physical observable.)
The first equation in (\ref{multicont}) gives after integration by part
\begin{eqnarray}\label{J8}
&&\!\!\! m^2 [u_1(w)-w] \nn\\
&&+ \int_0^1 \rmd w' R'''\big(u_1(w)-u_1(w')\big) u_1'(w') V(w') = 0 \ .\qquad
\end{eqnarray}
Taking a derivative and using again Eq.~(\ref{J5}), we finally arrive at the system of two equations:
\begin{align}\label{J9}
& u_1'(w) \\
&= \frac{m^2}{m^2 + R''''(0) \int_0^1 \rmd w'  u_1'(w') V(w') + 2 R'''(0^+) V(w) }\nn \\
& V(w) \left[ m^2 + R''''(0)  \int_0^1 \rmd w'  u_1'(w')  V(w')\right]\nn\\
&\qquad   + R'''(0^+) V(w)^2 = - \mu(w)\ . \label{J10}
\end{align}
Noting $\sigma = R'''(0)$ and $r_4 = R''''(0)$ we obtain, multiplying $u_1'(w)$ from (\ref{J9}) with $V(w)$  given by (\ref{J10})
\begin{equation}\label{J11}
u_1'(w) V(w)  = \frac{- m^2 \mu(w)}{[m^2 + r_4 A + 2 \sigma V(w)][m^2 + r_4 A + \sigma V(w)]}.
\end{equation}
 We have defined $A=\int_0^1 \rmd w'\,  u'(w') V(w')$.  We can now close the system of equations.
This gives the result quoted in the text with $r_4=-2 \sigma$. The action becomes
\begin{eqnarray} 
  -S_\lambda &=& \int_0^1 \rmd w\,\Big[-\mu'(w) - \frac{m^2}{2} V'(w)\Big] [u_1(w)-w]  \nn\\
&=&  \int_0^1 \rmd w\,\Big[\mu(w) + \frac{m^2}{2} V(w)\Big] [u_1'(w)-1] \nn \\
& =& - \int_0^1 \rmd w\, \left[\mu(w) + m^2 V(w)\right] - \frac{1}{2} r_4 A^2
\end{eqnarray}
as quoted in the text.

\tableofcontents 


\begin{thebibliography}{10}

\bibitem{UrbachMadisonMarkert1995}
J.~S. Urbach, R.~C. Madison  and J.~T. Markert,
\newblock {\em Interface depinning, self-organized criticality, and the
  {Barkhausen} effect},
\newblock Phys. Rev. Lett. {\bf 75} (1995)   276--279.

\bibitem{FisherDahmenRamanathanBen-Zion1997}
D.S. Fisher, K.~Dahmen, S.~Ramanathan  and Y.~Ben-Zion,
\newblock {\em Statistics of earthquakes in simple models of heterogeneous
  faults},
\newblock Phys. Rev. Lett. {\bf 78} (1997)   4885--4888.

\bibitem{DSFisher1998}
D.S. Fisher,
\newblock {\em Collective transport in random media: {From} superconductors to
  earthquakes},
\newblock Phys. Rep. {\bf 301} (1998)   113--150.

\bibitem{MoulinetGuthmannRolley2002}
S.~Moulinet, C.~Guthmann  and E.~Rolley,
\newblock {\em Roughness and dynamics of a contact line of a viscous fluid on a
  disordered substrate},
\newblock Eur. Phys. J. E {\bf 8} (2002)   437--443.

\bibitem{MoulinetRossoKrauthRolley2004}
S.~Moulinet, A.~Rosso, W.~Krauth  and E.~Rolley,
\newblock {\em Width distribution of contact lines on a disordered substrate},
\newblock Phys. Rev. E {\bf 69} (2004)   035103,
\newblock cond-mat/{\bf 0310173}.

\bibitem{CuleHwa1998}
D.~Cule and T.~Hwa,
\newblock {\em Static and dynamic properties of inhomogeneous elastic media on
  disordered substrate},
\newblock Phys. Rev. B {\bf 57} (1998)   8235--53.

\bibitem{BonamySantucciPonson2008}
D.~Bonamy, S.~Santucci  and L.~Ponson,
\newblock {\em Crackling dynamics in material failure as the signature of a
  self-organized dynamic phase transition},
\newblock Phys. Rev. Lett. {\bf 101} (2008)   045501.

\bibitem{SchmittbuhlMaloy1997}
J.~Schmittbuhl and K.J. Maloy,
\newblock {\em Direct observation of a self-affine crack propagation},
\newblock Phys. Rev. Lett. {\bf 78} (1997)   3888--91.

\bibitem{BlatterFeigelmanGeshkenbeinLarkinVinokur1994}
G.~Blatter, M.V. {Feigel'man}, V.B. Geshkenbein, A.I. Larkin  and V.M. Vinokur,
\newblock {\em Vortices in high-temperature superconductors},
\newblock Rev. Mod. Phys. {\bf 66} (1994)   1125.

\bibitem{LeDoussal2010Book}
Pierre~Le Doussal,
\newblock {\em Novel phases of vortices in superconductors},
\newblock in BCS: 50 years, L.N. Cooper and D. Feldman Editors World
  Scientific, Int. Journal of Modern Physics B, 24 (2010) 3855-3914. (2010).

\bibitem{DSFisher1986}
D.S. Fisher,
\newblock {\em Interface fluctuations in disordered systems: {$5-\epsilon$}
  expansion},
\newblock Phys. Rev. Lett. {\bf 56} (1986)   1964--97.

\bibitem{WieseLeDoussal2006}
K.J. Wiese and P.~Le Doussal,
\newblock {\em Functional renormalization for disordered systems: Basic recipes
  and gourmet dishes},
\newblock Markov Processes Relat. Fields {\bf 13} (2007)   777--818,
\newblock cond-mat/{\bf 0611346}.

\bibitem{NarayanDSFisher1993a}
O.~Narayan and D.S. Fisher,
\newblock {\em Threshold critical dynamics of driven interfaces in random
  media},
\newblock Phys. Rev. B {\bf 48} (1993)   7030--42.

\bibitem{NattermannStepanowTangLeschhorn1992}
T.~Nattermann, S.~Stepanow, L.-H. Tang  and H.~Leschhorn,
\newblock {\em Dynamics of interface depinning in a disordered medium},
\newblock J. Phys. II (France) {\bf 2} (1992)   1483--8.

\bibitem{ChauveLeDoussalWiese2000a}
P.~Chauve, P.~Le Doussal  and K.J. Wiese,
\newblock {\em Renormalization of pinned elastic systems: How does it work
  beyond one loop?},
\newblock Phys. Rev. Lett. {\bf 86} (2001)   1785--1788,
\newblock cond-mat/{\bf 0006056}.

\bibitem{LeDoussalWieseChauve2002}
P.~Le Doussal, K.J. Wiese  and P.~Chauve,
\newblock {\em 2-loop functional renormalization group analysis of the
  depinning transition},
\newblock Phys. Rev. B {\bf 66} (2002)   174201,
\newblock cond-mat/{\bf 0205108}.

\bibitem{LeDoussalWieseChauve2003}
P.~Le Doussal, K.J. Wiese  and P.~Chauve,
\newblock {\em Functional renormalization group and the field theory of
  disordered elastic systems},
\newblock Phys. Rev. E (2004)   026112,
\newblock cond-mat/{\bf 0304614}.

\bibitem{KoltonRossoGiamarchi2005}
A.B. Kolton, A.~Rosso  and T.~Giamarchi,
\newblock {\em Creep motion of an elastic string in a random potential},
\newblock Phys. Rev. Lett. {\bf 94} (2005)   047002,
\newblock cond-mat/{\bf 0408284}.

\bibitem{LeDoussalMiddletonWiese2008}
P.~{Le~Doussal}, A.A. Middleton  and K.J.\ Wiese,
\newblock {\em Statistics of static avalanches in a random pinning landscape},
\newblock Phys. Rev. E {\bf 79} (2009)   050101 (R),
\newblock arXiv:{\bf 0803.1142}.

\bibitem{LeDoussalWiese2008c}
P.~{Le~Doussal} and K.J. Wiese,
\newblock {\em Size distributions of shocks and static avalanches from the
  functional renormalization group},
\newblock Phys. Rev. E {\bf 79} (2009)   051106,
\newblock arXiv:{\bf 0812.1893}.

\bibitem{Burgers74}
J.M. Burgers,
\newblock {\em The non-linear diffusion equation},
\newblock Dordrecht, 1974.

\bibitem{BecKhanin2007}
Jeremie Bec and Konstantin Khanin,
\newblock {\em Burgers turbulence},
\newblock Phys. Rep. {\bf 447} (2007)   1--66.

\bibitem{BalentsBouchaudMezard1996}
L.~Balents, J.P. Bouchaud  and M.~M\'ezard,
\newblock {\em The large scale energy landscape of randomly pinned objects},
\newblock J. Phys. I (France) {\bf 6} (1996)   1007--20.

\bibitem{LeDoussal2006b}
P.~{Le Doussal},
\newblock {\em Finite temperature {Functional RG}, droplets and decaying
  {Burgers} turbulence},
\newblock Europhys. Lett. {\bf 76} (2006)   457--463,
\newblock cond-mat\slash{\bf 0605490}.

\bibitem{LeDoussal2008}
P.~{Le Doussal},
\newblock {\em Exact results and open questions in first principle {functional
  RG}},
\newblock Annals of Physics {\bf 325} (2009)   49--150,
\newblock arXiv:{\bf 0809.1192}.

\bibitem{LeDoussalMuellerWiese2010}
P.~Le Doussal, M.~M\"uller  and K.J. Wiese,
\newblock {\em Avalanches in mean-field models and the {Barkhausen} noise in
  spin-glasses},
\newblock EPL {\bf 91} (2010)   57004,
\newblock arXiv:{\bf 1007.2069}.

\bibitem{LeDoussalMuellerWiese2011}
P.~Le Doussal, M.~M\"uller  and K.J. Wiese,
\newblock {\em Equilibrium avalanches in spin glasses},
\newblock arXiv:\null {\bf 1110.2011} (2011).

\bibitem{RossoLeDoussalWiese2009a}
A.~Rosso, P.~{Le~Doussal}  and K.J.\ Wiese,
\newblock {\em Avalanche-size distribution at the depinning transition: A
  numerical test of the theory},
\newblock Phys. Rev. B {\bf 80} (2009)   144204,
\newblock arXiv:{\bf 0904.1123}.

\bibitem{LeDoussalWiese2011a}
P.~{Le Doussal} and K.J. Wiese,
\newblock {\em Distribution of velocities in an avalanche},
\newblock arXiv:\null {\bf 1104.2629} (2011).

\bibitem{KoltonLeDoussalWiesetobe}
A.~Kolton, P.~{Le Doussal}  and K.J. Wiese,
{\em Distribution of velocities in an avalanche: Numerics}
\newblock  to be published.

\bibitem{LeDoussalWiesetobe}
P.~{Le Doussal}  and K.J. Wiese,
{\em Distribution of velocities in an avalanche: The details}
\newblock to be published.

\bibitem{DobrinevskiLeDoussalWiesetobe}
A.~Dobrinevski, P.~{Le Doussal}  and K.J. Wiese,
{\em Non-stationary dynamics of the ABBM model}
\newblock to be published.

\bibitem{LeDoussalWiese2010c}
P.~{Le Doussal} and K.J. Wiese,
\newblock {\em Avalanches for a {$N$}-component field},
\newblock to be published.

\bibitem{LeDoussalRossoWiese2011}
P.~Le Doussal, A.~Rosso  and K.J. Wiese,
\newblock {\em Shock statistics in higher-dimensional {Burgers} turbulence},
\newblock EPL (2011)   14005,
\newblock arXiv:{\bf 1104.5048}.

\bibitem{ABBM}
B.~Alessandro, C.~Beatrice, G.~Bertotti  and A.~Montorsi,
\newblock {\em Domain-wall dynamics and {Barkhausen} effect in metallic
  ferromagnetic materials. {I. Theory}},
\newblock J. Appl. Phys. {\bf 68} (1990)   2901.

\bibitem{CarraroDuchon1998}
L.~Carraro and J.~Duchon,
\newblock {\em {\'Equation de {Burgers} avec conditions initiales \`a
  accroissements ind\'ependants et homog\`enes}},
\newblock Ann. Inst. Henri Poincar\'e {\bf 15} (1998)   431--458.

\bibitem{Bertoin1998}
J.~Bertoin,
\newblock {\em The inviscid {Burgers} equation with {Brownian} initial
  velocity},
\newblock Commun. Math. Phys. {\bf 193} (1998)   397--406.

\bibitem{Polyakov1995}
AM. Polyakov,
\newblock {\em Turbulence without pressure},
\newblock Phys. Rev. E {\bf 52} (1995)   6183--6188,
\newblock hep-th/{\bf 9506189}.

\bibitem{BouchaudMezardParisi1995}
J.P. Bouchaud, M.~Mezard  and G.~Parisi,
\newblock {\em Scaling and intermittency in {Burgers} turbulence},
\newblock Phys. Rev. E {\bf 52} (1995)   3656--3674,
\newblock cond-mat/{\bf 9503144}.

\bibitem{BernardGawedzki1998}
D.~Bernard and K.~Gawedzki,
\newblock {\em Scaling and exotic regimes in decaying {Burgers} turbulence},
\newblock J. Phys. A {\bf 31} (1998)   8735.

\bibitem{Valageas2009}
P.~Valageas,
\newblock {\em Some statistical properties of the {Burgers} equation with
  white-noise initial velocity},
\newblock J. Stat. Phys {\bf 137} (2009)   729,
\newblock arXiv:{\bf 0903.0956}.

\bibitem{ChabanolDuchon2004}
M.-L. Chabanol and J.~Duchon,
\newblock {\em Markovian solutions of inviscid {Burgers} equation},
\newblock J. Stat. Phys. {\bf 114} (2004)   525--534,
\newblock arXiv:nlin/{\bf 0302016}.

\bibitem{ChabanolDuchon2009}
M.-L. Chabanol and J.~Duchon,
\newblock {\em Levy solutions of a randomly forced {Burgers} equation},
\newblock arXiv:\null {\bf 0904.3397} (2009).

\bibitem{LeDoussalWieseMoulinetRolley2009}
P.~Le Doussal, K.J. Wiese, S.~Moulinet  and E.~Rolley,
\newblock {\em Height fluctuations of a contact line: {A} direct measurement of
  the renormalized disorder correlator},
\newblock EPL {\bf 87} (2009)   56001,
\newblock arXiv:{\bf 0904.1123}.

\bibitem{LeDoussalWiese2009a}
P.~{Le Doussal} and K.J. Wiese,
\newblock {\em Elasticity of a contact-line and avalanche-size distribution at
  depinning},
\newblock Phys. Rev. E {\bf 82} (2010)   011108,
\newblock arXiv:{\bf 0908.4001}.

\bibitem{MiddletonLeDoussalWiese2006}
A.A. Middleton, P.~{Le~Doussal}  and K.J. Wiese,
\newblock {\em Measuring functional renormalization group fixed-point functions
  for pinned manifolds},
\newblock Phys. Rev. Lett. {\bf 98} (2007)   155701,
\newblock cond-mat/{\bf 0606160}.

\bibitem{LeDoussalWiese2003b}
P.~Le Doussal and K.J. Wiese,
\newblock {\em Functional renormalization group at large ${N}$ for disordered
  elastic systems, and relation to replica symmetry breaking},
\newblock Phys. Rev. B {\bf 68} (2003)   174202,
\newblock cond-mat/{\bf 0305634}.

\bibitem{LeDoussalWiese2001}
P.~Le Doussal and K.J. Wiese,
\newblock {\em Functional renormalization group at large {$N$} for random
  manifolds},
\newblock Phys. Rev. Lett. {\bf 89} (2002)   125702,
\newblock cond-mat/{\bf 0109204v1}.

\bibitem{ChauveLeDoussal2001}
P.~Chauve and P.~Le Doussal,
\newblock {\em Exact multilocal renormalization group and applications to
  disordered problems},
\newblock Phys. Rev. E {\bf 64} (2001)   051102/1--27,
\newblock cond-mat/{\bf 0006057}.

\bibitem{FedorenkoLeDoussalWiese2006}
A.~Fedorenko, P.~{Le~Doussal}  and K.J. Wiese,
\newblock {\em Universal distribution of threshold forces at the depinning
  transition},
\newblock Phys. Rev. E {\bf 74} (2006)   041110,
\newblock cond-mat/{\bf 0607229}.

\bibitem{Burkhardt1993}
T.W. Burkhardt,
\newblock {\em Semiflexible polymer in the half plane and statistics of the
  integral of a brownian curve},
\newblock J. Phys. A {\bf 26} (1993)   L1157.


\bibitem{MajumdarRossoZoia2010}
S.~N. Majumdar, A.~Rosso  and A.~Zoia,
\newblock {\em Time at which the maximum of a random acceleration process is
  reached},
\newblock J. Phys. A {\bf 43} (2010)   115001.

\bibitem{WieseMajumdarRosso2010}
K.J. Wiese, S.N. Majumdar  and A.~Rosso,
\newblock {\em Perturbation theory for fractional {Brownian} motion in presence
  of absorbing boundaries},
\newblock Phys. Rev. E {\bf 83} (2011)   061141,
\newblock arXiv:{\bf 1011.4807}.

\bibitem{LeDoussalWiese2008a}
P.~Le Doussal and K.J. Wiese,
\newblock {\em Driven particle in a random landscape: disorder correlator,
  avalanche distribution and extreme value statistics of records},
\newblock Phys. Rev. E {\bf 79} (2009)   051105,
\newblock arXiv:{\bf 0808.3217}.

\bibitem{Wiese1998a}
K.J. Wiese,
\newblock {\em On the perturbation expansion of the {KPZ}-equation},
\newblock J. Stat. Phys. {\bf 93} (1998)   143--154,
\newblock cond-mat/{\bf 9802068}.

\bibitem{BertoinBook}
J.~Bertoin,
\newblock {\em L\'evy Processes},
\newblock Cambridge University Press, 1998.

\bibitem{DobrinevskiLeDoussalWiese2011}
A.~Dobrinevski, P.~{Le Doussal}  and K.J. Wiese,
\newblock {\em Interference in disordered systems: A particle in a complex
  random landscape},
\newblock Phys. Rev. E {\bf 83} (2011)   061116,
\newblock arXiv:{\bf 1101.2411}.

\bibitem{LeDoussalWiese2010b}
P.~{Le Doussal} and K.J. Wiese,
\newblock {\em Dynamics of avalanches},
\newblock to be published (2011).

\bibitem{RossoKrauthLeDoussalVannimenusWiese2003}
A.~Rosso, W.~Krauth, P.~Le Doussal, J.~Vannimenus  and K.J. Wiese,
\newblock {\em Universal interface width distributions at the depinning
  threshold},
\newblock Phys. Rev. E {\bf 68} (2003)   036128,
\newblock cond-mat{\slash\bf 0301464}.

\bibitem{LeDoussalWiese2003a}
P.~Le Doussal and K.J. Wiese,
\newblock {\em Higher correlations, universal distributions and finite size
  scaling in the field theory of depinning},
\newblock Phys. Rev. E {\bf 68} (2003)   046118,
\newblock cond-mat/{\bf 0301465}.

\bibitem{LeDoussalWiese2005a}
P.~Le Doussal and K.J.\ Wiese,
\newblock {\em 2-loop functional renormalization for elastic manifolds pinned
  by disorder in {$N$} dimensions},
\newblock Phys. Rev. E {\bf 72} (2005)   035101 (R),
\newblock cond-mat/{\bf 0501315}.

\end{thebibliography}
\end{document}